\documentclass[preprint, 11pt]{elsarticle}

\usepackage{amssymb,amsfonts}
\usepackage{graphicx}
\usepackage{epsfig}
\usepackage{epstopdf}
\usepackage{bm}
\usepackage{amssymb} 
\usepackage{amsmath, amsthm} 
\usepackage{graphicx} 
\usepackage{amsgen, amsfonts, amsbsy} 
\usepackage{mathrsfs} 
\usepackage{color}
\usepackage[normalem]{ulem}

\newcommand{\be}{\begin{equation}}
\newcommand{\ee}{\end{equation}}
\newcommand{\bea}{\begin{eqnarray}}

\newcommand{\eea}{\end{eqnarray}}

\journal{Physics Reports}

\def\inbar{\,\vrule height1.5ex width.4pt depth0pt}
\def\IR{\relax{\rm I\kern-.18em R}}
\def\IC{\relax\hbox{$\inbar\kern-.3em{\rm C}$}}

\begin{document}

\begin{frontmatter}

\title{Spherical inhomogeneous solutions of Einstein and scalar-tensor gravity: a map of the land}

\author[BU]{Valerio Faraoni\corref{CorrAuth}}
\ead{vfaraoni@ubishops.ca}

\author[BU,ETH]{Andrea Giusti}

\author[BU]{Bardia H. Fahim}

\address[BU]{Department of Physics \& Astronomy, Bishop's University, 2600 College St., Sherbrooke, Qu\'{e}bec, Canada J1M~1Z7} 											
\address[ETH]{Institute for Theoretical Physics, ETH Zurich, Wolfgang-Pauli-Strasse 27, 8093, Zurich, Switzerland}

\cortext[CorrAuth]{Corresponding author}

\begin{abstract}
We review spherical and inhomogeneous analytic solutions of the field 
equations of Einstein and of scalar-tensor gravity, including Brans-Dicke 
theory, non-minimally (possibly conformally) coupled scalar fields,  
Horndeski, and beyond Horndeski/DHOST gravity. The zoo 
includes both static and dynamic solutions, 
asymptotically flat, and asymptotically 
Friedmann-Lema\^itre-Robertson-Walker ones. We minimize overlap with 
existing books and reviews and we place emphasis on scalar field 
spacetimes and on geometries that are ``general'' within certain classes. 
Relations between various solutions, which have largely emerged during the 
last decade, are pointed out.
\end{abstract}


\end{frontmatter}

\newpage

\tableofcontents

\newpage
\section{Introduction} 
\label{sec:1} 

We have now known general relativity (GR) and its Schwarzschild solution describing a 
spherical isolated object for over a century, during which GR has been put to the test 
and has found astrophysical applications that were unconceivable at its beginnings. In the 
last few decades, other analytic solutions of Einstein 
theory have been discovered and the theory itself has been challenged (questioning GR, 
however, started early on with Weyl \cite{Weyl19} and Eddington \cite{Eddington23}). 
Attempts to quantize gravity and one-loop renormalization necessarily lead to 
deviations from the Einstein equations and to the introduction of extra degrees of 
freedom, extra fields, or higher order curvature terms \cite{UtyamaDeWitt62, Stelle77, 
Stelle78}.  Historically, the first model of inflation in the early 
universe, the Starobinsky model \cite{Starobinsky80}, incorporated some of these 
quadratic corrections by promoting the Einstein-Hilbert Lagrangian ${\cal R}$ (where 
${\cal R}$ is the Ricci scalar) to ${\cal R}+\alpha \, {\cal R}^2$. It is now known 
that this correction is equivalent to introducing a massive scalar field in addition to 
gravity. So, at least the Starobinsky inflationary scenario includes corrections to the 
Einstein-Hilbert action \cite{Starobinsky80}. 

The surprising discovery, made with Type Ia supernovae \cite{Riesseal98, Riessetal99, 
Riessetal04, Eisensteinetal05, Astieretal06, Spergeletal07} that the present expansion 
of the universe is accelerated has led cosmologists to postulate a completely {\em ad 
hoc} dark energy, which is possibly a cosmological constant $\Lambda$ suffering from 
the notorious cosmological constant problems \cite{Weinberg89, Carroll2001a}. 
Quintessence models of dark energy (\cite{ PeeblesRatra88, RatraPeebles1988, 
Wetterich1988, OstrikerSteinhardt1995, Caldwelletal1998, Carroll1998, Bahcalletal1999, 
Wangetal2000}, see \cite{AmendolaTsujikawabook} for a review) are not well-motivated 
theoretically. As a consequence, although fitting most data, the standard model of 
cosmology, the $\Lambda$-Cold Dark Matter ($\Lambda$CDM) model, appears unnatural. In 
fact, if the cosmic acceleration is attributed to a cosmological constant, it suffers 
from the fine-tuning and coincidence problems \cite{AmendolaTsujikawabook}.  
If $\Lambda$CDM is taken to 
mean cold dark matter plus dark energy, it remains problematic because dark energy was 
introduced {\em ad hoc} in the theoretical literature to explain the cosmic 
acceleration and its nature is completely unknown, as testified by the plethora of 
vastly different dark energy models in the literature. In this sense, its introduction 
seems unnatural. In any case, the presence of a speculative and exotic dark energy of 
completely unknown nature remains deeply unsatisfactory.
 
Starting in 2003 \cite{CCT,CDTT}, this situation 
has driven many authors to explain the present cosmic acceleration by 
modifying Einstein gravity at large scales instead of invoking an {\em ad 
hoc} dark energy. This is the most pressing need to explore modified 
theories of gravity, among which the so-called $f({\cal R})$ theories 
(\cite{CCT,CDTT}, see \cite{review1,review2, review3} for reviews) 
are very popular. This fact has caused a 
resurgence of interest in 
modified gravity and in particular in scalar-tensor gravity, a broader 
category known since at least 1961 \cite{BD} in which $f({\cal R})$ 
theories fall as a subfamily. The discovery of gravitational waves with 
the {\em LIGO} and {\em VIRGO} laser interferometric detectors 
\cite{LIGO1,LIGO2,LIGO3} in 2015 and the 2019 Event Horizon Telescope 
imaging of the near-horizon region of the M87 black hole \cite{EHT1,EHT2, 
EHT3, EHT4, EHT5, EHT6} have provided unprecedented tools and motivation 
to study the detailed physics near black holes and compact objects in both 
GR and alternative gravity. The 
comparison of observations and theoretical predictions requires the 
knowledge of analytical solutions in these theories and of the 
phenomenology of test particles (massive and massless) and fluids in 
these geometries. Without  these background geometries it is impossible to compute 
strong field effects and these solutions must necessarily be stable ones. What is 
more, due to the unexpectedly high black hole masses measured in the {\em LIGO} and 
{\em 
VIRGO} events, primordial black holes have attracted again attention 
\cite{SasakiSuyamaTanaka16, CarrSilk18}. Their importance is that, if they 
abound in the universe, they can be significant dark matter components. 
This motivates the study of toy models ({\em i.e.}, analytic solutions) in 
which such objects, if they exist, evolve on time scales not completely 
negligible with respect to the Hubble scale ({\em e.g.}, 
\cite{Boehm:2020jwd,RuizMolinaLima20}).

In summary, this renewed activity in cosmology and astrophysics has led 
to 
the need of a better understanding of the physics of black holes and other 
objects of astrophysical or mathematical interest in these theories, and 
to deeper studies of their analytic solutions. The search for analytic 
solutions of a physical theory unveils its physical aspects and 
promotes 
physical insight. Exact solutions of the field equations of non-linear 
theories are difficult to find, and they are usually found only by 
simplifying the problem which, in practice, means by assuming symmetries. 
Free from as many complications as possible, it is easier to gain physical 
insight by examining symmetric solutions which allow one to focus on a 
few aspects of the theory at a time. Certainly this has been the case for 
the Schwarzschild solution of GR, which has led to the discovery of black 
holes and to the investigation of many astrophysical and fundamental 
physics 
phenomena and, indirectly (through black hole mergers), to the discovery 
of 
gravitational waves. When trying to make sense of a theory of gravity, one 
of the first aspects to examine consists of the available exact solutions 
of its 
field equations and, among these, its spherically symmetric solutions 
describing black holes, stars, and possibly other objects. It is 
unavoidable to compare these solutions with the corresponding one of GR, 
{\em i.e.}, the Schwarzschild geometry \cite{Schwarzschild16}.

Historically, there have been attempts to find exact spherically symmetric 
solutions of the field equations of theories of gravity describing 
increasingly complicated physical situations: vacuum, vacuum with a 
cosmological constant $\Lambda$, electrovacuum, perfect fluids, imperfect 
fluids, scalar fields in the context of GR, and then situations in which 
the scalar field has gravitational nature and is an unavoidable ingredient 
of the theory of gravity, as in scalar-tensor theories. In practice, 
solving the field equations directly is challenging and many attempts to 
generate new analytic solutions have relied on the knowledge of previously 
existing solutions in different physical contexts. These studies have 
unveiled relations between different geometries and between different 
theories (for example, solution-generating techniques have employed 
conformal transformations relating GR with scalar-tensor gravity through 
the use of different conformal frames).

Here we review analytic solutions of GR (part~I) and of scalar-tensor 
gravity (part~II) which are spherically symmetric, asymptotically or 
non-asymptotically flat, static or dynamical. We will not repeat the 
discussion of the most well-known solutions that are found in 
standard books and reviews (especially those sourced by a perfect or an 
imperfect fluid), nor will we discuss at length the mathematical structure 
of the solutions, their algebraic classification, conformal 
diagrams, or their causal structure. For these aspects, we refer the 
reader to the monumental 
work~\cite{StephaniKramerMacCallumHoenselaersHerlt} on analytic solutions of 
the Einstein equations, and to \cite{Krasinski97, GriffithsPodolsky09, 
AHbook, Faraoni:2018xwo, TretyakovaLatosh18} for other aspects, more 
detailed discussions, and recent developments. Here, instead, we focus on 
exact solutions within certain classes, pointing out the relations between 
various families of exact solutions that have emerged during the last few 
decades.

Of course, one can write down a metric tensor and run the Einstein 
equations from left to right \`a la Synge, computing an effective 
energy-momentum tensor which, in general, violates all the energy 
conditions \cite{Waldbook} and is completely unphysical. We do not 
consider such solutions unless the effective energy momentum tensor that 
sources them is physically reasonable.

Solutions containing a scalar field occupy a prominent position in this 
review. This scalar could be a matter field in Einstein theory (as would 
be, for example, the recently discovered Higgs boson of the Standard Model of 
particle 
physics \cite{Higgs1, Higgs2}). Alternatively, the scalar field could be a 
gravitational field in an alternative theory of gravity, such as 
the Brans-Dicke or the dilaton fields. Among the wide spectrum of 
alternative theories of gravity, we restrict ourselves to Brans-Dicke 
gravity \cite{BD}, its scalar-tensor generalizations \cite{Bergmann68, 
Wagoner70, Nordtvedt70} including the wide class of $f({\cal R})$ 
theories \cite{review1, review2, review3}, Horndeski gravity 
\cite{Horndeski:1974wa}, and degenerate higher order scalar-tensor theories 
\cite{Gleyzes:2014dya,Gleyzes:2014qga}. We 
exclude torsion and, in order to keep the material contained, we restrict 
to spherically symmetric solutions of the field equations and to four 
spacetime dimensions (for higher-dimensional solutions, 
see~\cite{EmparanReall08,Horowitzbook} and the references therein).

Certain dynamical solutions in this review are believed to represent black 
holes. In the dynamical case the black hole boundary, and the black hole 
concept itself, are not defined by the usual notion of event horizon, 
which is a null surface and a causal boundary ({\em i.e.}, a connected 
component of the boundary of the causal past of future null infinity 
\cite{Waldbook}). The teleological notion of event horizon requires the 
knowledge of the entire causal structure of spacetime and is impractical 
in dynamical situations and in numerical relativity. Instead, dynamical 
black holes are defined quasilocally by apparent horizons and marginally 
trapped surfaces. A drawback of this definition (see 
\cite{Booth:2005qc,Nielsen:2008cr,AHbook,Faraoni:2018xwo} for reviews) is 
that apparent horizons depend on the choice of the spacetime foliation, as 
exemplified dramatically by the fact that the Schwarzschild spacetime 
admits non-symmetric foliations without apparent horizons 
\cite{Wald:1991zz,Schnetter:2005ea}. However, in the presence of spherical 
symmetry (to which we restrict here), all spherically symmetric foliations 
produce the same apparent horizons \cite{Faraoni:2016xgy}.

We follow the notation of Ref.~\cite{Waldbook}. The metric signature is 
$({-}{+}{+}{+})$, we use geometric units in which the speed of light $c$ and 
Newton's constant $G$ assume the values $c=G=1$, $\kappa \equiv 8\pi G$, 
and $\Box \equiv g^{ab}\nabla_a \nabla_b$ is d'Alembert's operator. The 
Riemann and Ricci tensors are given in terms of the Christoffel symbols 
$\Gamma_{ab}^{d}$ by
\be
{{\cal R}_{abc}}^{d}=\Gamma_{ac
,b}^{d}-\Gamma_{bc , a}^{d}
+\Gamma_{ac}^{e} \Gamma_{eb}^{d}-
\Gamma_{bc}^{e}\Gamma_{ea}^{d} \,,
\ee
\be
{\cal R}_{ac} \equiv {{\cal R}_{abc}}^b = 
\Gamma^{b}_{ac ,b}-\Gamma^{b}_{bc ,a}+
\Gamma^{d}_{ac}\Gamma^{e}_{de}-
\Gamma^{d}_{bc}\Gamma^{b}_{d a} \,, 
\ee
and $ {\cal R} \equiv g^{ab} {\cal R}_{ab}$ is the Ricci curvature.

Let us define some terminology that will be used in this work. A 
perfect fluid is identified by the energy-momentum tensor
\be
T_{ab}=\left( P+\rho \right) u_a u_b +Pg_{ab} \,,
\ee
where $u^c$ is the fluid four-velocity (normalized to $u^c u_c=-1$), 
$\rho$ is its energy density, and $P$ is the isotropic pressure. In 
practice, $P$ and $\rho$ are not unrelated and there 
will be an equation of state relating them but, technically, 
this is not required by the definition above. This fact opens the 
possibility 
that, when looking for formal analytic solutions of the field equations 
sourced by a fluid, one can adjust $P$ and $\rho$, which depend on the 
spacetime point, so that a certain geometry  can be engineered to 
satisfy the field equations. This procedure 
detracts from the physics, in the sense that it would be much preferable 
to assign {\em a priori} an equation of state valid everywhere, for 
example a 
barotropic equation of state of the form $P=w\rho$ with $w=$~constant, 
and then solve the field equations. In practice, this method is not always 
followed  or does  not produce new 
solutions. The less than satisfactory procedure of determining the equation of state 
{\em a posteriori} is not unique to 
spherical solutions (it is used, for example, already in cosmology where 
the field equations are simpler). There are 
instances where this procedure does not cause harm, for example when one 
decides to regard a minimally coupled scalar field as an effective 
perfect fluid. In this case, the equation of state $P_\text{eff} = w 
\rho_\text{eff}$ exhibits a dynamical equation of state 
parameter $w=w \left( x^{\alpha} \right)$ depending on the spacetime position. 
Other times, the equation of 
state of the fluid does not originate in a  field theory but is derived 
in order to satisfy the equations and, while it may reduce to a satisfactory 
equation of state in certain regions of spacetime, it will remain rather 
{\em ad hoc} in other regions. This is the case, for example,  for the 
McVittie  solution of the Einstein equations describing a central object 
embedded in a cosmological space, that will be discussed in 
Sec.~\ref{sec:7}. But then it turns out that the McVittie solution is a 
also a  solution of a different theory of gravity that does not require an 
{\em ad hoc} fluid equation of state. This new 
perspective somehow rescues the McVittie geometry from the limbo of 
solutions with questionable motivation (the borders of which are rather 
subjective). The possibility of having different sources for a certain 
GR geometry, or that the latter solves a different theory of gravity, 
 remains unexplored for most of the solutions reported in this review.     

Another term that is widely used in this work should be defined. We 
refer to  {\em vacuum} to denote the absence of a proper matter 
stress-energy tensor. In the Einstein equations
\be
{\cal R}_{ab} -\frac{1}{2} \, g_{ab} {\cal R} +\Lambda g_{ab}=8\pi 
T_{ab}^\text{(m)} \,,
\ee
the term containing the cosmological constant $\Lambda$ can be taken to 
the right hand side and regarded as an effective stress-energy tensor 
$T_{ab}^{(\Lambda)}=- \frac{\Lambda}{8\pi} \, g_{ab}$: we still refer to 
this 
situation as ``vacuum with a cosmological constant''. Similarly, in 
scalar-tensor gravity, one can write the field equations in the form of 
effective  Einstein equations
\be
{\cal R}_{ab} -\frac{1}{2} \, g_{ab} {\cal R} =
8\pi T_{ab}^\text{(m)} + 8\pi T_{ab}^\text{(eff)} \,,
\ee
where $T_{ab}^\text{(m)}$ describes a ``real'' matter source while 
$T_{ab}^\text{(eff)}$ contains all the terms describing extra degrees 
of freedom or higher curvature terms which form an effective (as 
opposed to ``real'') stress-energy tensor. In this context, ``vacuum scalar-tensor 
gravity'' denotes again the situation 
$T_{ab}^\text{(m)} =0$ even though $T_{ab}^\text{(eff)} \neq 0$.


\part{Spherical solutions of general relativity}

We begin by reviewing spherical solutions of the Einstein equations. 
Proceeding in order of increasing complexity, one is naturally led to 
consider vacuum, vacuum with a cosmological constant, and then simple 
forms of matter:  electrovacuum, a scalar field, a perfect, and then an 
imperfect fluid.

Without loss of generality, the spherically symmetric line element in four 
spacetime dimensions can be written in polar coordinates $\left( t, 
r, \vartheta, \varphi \right)$ as (see Sec.~11.7 of Ref.~\cite{Weinberg}) 
\be 
ds^2=-A^2(t,R)dt^2+B^2(t,R) dR^2 +R^2 d\Omega_{(2)}^2 \,, 
\label{lineelement} 
\ee 
using the areal radius $R$ as the radial  coordinate, and where 
$d\Omega_{(2)}^2 \equiv d\vartheta^2 +\sin^2 \vartheta \, d\varphi^2$ is 
the line 
element on the unit 2-sphere. 

The situation $A^2B^2=1$ is of special interest \cite{BondiKilmister60, 
French77,Jacobson07} and highlights special algebraic properties of the 
geometry. Jacobson has shown \cite{Jacobson07} (also in higher dimension) 
that it is equivalent to require that the double projection of the Ricci 
tensor ${\cal R}_{ab} \ell^a \ell^b $ onto radial null vectors $\ell^a$ vanishes. 
Another characterization is that the restriction of the Ricci tensor to 
the $\left( t,R\right)$ subspace be proportional to the restriction of the 
metric $g_{ab}$ to the same subspace \cite{Jacobson07}. A third 
characterization is that the areal radius $R$ is an affine parameter along 
radial null geodesics \cite{Jacobson07}.  GR geometries satisfying these 
conditions are varied, ranging from vacuum, electrovacuum with either 
Maxwell or non-linear Born-Infeld electrodynamics, to the ``string 
hedgehog'' global monopole \cite{hedgehog}.

A convenient form of the gauge~(\ref{lineelement}) is 
\cite{NielsenVisser,AbreuVisser}
\begin{eqnarray}
ds^2 &=& -\, \mbox{e}^{-\Phi(t,R)} \left( 1-\frac{2M_\text{MSH}(t,R)}{R}
\right) dt^2 \nonumber\\
&&\nonumber\\
&\, & + \frac{dR^2}{ 1-2M_\text{MSH}(t,R)/R } +R^2 d\Omega_{(2)}^2 
\,,\label{NielsenVissergauge}
\end{eqnarray}
where $M_\text{MSH}(R)$ is the Misner-Sharp-Hernandez mass
\cite{MSH1,MSH2} enclosed in a sphere of radius $R$.

In view of astrophysical applications, the motion of massive particles and 
fluids around a black hole is of great interest.  Since these motions can 
be quite complicated, pseudo-Newtonian potentials have been introduced 
that give an effective simplified description of timelike geodesics around 
Schwarzschild and Kerr black holes \cite{PaczynskiWiita80, Ragtime, Marek, 
pseudohistorical1,pseudohistorical2, pseudohistorical3, pseudohistorical4, 
pseudohistorical5, pseudohistorical6, pseudohistorical7, 
pseudohistorical8, pseudohistorical9, pseudohistorical10, 
TejedaRosswog14}. The Paczynski-Wiita pseudopotential for the 
Schwarzschild black hole~(\ref{Schwarzschild}) is simply 
\cite{PaczynskiWiita80,Marek}
\be
\phi_\text{PW}(R) = -\frac{m}{R-2m} \,.
\ee

This potential reproduces exactly the radius of the innermost stable 
circular orbit (``ISCO'' or ``marginally stable orbit''), that of the 
marginally 
bound orbit, and the form of the Keplerian angular momentum $L(R)$. It 
reproduces, accurately but not exactly, the Keplerian angular velocity and 
the 
radial epicyclic frequency \cite{PaczynskiWiita80,Marek}. Therefore, the 
structure of the phase space of timelike geodesics in the Schwarzschild 
metric does not differ much from the one obtained with the  
pseudopotential \cite{WitzanySemerakSukova15}.

The motion of massive test particles and fluids around black holes offers 
potential tests of the theory of gravity using  observations of 
these motions near a black hole horizon. A pseudopotential can be 
introduced for any static and spherically symmetric metric of the 
form~(\ref{lineelement}) with $A=A(R)$ and $B=B(R)$, obtaining 
\be 
\phi(R)=\frac{1}{2} \left[ 1- \frac{1}{A^2(R)} \right] 
\ee 
in the approximation of particle velocities $ v\ll 1$ \cite{Marek}. If the 
further assumption $A^2(R)B^2(R)=1$ holds, then this pseudopotential can be 
written as \cite{FaraoniKeetLapierre16} 
\be 
\phi(R) = - \frac{ M_\text{MSH}(R)}{ R-2M_\text{MSH}(R)} \,,
\ee
as follows immediately from the gauge~(\ref{NielsenVissergauge}) with 
$\Phi=0$.

\section{Vacuum: the Jebsen-Birkhoff theorem and its generalization to 
$\mathbf{ \Lambda\neq 0}$}
\label{sec:2}

As is well-known from any GR textbook, the Jebsen-Birkhoff theorem 
\cite{Jebsen21, Birkhoff23} establishes that the unique vacuum, 
spherically symmetric, and asymptotically flat solution of the Einstein 
equations is the Schwarzschild geometry 
\cite{Schwarzschild16}\footnote{In spacetime 
dimension $d>4$, the uniqueness of the corresponding 
``Schwarzschild-Tangherlini'' solution was proved by Bronnikov and 
Melnikov \cite{Bronnikov1}, 
and by  Gibbons {\em et al.} \cite{STunique1,STunique2}.}
\be
ds^2 =-\left( 1-\frac{2m}{R} \right)dt^2 +\frac{dR^2}{1-2m/R} +R^2 
d\Omega_{(2)}^2 \,, \label{Schwarzschild}
\ee
and the generalization of the theorem to electrovacuum establishes the 
uniqueness of the Reissner-Nordstr\"om solution under the same assumptions 
(see \cite{LivingReviews} for a review of GR black holes). A weak 
form of the Jebsen-Birkhoff theorem states that the solution is static for 
special forms of the matter energy-momentum tensor \cite{Das60, Isaev76, 
Bronnikovetal76, BronnikovKovalchuk80}. In 
the literature, one finds also almost-Birkhoff theorems that study small 
deviations from spherical symmetry or from vacuum 
\cite{almostBirkhoff1,almostBirkhoff2, almostBirkhoff3, 
NziokiGoswamiDunsby2014}.

The Jebsen-Birkhoff theorem can be extended to vacuum with a cosmological 
constant $\Lambda$ in a straightforward way concluding that, in this case, 
the unique spherical solution of the vacuum Einstein equations is the 
Kottler/Schwarzschild-de Sitter (KSdS) geometry \cite{Kottler} if 
$\Lambda>0$ and the asymptotics are de Sitter, or is the 
Schwarzschild-anti-de Sitter (SAdS) geometry if, instead, $\Lambda<0$ and 
the asymptotics are anti-de Sitter. In the following, we refer explicitly 
to the KSdS case, but the discussion for $\Lambda<0$ and SAdS is 
straightforward. The standard textbook discussion of the Jebsen-Birkhoff 
theorem is obtained as a special case by setting $\Lambda=0$.

In locally static (Schwarzschild-like or ``curvature'')  coordinates 
$\left( T,R, \vartheta, \varphi \right)$, the KSdS metric reads 
\begin{eqnarray} 
ds^2 &=&-\left( 1-\frac{2m}{R}-H^2R^2 \right) dT^2 
+\frac{dR^2}{1-\frac{2m}{R}-H^2R^2} 
+R^2 d\Omega_{(2)}^2 \,, 
\label{KSdS} 
\end{eqnarray}
where $H=\sqrt{\Lambda/3}$ and $m$ are positive constants.   

Surprisingly, modern relativity textbooks do not make reference to the 
Jebsen-Birkhoff theorem when $\Lambda \neq 0$, and the literature is 
ambiguous to this regard.  Occasionally, there are explicit statements of 
the uniqueness of the S(A)dS space ({\em e.g.}, \cite{Schmidt, 
NziokiGoswamiDunsby2014, FabianLasenby2015}) and a proof of the extension 
of the theorem to $\Lambda\neq 0$ was given in Synge's 1960 textbook 
\cite{SyngeGR} (without mention of Kottler's 1918 paper \cite{Kottler}).  A 
simple proof in null coordinates was given by Schleich and Witt
\cite{SchleichWitt} and there are mathematically more sophisticated 
proofs 
\cite{BoucherGibbonsHorowitz84, Kodama04, LeFloch:2010hf, AlamYu15}.  As 
for the 
Schwarzschild solution, uniqueness is associated with stability of 
the KSdS and SAdS solutions with respect to perturbations, as established 
in 
\cite{GuvenNunez90, BalbinotPoisson90, MellorMoss90, OtsukiFutamase91}. 
However, the evidence of the uniqueness of KSdS and SAdS is downplayed in 
the literature and papers exist that erroneusly state the existence of 
spherical solutions of the vacuum Einstein equations with $\Lambda>0$ 
alternative to KSdS \cite{Abbassi99,Abbassi02,Meissner09}, which would 
contradict uniqueness. The problem  is
that the line element is written in rather complicated coordinates and is 
not recognized as the KSdS line element, or an incorrect coordinate 
transformation leads to a line element formally different from the KSdS 
one. Furthermore, there are solutions of the Einstein 
equations that represent central inhomogeneities embedded in 
Friedmann-Lema\^itre-Robertson-Walker (FLRW) spaces 
and that {\it apparently} reduce to alternatives to the KSdS solution in 
the 
special case when the FLRW
``background''\footnote{We use quotation marks because, due to 
the non-linearity of the field equations, it is impossible to split a 
metric into a background plus a deviation from it (with the exception of 
Kerr-Schild metrics \cite{StephaniKramerMacCallumHoenselaersHerlt}).}   
reduces to 
de Sitter \cite{CasteloFerreira09, CasteloFerreira10, CasteloFerreira14, 
CasteloFerreira13}. However, these are solutions with matter sources and 
not $\Lambda$-vacuum solutions. These putative alternatives have been 
shown to be either KSdS disguised by unconventional coordinates, or 
genuinely different solutions in the presence of matter 
\cite{FaraoniCardiniChung18}.

\subsection{Uniqueness of the KSdS and SAdS metrics}

Here we prove the uniqueness of the KSdS and SAdS spacetimes using a 
particular gauge but the result is, of course, gauge-independent.  The 
vacuum Einstein equations 
\be 
G_{ab}=-\Lambda g_{ab} 
\ee 
yield, in the gauge~(\ref{lineelement}), 
\begin{eqnarray}
&& \frac{\dot{B} }{RB}=0 \,, \label{JB1}\\ 
&&\nonumber\\
&&\frac{2B'}{B^3 R}-\frac{1}{B^2 R^2} +\frac{1}{R^2} = \Lambda \,, 
\label{JB2}\\
&&\nonumber\\
&& \frac{2A'}{A R}-\frac{B^2}{R^2} +\frac{1}{R^2} =-\Lambda B^2
\,, \label{JB3}\\
&&\nonumber\\
&& \frac{A'B}{A}- B' - \frac{R B^2
\ddot{B}}{A^2} +\frac{R \dot{A}\dot{B} B^2}{A^3} -\frac{RA' B'}{A} +\frac{ 
RA'' B}{A} =-\Lambda R B^3\,, \qquad \label{JB4} 
\end{eqnarray} 
where an overdot 
and a prime denote differentiation with respect to $t$ and $R$, 
respectively.\footnote{The $\vartheta$-$\vartheta$ and 
$\varphi$-$\varphi$ components of the Einstein equations provide 
the same information because of the symmetry.} Equation~(\ref{JB1}) 
implies that $B=B(R)$ and, setting $\dot{B}$ and $\ddot{B}$ to zero 
in Eq.~(\ref{JB4}) as a consequence, Eq.~(\ref{JB2}) yields 
\be 
\left(  \frac{R}{B^2} \right)'= 1 -\Lambda R^2 \,, 
\ee 
with solution 
\be 
B^2(R) = \frac{1}{1+\frac{C}{R}-\frac{\Lambda R^2}{3} } \,, 
\ee 
with $C$ is an 
integration constant. In the limit $\Lambda \rightarrow 0$, one must 
recover the Schwarzschild line element~(\ref{Schwarzschild}) for a mass 
$m$, which fixes the 
integration constant to $C=-2m$ and 
\be 
B^2= \frac{1}{1-\frac{2m}{R}- 
\frac{\Lambda R^2}{3} } \,. 
\ee 
Now Eq.~(\ref{JB3}) yields 
\be 
\left( \ln  A^2 \right)'=\left[ \ln \left( 1-\frac{2m}{R}- \frac{ \Lambda 
R^2}{3}  \right) \right] ' \,, 
\ee 
with general solution 
\be 
A^2(R)=  \, \mbox{e}^{D(t)} \left( 1-\frac{2m}{R} - \frac{\Lambda R^2}{3} 
\right) \,, 
\ee 
where $D(t)$ is an integration function of time. By rescaling the time 
coordinate so that $ dT= \, \mbox{e}^{D(t)/2 } dt $, 
the line element necessarily becomes 
\begin{eqnarray} 
ds^2 = -\left( 1-\frac{2m}{R}- 
\frac{\Lambda R^2}{3} \right)dT^2 +\frac{dR^2}{ 1-\frac{2m}{R}- 
\frac{\Lambda R^2}{3} } +R^2 d\Omega_{(2)}^2 \,,
\end{eqnarray} 
which is the KSdS geometry if $\Lambda >0$ (and then 
$H=\sqrt{\Lambda/3}$), or the SAdS one if $\Lambda <0$. Therefore, KSdS 
[SAdS] is the unique vacuum spherical solution if $\Lambda>0$ ~[if 
$\Lambda<0$]. It reduces to the Schwarzschild geometry if $\Lambda=0$.

\subsection{Simultaneous baldness and cosmic baldness}

Well-known cosmic no-hair theorems state that, with a few exceptions ({\em 
i.e.}, overdense Bianchi models that collapse before the cosmological 
constant has a chance to dominate the cosmic dynamics), de Sitter space 
behaves as an attractor in the late-time dynamics of the universe 
\cite{Waldtheorem1,Waldtheorem2,Waldtheorem3}. In an analogy, no-hair 
theorems establish the uniqueness of the Schwarzschild solution among GR 
black holes and exclude the possibility of fields living in their 
exterior, since they would make the geometry deviate from Schwarzschild 
\cite{Chase70,RuffiniWheeler71, Bekenstein72a, Bekenstein72b, 
Bekenstein72c, Bekenstein96,Teitelboim72,Zannias95,Bekenstein95, 
Saa96,Bronnikov01,HerdeiroRadu15,Sotiriou15}. Moreover, the KSdS solution 
is unique, as shown above, and is perturbatively stable 
\cite{GuvenNunez90, BalbinotPoisson90, MellorMoss90, OtsukiFutamase91}. 
Since the KSdS geometry combines black hole and de Sitter physics, it is 
natural to expect that simultaneous no-hair and cosmic no-hair theorems exist and 
that the KSdS spacetime acts as a late-time attractor for the dynamics of 
time-dependent spherical spacetimes containing a central inhomogeneity in 
the presence of a positive $\Lambda$. This idea has not been explored 
much, probably due to the difficulty of handling the Einstein equations in 
the presence of matter.  As a starter in this direction, a 
non-perturbative result in the presence of an imperfect fluid was derived 
in \cite{FaraoniCardiniChung18}.

Consider the Einstein equations with matter 
\be 
G_{ab}=-\Lambda g_{ab}+8\pi T_{ab} \label{efe} 
\ee 
and assume spherical symmetry, then the line element 
is~(\ref{lineelement}). Further assume that the geometry is asymptotically 
de Sitter, {\em i.e.}, that there is a de Sitter-like cosmological horizon 
(with areal radius $R_\text{H}$) and the solution of~(\ref{efe}) reduces 
to~(\ref{KSdS}) as $R \rightarrow R_\text{H}^{-}$. Then, assume that the 
energy-momentum tensor in the right hand side of the Einstein equations
\begin{eqnarray}
&& \frac{\dot{B} }{BR}=4\pi T_{01} \,, \label{SS1}\\ 
\nonumber\\ 
&& A^2  \left( \frac{2B'}{B^3 R}-\frac{1}{B^2 R^2} +\frac{1}{R^2}
\right) = \Lambda A^2 + 8\pi T_{00} \,, \label{SS2}\\
&&\nonumber\\
&& \frac{2A'}{A R}-\frac{B^2}{R^2} +\frac{1}{R^2} =-\Lambda B^2
+8\pi T_{11} \,, \label{SS3}\\
&&\nonumber\\\
&&  \frac{A'B}{A}- B' - \frac{R B^2 \ddot{B}}{A^2} +\frac{R \dot{A}\dot{B} 
B^2}{A^3} -\frac{RA' B'}{A} \nonumber\\
&&\nonumber\\\
&& +\frac{  RA'' B}{A}  = \left( -\Lambda R^2 +8\pi T_{22} \right) 
\frac{B^3}{R} \,, \label{SS4} 
\end{eqnarray} 
describes an imperfect fluid with constant equation of state and a purely 
spatial radial heat flow with flux density $q^a$,
\begin{eqnarray} 
T_{ab} &=& \left( P+\rho \right) u_a u_b +P g_{ab} 
+q_a u_b +q_b u_a \,,\label{imperfect}\\
&&\nonumber\\
P&=&w\rho \,, \;\;\;\;\; w=\mbox{const.} \,,\\
&&\nonumber\\
u^a u_a &=& -1 \,, \;\;\;\;\;\; q^c q_c >0 \,, \;\;\;\; q^c u_c=0 \,. 
\end{eqnarray} 
As shown in \cite{FaraoniCardiniChung18}, either the matter fluid 
reduces to a cosmological constant at late times, in which case the vacuum 
uniqueness theorem for KSdS holds, or else both $\rho$ and $P=w\rho$ 
become subdominant and $\Lambda$ comes to dominate the expansion at late 
times while $\rho$ and $P$ become unimportant---then the solution reduces 
to KSdS.

\section{No-hair theorems}
\label{sec:3} 

No-hair theorems have a long history. Attempts to evade the 
Jebsen-Birkhoff theorem include the possibility of a matter source in the 
black hole exterior, for example a fluid or a scalar field. However, it is 
not so easy because this matter becomes physically pathological. For 
example, according to an early no-hair result by Chase \cite{Chase70}, a 
four-dimensional static, massless, asymptotically flat scalar field 
solution with a simply connected event horizon has the squared norm of the 
timelike Killing vector diverging there. The only solution is 
Schwarzschild, corresponding to $\phi\equiv 0$ outside the horizon, or 
the Fisher-Janis-Newman-Winicour-Buchdahl-Wyman (``FJNWBW'') 
solution containing a naked singularity, which is described below, see 
Eqs.~(\ref{FJNWBW}) and (\ref{F1}). The remaining possibility of the 
scalar field being singular on the horizon is not mentioned explicitly by 
Chase \cite{Chase70}, who views the central naked singularity as a horizon 
shrunk to a point.

Christodoulou has shown \cite{Christodoulou99} that the naked singularity 
of the FJNWBW geometry is unstable to gravitational collapse.\footnote{See 
Ref.~\cite{LiuLi17} for a version of Christodolou's theorem with slightly 
relaxed assumptions.} In numerical studies of scalar field collapse to a 
black hole in spherical symmetry, the scalar field either collapses below 
the Schwarzschild horizon forming a Schwarzschild black hole, or disperses 
to infinity \cite{GoldwirthPiran87}. Exact scalar field solutions 
describing other configurations ({\em i.e.}, naked singularities) exist, 
but they are unstable and do not form in numerical collapse.\footnote{See 
Refs.~\cite{Joshi93,Joshi13} for an overview of gravitational collapse.} 
They are discussed in the next section.

No-hair theorems for a scalar field with potential $V(\phi)$ that is 
monotonically increasing and convex have been proved by Bekenstein  
\cite{Bekenstein72c, Bekenstein72a, Bekenstein72b}. A generalization to 
general positive potentials $V(\phi)$ was given in \cite{Heusler92, 
Sudarsky95}, followed by a version of the no-hair theorem for scalar 
multiplets \cite{Bekenstein95} (and by a no-hair theorem for nonminimally 
coupled scalars, which are discussed later \cite{MayoBekenstein96, 
Bekenstein96}). The no-hair theorems, however, fail in Einstein-Yang-Mills 
theory \cite{VolkovGaltsov89, Bizon90, KunzleAlam90}, in Einstein-Skyrme 
theory \cite{BizonChmaj92, DrozHeuslerStraumann91}, and in situations 
where the above scenarios couple with dilaton 
\cite{LavrelashviliMaison93, ToriiMaeda93} or Higgs 
\cite{GreeneMathurONeill, LeeNairWeinberg92} fields.

From the physical point of view, the parameters charactering 
stationary black holes are conserved quantities associated with a Gauss 
law: mass, angular momentum, electric and magnetic charges. Therefore, the 
Maxwell field satisfying the Gauss law leads to the absence of Maxwell 
hair and to the uniqueness of the Reissner-Nordstr\"om solution.  
Moreover, according to theorems by Israel \cite{Israel67,Israel68}, 
staticity implies spherical symmetry for electrovacuum GR, from which the 
uniqueness of the Reissner-Nordstr\"om black hole follows. More in 
general, Hawking's strong rigidity theorem 
\cite{Hawking:1973uf,Hawkingprevious} states that a stationary black 
hole 
in GR is either spherical or axisymmetric when the null energy condition 
is satisfied (this theorem underlies the proof of Hawking's original 
no-hair theorem \cite{Hawkingtheorem} for Brans-Dicke black holes). The 
Hawking and Israel theorems, however, do not extend to non-Abelian 
Yang-Mills fields. In this respect a scalar field is similar: since it has 
no associated Gauss law, it is in principle possible to have scalar hair.

Non-Abelian gauge theories with gravity typically violate the no-hair 
theorem, especially when they possess solitons in the limit to Minkowski 
spacetime. Non-Abelian Yang-Mills fields in GR exhibit a short-ranged 
gauge field which is not killed by setting to zero the conserved charges. 
In addition, non-Abelian Yang-Mills black holes in GR can be static but 
still deviate from spherical symmetry, thus evading the Israel theorems 
and allowing for hair. Since these fields violate some of the ingredients 
necessary for the proofs of the Hawking and Israel theorems, which in turn 
are needed to prove no-hair results, it is not too surprising that they 
exhibit hair. Skyrme fields exhibit some of these features of non-Abelian 
Yang-Mills fields and admit hair as well (see \cite{VolkovGaltsov89, 
Bizon90, KunzleAlam90,BizonChmaj92, DrozHeuslerStraumann91} for more 
detailed discussions).

Even in scalar-tensor gravity with a simple Brans-Dicke-like field, 
it becomes more difficult to make statements when the black 
holes are not asymptotically flat. A no-hair theorem for asymptotically 
anti-de Sitter black holes has been proved when $\phi$ settles 
asymptotically in a global negative minimum of the potential $V(\phi)$ 
\cite{SudarskyGonzalez03}.  Scalar field potentials not bounded from 
below, although unphysical in GR and scalar-tensor gravity, can become 
physical in supergravity and string theories 
\cite{HertogHorowitzMaeda03}---the reader should bear in mind that, in 
this review, we restrict ourselves to GR and scalar-tensor gravity.

\section{Spherical scalar field solutions}
\label{sec:4} 

The next step in the search for spherically symmetric solutions consists 
of introducing a relatively simple form of matter: a scalar field 
minimally coupled to the Ricci curvature and described by the 
stress-energy tensor
\be 
T_{ab}^{(\phi)}= \nabla_a \phi \nabla_b \phi - 
\frac{1}{2} \, g_{ab} \, \nabla^c \phi \nabla_c \phi - \frac{V(\phi)}{2} 
\, g_{ab} \,, 
\ee 
where $V(\phi)$ is the scalar field potential. The action is
\begin{eqnarray}
 S=\int d^4x\sqrt{-g}\left[\frac{ {\cal R} }{2\kappa} -\frac{1}{2} \, 
\nabla_{a}\phi\nabla^{a}\phi -V\left(\phi \right)\right] \,, 
\end{eqnarray} 
and the coupled Einstein-Klein-Gordon equations are 
\begin{eqnarray}
&& {\cal R}_{ab}-\frac{1}{2} \, g_{ab}\, {\cal R} =8\pi T_{ab}^{(\phi)} 
\,,\\ 
&&\nonumber\\
&& \Box \phi -\frac{dV}{d\phi}=0 \,.
\end{eqnarray} 
Assuming spherical symmetry, staticity, and the line element in the 
gauge~(\ref{lineelement}), these equations take the form
\begin{eqnarray}
&& \frac{\dot{B} }{B}=4\pi R \dot{\phi}\phi' \,, \label{SSF1}\\ 
&&\nonumber\\
&& \frac{1}{R^2} + \frac{2B'}{B^3 R}-\frac{1}{B^2 R^2} =
 \frac{8\pi}{A^2} \left[ \dot{\phi}^2 + \frac{A^2}{2} \left(-\, 
\frac{\dot{\phi}^2}{2} +\frac{ \phi'^2}{2} \right) \right. 
 +VA^2 \Bigg] \,, \quad  \label{SSF2}\\
&&\nonumber\\
&& \frac{2A'}{A R}-\frac{B^2}{R^2} +\frac{1}{R^2} =
8\pi \left[ \phi'^2 - \frac{B^2}{2} \left( -\, \frac{\dot{\phi}^2}{A^2} 
+\frac{\phi'^2}{B^2} \right) \right. -VB^2 \Bigg] \,,  \label{SSF3}\\
&&\nonumber\\\
&&  \frac{R}{B^3} \left( \frac{A'B}{A}- B' - \frac{R B^2\ddot{B}}{A^2}
+\frac{R \dot{A}\dot{B} B^2}{A^3} -\frac{RA' B'}{A^3} 
\right.\nonumber\\
&&\nonumber\\
&&  +\frac{ RA'' B}{A} \Bigg) = 8\pi \left[ -\frac{R^2}{2} \left( 
-\, 
\frac{ \dot{\phi}^2}{A^2} +
\frac{ \phi'^2}{B^2} \right) -VR^2 \right] \,.\label{SSF4} 
\end{eqnarray}

\subsection{Static case: FJNWBW geometry}

Naturally, one first searches for solutions with a free scalar field ({\em 
i.e.}, $V(\phi) \equiv 0$), in which case the field equations simplify to 
\begin{eqnarray}
&& {\cal R}_{ab} = 8\pi \nabla_a \phi \nabla_b \phi 
\,,\label{masslessphi1}\\ 
&&\nonumber\\
&& \Box \phi =0 \,.\label{masslessphi2}
\end{eqnarray} 
Their general solution which is static, spherically symmetric, and 
asymptotically flat is the Fisher solution \cite{Fisher48}, also known as 
Janis-Newman-Winicour-Buchdahl-Wyman solution  (herafter 
referred to as the FJNWBW geometry). This solution was 
rediscovered several times \cite{BergmanLeipnik57, JanisNewmanWinicour68, 
Buchdahl72, Wyman81, AgneseCamera82, Dionysiu82, AgneseCamera85, 
Virbhadra97} by authors who did not recognize it in different coordinate 
systems, or were unaware of its previous derivation. A simple proof of its 
uniqueness is given by Wyman in \cite{Wyman81}. It does not describe 
a black hole: it has no horizons and represents a central naked 
singularity at areal radius $R=0$. Roberts rederived this 
solution in null coordinates \cite{Roberts3} whereas Agnese and La Camera studied it in the 
Newman-Penrose formalism \cite{AgneseCamera85}. The line element and scalar field are 
\cite{Fisher48, Wyman81}
\begin{eqnarray}\label{FJNWBW} 
ds^2 &=& - \, \mbox{e}^{ \alpha/r} dt^2 + \, \mbox{e}^{-\alpha/r} \left( 
\frac{ 
\gamma/r}{\sinh( \gamma/r)}\right)^4 dr^2 \nonumber\\
&&\nonumber\\
&\, & + \, \mbox{e}^{-\alpha/r} \left( \frac{ \gamma/r}{\sinh(
\gamma/r)}\right)^2 r^2 d\Omega_{(2)}^2 \,,\\
&&\nonumber\\
\phi &=& \frac{\phi_*}{r} \,, \;\;\;\;\;\;\;\;\;\;\quad  
\phi_*=\frac{-\sigma}{4\sqrt{\pi}} \,, \label{F1} 
\end{eqnarray} 
where $\alpha $, $\gamma$, and $\phi_{*}$ are constants, $\sigma$ is a 
scalar charge, 
and one can take $\gamma \geq 0$ without loss of generality. These three 
constants are related by \cite{Wyman81}
\be\label{WymanRelation} 
4\gamma^2=\alpha^2+2 \sigma^2 \,. 
\ee 
If $\sigma=0$ the constants $\alpha$ and $\gamma$ both vanish, the scalar 
field vanishes, and the solution (\ref{FJNWBW}) reduces to the Minkowski 
metric. In the notation of Wyman \cite{Wyman81} that we follow, the 
relation (\ref{WymanRelation}) between the constants $\alpha$, $\gamma$ 
and $\sigma$ does not allow one to see this fact. However, in the original 
derivation of Eq.~(\ref{FJNWBW}) in \cite{Wyman81}, one sees that the 
vanishing of $\sigma$ implies the vanishing of both $\alpha$ and $\gamma$.

The areal radius is 
\be 
R(r)= \gamma \, \frac{ 
\mbox{e}^{-\frac{\alpha}{2r}} }{\sinh \left( \gamma/r \right)} \rightarrow 
0^{+} \;\; \mbox{as} \; r\rightarrow 0^{+} 
\ee 
and there is a spacetime singularity at $R=0$ (or $r=0$) since the 
Ricci scalar
\be 
{\cal R}=8\pi \nabla^c \phi \nabla_c\phi = \frac{8\pi \phi_*^2}{\gamma^4} 
\, \mbox{e}^{\alpha/r} \, \sinh^4 \left( \gamma/r\right)
\ee 
diverges as $r\rightarrow 0$. This singularity is naked. In fact, horizons 
that could cover this singularity correspond to roots of the equation 
$\nabla^c R\nabla_c R=0$ (single roots corresponding to black hole 
horizons and double roots to wormhole throats \cite{MSH1, NielsenVisser, 
AHbook}), and no such root 
exists.\footnote{In particular, no wormhole throats exist, contrarily to 
claims in the literature \cite{FormigaAlmeida11}.} To see this, one 
computes
\be 
\nabla^c R \nabla_c R = \frac{1}{16\gamma^2} \left[ \left( \alpha+2\gamma 
\right) \mbox{e}^{ \gamma/r} -\left(\alpha-2\gamma\right) 
\mbox{e}^{-\gamma/r } \right]^2 
\ee 
and notices that the equation $\nabla^c R \nabla_c R =0 $ locating all the 
possible horizons is equivalent to
\be 
\mbox{e}^{2\gamma/r}= 
\frac{2\gamma-\alpha}{2\gamma+\alpha} \,. 
\ee 
The left hand side is never smaller than unity, while the right hand side 
is always smaller, hence there are no roots and no horizons covering the 
central singularity. This singularity is timelike. In fact, surfaces of 
constant radius have normal $N_{\mu}=\nabla_{\mu} \left( 
r-r_0\right)=\delta_{\mu 1}$ with magnitude squared 
\be
 N^c N_c= g^{11}= 
\left( r/\gamma\right)^4 \, \mbox{e}^{ \alpha/r} \sinh^4 \left( 
\gamma/r\right) \,,
\ee
 which is always positive, therefore $N^c$ is spacelike 
and the surface $r=r_0$ is timelike (including in the limit 
$r_0\rightarrow 0$).
 
When $\gamma\neq 0$, by performing the two consecutive coordinate 
transformations 
\begin{equation}  
\mbox{e}^{\gamma/r} = \frac{1+B/\rho}{1-B/\rho} \,,\qquad \bar{r} = \rho 
\left( 1+\frac{B}{\rho} \right)^2 \,, 
\end{equation} 
setting $\eta=2B=\sqrt{m^2+\sigma^2} \,, m/\eta =-\alpha/(2\gamma)$, and 
rescaling the time coordinate by a factor $|\gamma/(2B) |$, the FJNWBW 
solution becomes
\begin{eqnarray} 
ds^2 &=& -\left( 1-\frac{2\eta}{r} 
\right)^{m/\eta} dt^2 + \left( 1-\frac{2 \eta }{r} \right)^{-m/\eta} dr^2
\nonumber\\
&&\nonumber\\
&\, & + \left( 1-\frac{2\eta}{r} \right)^{1- m/\eta} r^2 d\Omega_{(2)}^2 
  \,,\label{becomes1}\\
&& \nonumber\\
\phi(r) &=& \frac{\sigma}{2\eta} \, \ln \left( 1-\frac{2\eta}{r} \right) 
\,. \label{becomes2}
\end{eqnarray} 
This is the most well-known form of the FJNWBW solution (however, since it 
requires $\gamma\neq 0$, it is not the most general). Taking this form at 
face value, one could argue for the existence of non-trivial solutions 
with non-zero scalar charge $\sigma$ but vanishing mass $m$, as is 
sometimes done in the literature. However, this possibility is forbidden 
by the fact that this form of the solution is only valid for $\gamma \neq 
0$. The solution in this form is invariant under the symmetry
\begin{eqnarray} 
\frac{m}{\eta} & \rightarrow & 
\frac{m'}{\eta'}=-\frac{m}{\eta} \,,\\
 &&\nonumber\\
r & \rightarrow & r'= 2\eta -r \,,\\
 &&\nonumber\\
\phi & \rightarrow & \phi'=- \phi \,. 
\end{eqnarray} 
In the special case $\gamma=0$, the FJNWBW metric reduces to the 
Papapetrou-Yilmaz or ``anti-scalar'' geometry\footnote{See 
\cite{Matt18} for a recent perspective on this ``exponential 
metric''.} \cite{Papapetrou54, Yilmaz58, Yilmaz71, MakukovMychelkin18}
\begin{eqnarray} 
ds^2 &=& 
- \, \mbox{e}^{\alpha/r}dt^2+ \, \mbox{e}^{-\alpha/r} \left( dr^2 +r^2 
d\Omega_{(2)}^2 \right) \,,\\
&&\nonumber\\
\phi & = & \frac{\phi_0}{r} \,. 
\end{eqnarray} 
In this case the areal radius is $R(r)=r \, \mbox{e}^{-\frac{\alpha}{2r}} 
$ 
and the equation locating the horizons is
\be 
\nabla^c R\nabla_c R = \left( 1+ \frac{\alpha}{2r} 
\right)^2=0 \,. 
\ee 
If $\alpha>0$ there are no real roots and no horizons, $R(r) \rightarrow 
0^+$ as $r\rightarrow 0^+$, and the Ricci scalar 
\be 
{\cal R}=8\pi 
\nabla^c \phi \nabla_c \phi = -8\pi \phi_0^2 \, \frac{ 
\mbox{e}^{\alpha/r}}{r^4} 
\ee 
diverges as $R\rightarrow 0^{+}$: there is a naked central 
singularity. The Misner-Sharp-Hernandez mass \cite{MSH1,MSH2}
\be
M_\text{MSH}= -\frac{\alpha}{2} \, \mbox{e}^{- \frac{\alpha}{2r} } \left( 
1+\frac{\alpha}{4r} \right)
\ee
is negative everywhere. If $\alpha<0$, instead, there is only one double 
root $r_* = 
|\alpha|/2$ of $\nabla^c R \nabla_c R=0$, corresponding to a wormhole 
throat at areal radius $R_* = |\alpha| \, \mbox{e} /2$  \cite{Matt18}. In 
this case, $M_\text{MSH}$ is positive for $R> |\alpha| 
\, \mbox{e}^2 /4$.     

The radial null geodesics of the FJNWBW geometry were studied in  
\cite{Formiga11}. Numerical studies of scalar field collapse generically 
lead to black holes and not to naked singularities (\cite{Christodoulou87, 
Christodoulou94}, see\footnote{Critical phenomena occuring in scalar 
field collapse were discovered by Choptuik in \cite{Choptuik93} 
(see also \cite{Brady95, Gundlach95} and see \cite{Gundlach99} 
for a review).} \cite{GoldwirthPiran87} for a review), which leads to 
seeing the FJNWBW solution as unphysical. As expected from these numerical 
studies, the FJNWBW geometry is indeed unstable, as shown long ago in  
\cite{Abe88}.
  
The FJNWBW geometry is the limit of the Garfinkle-Horowitz-Strominger 
black hole solution \cite{GarfinkleHorowitzStrominger91} of 
Einstein-Maxwell-dilaton gravity for vanishing mass $M$ and electric 
charge $Q$, keeping the ratio $Q^2/M$ finite, in which the horizon shrinks 
to a point. An electrovacuum version of the FJNWBW solution of the coupled 
Einstein-Maxwell-massless scalar field equations was given in 
\cite{AgneseCamera85}. The $d$-dimensional FJNWBW solution for 
$d\geq 4$ is discussed by Abdolrahimi and Shoom in 
\cite{AbdolrahimiShoom10}. Conformal diagrams for the FJNWBW 
geometry are constructed in \cite{AbdolrahimiShoom10}.

The energy conditions satisfied by the scalar field source of the 
FJNWBW metric, and the nature of the naked singularity, were discussed in 
\cite{Virbhadra:1995iy}. Gravitational lensing by the FJNWBW naked 
singularity, including image magnification and time delay between images, 
were studied in \cite{Virbhadra:1998dy,Virbhadra:2002ju,Virbhadra:2007kw}.

\subsection{Solutions with $V(\phi) \neq 0$}

No-hair theorems hold for a scalar field $V(\phi)$ that is monotonically 
increasing and convex \cite{Bekenstein72c, Bekenstein72a, Bekenstein72b}, 
but not much is known for other types of self-interaction potential. 
Allowing for a scalar field potential makes the integration of the field 
equations more difficult.

Various theorems were formulated which relate the existence of black 
holes and wormholes with the potential $V(\phi)$ 
\cite{Bekenstein96,Bekenstein:1998aw,Mazur:2000pn}; generally, they place 
necessary (but not sufficient) restrictions on the shape or sign of $V$ for the 
existence of black holes and other static spherical solutions. Bronnikov 
and Fabris 
\cite{Bronnikov:2005gm} classified the regular static and spherically 
symmetric configurations for canonical and phantom scalar fields with a 
potential $V(\phi)$. The action is now
\be
S= \int d^4x \sqrt{-g} \left[ \frac{ {\cal R}}{2} -\frac{\epsilon}{2} \, 
\nabla^c \phi \nabla_c \phi -V(\phi) \right] \,,
\ee
with $\epsilon=+1$ for a canonical and $\epsilon=-1$ for  a phantom 
scalar and static spherical solutions are sought for in the form
\be
ds^2=-A(r) dt^2 +\frac{dr^2}{A(r)} +R^2(r) d\Omega_{(2)}^2  \,, \quad 
\quad \phi=\phi(r)   \,.\label{BFlineelement}
\ee
For $\epsilon=+1$, it was shown \cite{Bronnikov:2005gm} that $ d^2R/dr^2 \leq 
0$, which prevents wormholes (because the areal radius $R(r)$ cannot have 
a 
positive minimum $R_0>0$) and stars with finite radius, as well as regular 
black holes. For $\epsilon=-1$, there is no such restriction and classic 
wormhole solutions for free scalars \cite{osti4570317,Bronnikov2} are 
recovered.  In total, Bronnikov and Fabris classify the possible regular 
configurations in sixteen different classes. These include solutions with 
a regular centre at areal radius $R_0$ and asymptotics that are Minkowski, 
de Sitter, or anti-de Sitter as $R\rightarrow + \infty$ and $r\rightarrow 
+ \infty$; solutions with $-\infty < r < +\infty$ which are wormholes 
separating two asymptotic spacetime regions; and Kantowski-Sachs 
homogeneous anisotropic spaces where $A(r)<0$ and the coordinates $t$ and 
$r$ swap their timelike and spacelike character.

An example of a phantom solution is given in 
\cite{Bronnikov:2005gm}, which is of the form~(\ref{BFlineelement}) with  
\begin{eqnarray}
R(r) &=& \sqrt{ r^2+b^2} \,,\\
&&\nonumber\\
A(r) &=& R^2 \left\{\frac{c}{b^2} +\frac{1}{R^2(r)} + \frac{r_0}{b^3} 
\left[ \frac{br}{R^2(r)} +\arctan \left( \frac{r}{b} \right) \right] 
\right\} \,,\\
&&\nonumber\\
\phi(r) &=&\pm \sqrt{2} \arctan \left( \frac{r}{b} \right) +\phi_0 \,,
\end{eqnarray} 
where $b >0,c, r_0$, and $\phi_0 $ are constants. The corresponding 
potential is not one of the usual physically motivated ones but is 
constructed {\em ad hoc} for this solution and is
\begin{eqnarray}
V( \phi) &=& -\frac{c}{b^3} \left[ 3-2\cos^2 \left( \frac{\phi}{\sqrt{2} } 
\right) \right] \nonumber\\
&&\nonumber\\
&\, & -\frac{r_0}{b^3} \left\{ \frac{3}{2} \sin\left( \sqrt{2} \, \phi 
\right) + 
\frac{\phi}{\sqrt{2}} \left[ 3-2 \cos^2 \left( \frac{\phi}{\sqrt{2} } 
\right) \right] \right\} \,.
\end{eqnarray}
In spite of this rather contrived potential,  this is a concrete example 
of a phantom wormhole solution. The 
special case $r_0=c=0$ gives $A=1$ and $V=0$ and reproduces the Ellis 
wormhole \cite{osti4570317}.  The classification of Bronnikov 
and Fabris 
\cite{Bronnikov:2005gm} extends to more general k-essence theories with 
Lagrangian ${\cal L}= P(X)-V(\phi)$, where $X\equiv - \nabla^c \phi 
\nabla_c\phi$ and the regular function $P$ is monotonic.

Another analytic solution of the Einstein-Klein-Gordon equations 
that is static, spherical, asymptotically flat, and has $\phi\rightarrow 
0$, $V(0)=0$, $V'(0)=0$ and $V''(0)=0$ was proposed in  
\cite{CadoniFranzin15}. It represents a black hole, but the complicated 
potential $V(\phi)$ has a vertical asymptote where it diverges to minus 
infinity, which makes the solution unphysical. Furthermore, the asymptotic 
value $\phi=0$ of the potential is not a minimum but an inflection point. 
For a particular value of a parameter, this geometry reproduces a previous 
solution of Anabal\'{o}n and Oliva \cite{AnabalonOliva12} which has been shown 
to be unstable with respect to radial linear perturbations 
\cite{AnabalonDeruelle13}.

\subsection{Time-dependent solutions}

In the cases when the solution of the Einstein-scalar field equations is 
not asymptotically flat or is time-dependent, there is no known general 
spherical solution and only special solutions are known. We report them in 
the following.

\subsubsection{Wyman's ``other'' solutions}

In addition to the well-known FJNWBW solution, in \cite{Wyman81} Wyman 
presented another class of static spherically symmetric solutions with a 
massless free scalar field which depends on time. Instead of looking for 
static solutions with scalar field $\phi=\phi(r)$, the Einstein equations 
can also be solved by assuming a static metric and $\phi=\phi(t)$, 
which is what originates Wyman's ``other'' solutions. In the context of 
the equivalence between a scalar field and a fluid, this class can be seen 
as stiff fluid solutions. In most of these solutions, the metric 
coefficients are given in the form of power series, which leads to 
formulae too clumsy to be useful in practice (for example, in curvature 
calculations). The same 
consideration applies to their 
generalizations to a non-vanishing cosmological constant given by Varela  
\cite{Varela87}. However, a particularly simple solution is given 
explicitly by Wyman \cite{Wyman81}:
\begin{eqnarray} 
ds^2
&=& - \kappa\, r^2 dt^2 +2dr^2 +r^2 d\Omega_{(2)}^2 \,,
\label{Wymanother1}\\
&&\nonumber\\
\phi(t) &=& \phi_0 \, t \,. \label{Wymanother2} 
\end{eqnarray} 
Substitution into the Einstein equations with a 
free scalar field as a source yields $\phi_0=1/\sqrt{\kappa}$.  This 
solution was generalized to the case $\Lambda\neq 0$ by  
Sultana \cite{Sultana15} and reproduces a special case of an older   
family of perfect fluid solutions found by 
Ibanez and Sanz \cite{IbanezSanz82}.  The geometry~(\ref{Wymanother1}) 
is not asymptotically 
flat and has a singularity at $r=0$, where the Ricci scalar ${\cal R}=8\pi 
\nabla^c\phi \nabla_c\phi = -1/r^2$ and the Ricci squared 
${\cal R}_{ab} {\cal R}^{ab}=1/r^4$ diverge. Since $\nabla^cr \nabla_c r = 
1/2$, 
there 
are no horizons (which are the roots of $\nabla^c r \nabla_c r=0$) and the 
central singularity is naked. The same conclusions apply to Sultana's 
generalization \cite{Sultana15} with $\Lambda \neq 0$. The 
Misner-Sharp-Hernandez mass \cite{MSH1,MSH2} $M_\text{MSH}=r/4$ contained 
in a 
sphere of radius $r$ is always positive.

Both Wyman's solution~(\ref{Wymanother1}), (\ref{Wymanother2}) and its 
generalization by Sultana for $\Lambda \neq 0$ were mapped into the Jordan 
frame of a conformally coupled scalar field to generate exact solutions, 
following a technique made popular by Bekenstein for a free scalar and 
adapted, {\em e.g.}, by Abreu {\em et al.} \cite{Abreuetal94} to 
the case of a non-zero potential. The same conformal mapping tecnique 
generates time-dependent spherical solutions of Jordan frame Brans-Dicke 
gravity (see below) and of $f({\cal R})={\cal R}^2$ gravity 
\cite{AliBehnaz}.

The Wyman geometry~(\ref{Wymanother1}) is reproduced as a special case of a more general 
static line element obtained by Carloni and Dunsby \cite{CarloniDunsby16}, but for a 
scalar-tensor theory in which the scalar field is only radius-dependent and is subject to a 
power-law potential.
 
The next two solutions of the Einstein equations describe spherical inhomogeneities embedded 
in a scalar field-fueled FLRW universe, the first for a free field and the second for 
exponential potential.  They have also been used to generate solutions of scalar-tensor 
gravity using conformal transformations.

\subsubsection{Husain-Martinez-Nu\~nez solution}

The scalar field Husain-Martinez-Nu\~nez (HMN) solution of the Einstein 
equations 
\cite{HusainMartinezNunez} is dynamical. It describes a spherical 
inhomogeneity embedded in a spatially flat FLRW ``background'' and sourced 
by a free scalar field. This solution is \cite{HusainMartinezNunez} 
\begin{eqnarray} 
ds^2 &=& \left( A_0 \eta +B_0 \right) \left[ - \left( 
1-\frac{2C}{r}\right)^{\alpha} d\eta^2 +\frac{dr^2}{\left( 
1-\frac{2C}{r}\right)^{\alpha}
} \right. \nonumber\\ &&\nonumber\\ &\, & \left. + r^2 \left( 
  1-\frac{2C}{r}\right)^{1-\alpha} d\Omega_{(2)}^2 \right] \,, 
  \label{HMNconformaltime}\\
&&\nonumber\\
\phi(\eta, r ) &=& \pm \frac{1}{4\sqrt{\pi}} \, \ln \left[ D\left( 
1-\frac{2C}{r}\right)^{\alpha/\sqrt{3}} \left( A_0 \eta +B_0 
\right)^{\sqrt{3}} \right] \,,\nonumber\\
&& \label{HMNscalar} 
\end{eqnarray} 
where $\eta >0 $ is the conformal time of the FLRW ``background'' (with 
the Big Bang at $\eta=-B_0/A_0$), $A_0 , B_0 , C$, and $D$ are 
non-negative constants, and the parameter $\alpha =\pm \sqrt{3}/2 $ can 
assume only two values. If $A_0=0$ the HMN geometry reduces to 
the FJNWBW one, and the constant $B_0$ can be set to zero if $A_0\neq 0 $, 
which we do in the following.

The HMN line element is conformal to the FJNWBW one~(\ref{FJNWBW}), with 
the scale factor of the ``background'' FLRW space as the conformal factor, 
$ \Omega=\sqrt{A_0 \eta }$, and with only two possible values of the 
Fisher exponent parameter. The sign in Eq.~(\ref{HMNscalar}) is 
independent of the sign of $\alpha$. The HMN metric approaches FLRW as $ 
r\rightarrow +\infty$ and it is exactly FLRW if $ C =0$ (in which case 
$A_0$ can be eliminated by rescaling $\eta$).

There is a central singularity. In fact, both the scalar field $\phi$ and 
the Ricci scalar 
\begin{eqnarray}
{\cal R} &=& 8\pi \nabla^c\phi \nabla_c \phi = 
\frac{2\alpha^2 C^2 \left( 1-\frac{2C}{r}\right)^{\alpha-2}}{ 3 r^4 A_0 
\eta } \nonumber\\
&&\nonumber\\
&\, & - \frac{ 3A_0^2}{ 2 \left( A_0 \eta \right)^3 \left( 
  1-\frac{2C}{r}\right)^{\alpha} } \label{HMNRicci}
\end{eqnarray}
diverge at $ r=2C$ for both values of $\alpha$, and $r=2C$ corresponds to 
areal radius $R=0$. The areal radius is 
\be\label{HMNarealradius} 
R(\eta,  r)= \sqrt{A_0 \eta} \, r \left( 
1-\frac{2C}{r}\right)^{\frac{1-\alpha}{2}} 
\ee 
and only the range $ 2C \leq r<+\infty $ is physically meaningful, with 
$r=2C$ corresponding to areal radius $R=0$.

The HMN line element can be rewritten in terms of the comoving time $t$ of 
the FLRW ``background'' as
\begin{eqnarray}
 ds^2 &=& - \left( 1-\frac{2C}{r}\right)^{\alpha} dt^2 +a^2(t) \nonumber\\
&&\nonumber\\
&  \times &  \left[ 
\frac{ dr^2}{\left( 1-\frac{2C}{r}\right)^{\alpha} }
+ r^2 \left( 1-\frac{2C}{r}\right)^{1-\alpha} d\Omega_{(2)}^2 \right] \,, 
\label{HMNcomoving}\\
&&\nonumber\\
\phi( t, r ) &=& \pm \frac{1}{4\sqrt{\pi}} \, \ln \left[ D\left( 
1-\frac{2C}{r}\right)^{\alpha/\sqrt{3}} a^{2\sqrt{3}}(t) \right] 
\,,
\end{eqnarray} 
where 
\be 
a(t)=\sqrt{A_0\eta}= a_0 \, t^{1/3} \,, 
\;\;\;\;\; a_0=\left( \frac{3A_0}{2} \right)^{1/3} 
\ee 
(the scale factor dictated by the stiff equation of state $ P=\rho $ of a 
free massless scalar field in a FLRW universe). In terms of the areal 
radius $ R(t,r)=a(t)r A^{\frac{1-\alpha}{2}}(r) $, of a new time 
coordinate $T$ (introduced to eliminate a cross-term in $dtdR$), and of 
$A(r)\equiv 1-2C/r$, one obtains \cite{AHbook}
\begin{eqnarray} 
&& ds^2 =-  A^{\alpha}(r) \left[ 1- \frac{H^2R^2 A^{2(1-\alpha)}(r)  
}{B^2(r)} \right]  F^2dT^2 \nonumber\\
&&\nonumber\\
& & +\frac{H^2 R^2 A^{2-\alpha}(r)}{B^2(r)} \left[ 1+
\frac{ A^{1-\alpha}(r)} { B^2(r) -H^2R^2 A^{2(1-\alpha)}(r)} \right] dR^2  
\nonumber\\
&&\nonumber\\
& & +R^2 d\Omega_{(2)}^2 \,,
\end{eqnarray} 
where $ B(r) \equiv 1- (\alpha+1) C/r $ and $F$ is an integrating factor 
guaranteeing that $dT$ is an exact differential. The apparent horizons of 
the HMN geometry, located by $ g^{RR}=0$, are the roots of
\be\label{HMNAH} 
B(r)=H(t)R\left( A(r) \right)^{1-\alpha} \,, 
\ee 
where $ r=r(t,R) $, equivalent to \cite{HusainMartinezNunez} 
\be 
\frac{1}{\eta} = \frac{2}{r^2} \Big[ r-(\alpha+1)C \Big] \left( 
1-\frac{2C}{r} \right)^{\alpha -1} \,. \label{HMNoriginalAH} 
\ee
If $ \alpha=\sqrt{3}/2 $,  
there exists a critical time $t_*$ such that, between the Big Bang 
and $t_*$  there is only one, expanding, apparent horizon. The equation 
locating the apparent horizons can only be solved numerically and, as a 
consequence, there is no analytic expression for this time  $t_*$, which 
depends on the mass parameter $C$ and on $A_0$. At $t=t_*$, two 
additional 
apparent horizons appear, one of which expands forever, while the other (a 
black hole apparent horizon) contracts, then merges with the first 
apparent horizon which in the meantime has been growing. As they meet, 
these two apparent horizons merge and disappear, causing a timelike naked 
singularity to appear at $R=0$ in the FLRW ``background'' 
\cite{HusainMartinezNunez,AHbook}. This naked singularity inhabits the 
universe for the rest of its history and the scalar field is regular on 
the apparent horizons. This dynamics of apparent horizon radii has been 
dubbed ``S-curve phenomenology'' (because of the shape of the horizon 
radii versus time) and recurs in solutions of GR and of 
Brans-Dicke theory \cite{AHbook}. The apparent horizons of the HMN 
solution are spacelike \cite{HusainMartinezNunez, AHbook}.
  
Apparent horizons appearing or disappearing in pairs constitute a typical phenomenology of 
time-dependent black holes in GR and in alternative theories of gravity \cite{AHbook}. 
Geometrically, it can be understood by the fact that the surface 
$R_\text{AH}=R_\text{AH}(t)$ defining the apparent horizons can intersect more than once 
slices of comoving time, producing one, two, or three horizon radii. For some critical value 
of time, two of these roots coincide and the number of apparent horizons before and after 
$t_*$ differ by two, which is associated with apparent horizons appearing (or disappearing) 
in pairs.

If $ \alpha=-\sqrt{3}/2 $, instead, only one expanding cosmological apparent horizon is 
present at all times, with a naked singularity at $R=0$.

\subsubsection{Fonarev spacetime}

The Fonarev solution \cite{Fonarev} of the coupled Einstein-Klein Gordon 
equations generalizes the free field Husain-Martinez-Nu\~nez solution to 
the case of an exponential potential
\be 
V(\phi)=V_0 \,  \mbox{e}^{-\lambda \phi}  \,, \label{eq:expotential} 
\ee with $ \lambda$ and $V_0$ positive 
constants (this potential has been investigated at length in FLRW 
cosmology). 
The Einstein field equations are now  
\be 
{\cal R}_{ab}=8\pi \left( 
\nabla_a\phi \nabla_b \phi +g_{ab}V \right) \,. \ee The spherical and 
time-dependent Fonarev solution is 
\begin{eqnarray}  
ds^2 & = & a^2 \left(\eta\right) \left[ -f^2\left(r\right) 
d\eta^2+\frac{dr^2}{f^2\left(r\right)} +S^2 \left(r\right) d\Omega^2_{(2)} 
\right] \,, \label{Fonarev1}\\
&&\nonumber\\
\phi \left( \eta, r \right) &=&
 \frac{1}{\sqrt{\lambda^2+2}}\ln \left(1-\frac{2w}{r}\right) +\lambda\ln a 
 + \frac{1}{\lambda} \ln \left[ \frac{V_0\left(\lambda^2-2\right)^2 
  }{2A_0^2 \left(6-\lambda^2\right)} \right] \, , \nonumber\\
&& 
\end{eqnarray} 
where 
\begin{eqnarray} 
f(r) &=&\left(1-\frac{2w}{r}\right)^{\alpha/2} \,,\;\;\;\;\; \alpha
=\frac{\lambda}{\sqrt{\lambda^2+2}} \,, \\
&&\nonumber\\
S(r)  & = & r \left(1-\frac{2w}{r}\right)^{\frac{1-\alpha}{2} } \, ,\\
&&\nonumber\\
a(\eta) &=& A_0 |\eta|^{\frac{2}{\lambda^2-2}} \,, 
\end{eqnarray} 
with $w$ and $ A_0$ constants. $\eta$ is the conformal time of the FLRW 
``background'' and we set $A_0=1$ for simplicity. In the limit $ 
w\rightarrow 0 $, the metric (\ref{Fonarev1})  reduces to the spatially 
flat FLRW geometry with a non-trivial scalar field in the 
exponential potential~(\ref{eq:expotential}). In this limit the scalar 
field reduces to 
\be
\phi=\lambda \ln a(t) +\phi_0 \,, \quad\quad \phi_0 = \frac{1}{\lambda} 
\ln \left[ 
\frac{V_0 \left(\lambda^2-2\right)^2   }{
2A_0^2 \left(6-\lambda^2\right)} \right] \, , 
\ee
which is equivalent to a perfect fluid with energy density and pressure
\begin{eqnarray} 
\rho_{(\phi)} & =& \frac{\dot{\phi}^2}{2} +V(\phi) = \frac{\lambda^2 
H^2}{2} + \frac{V_0 \, \mbox{e}^{-\lambda \phi_0} }{a^{\lambda^2} } \,,\\
&&\nonumber\\
P_{(\phi)} & =& \frac{\dot{\phi}^2}{2} -V(\phi) = \frac{\lambda^2 
H^2}{2} - \frac{V_0 \, \mbox{e}^{-\lambda \phi_0} }{a^{\lambda^2} } \,,
\end{eqnarray}
and effective dynamical equation of state 
\be
w \equiv \frac{  P_{(\phi)} }{\rho_{(\phi)} }= 
\frac{ \lambda^2 H^2 a^{\lambda^2}- 2V_0 \, \mbox{e}^{-\lambda \phi_0} }{
\lambda^2 H^2 a^{\lambda^2} +2V_0 \, \mbox{e}^{-\lambda \phi_0} } \,.
\ee
The Friedmann equation  $H^2=8\pi \rho/3$ is then immediately 
integrated, giving the power law
\be
a(t) = a_0 \, t^{2/\lambda^2}
\ee
for the scale factor, where
\be
a_0= \left[ \lambda^2 \sqrt{ \frac{ 2\pi V_0 \, \mbox{e}^{-\lambda 
\phi_0}}{3-4\pi \lambda^2} } \right]^{2/\lambda^2} \,.
\ee

In the other limit $ a \equiv 1$ and $ \alpha=1 $, the Fonarev geometry    
becomes the Schwarzschild spacetime (however, one cannot obtain the 
special value $ \alpha=1 $ if the condition $ 
\alpha=\frac{\lambda}{\sqrt{\lambda^2+2}} $ holds). The Fonarev line 
element approaches asymptotically a spatially flat FLRW one as $ 
r\rightarrow +\infty $. The Husain-Martinez-Nu\~nez metric corresponds to 
the special parameter values $ \lambda=\pm \sqrt{6} $ and $ V_0=0 $.

Kastor and Traschen have shown in \cite{KastorTraschen17} that the Fonarev 
geometry arises as a limit of a family of charged dilaton black holes 
important in string theory \cite{dilatonBH1,dilatonBH2} (indeed, the 
exponential potential is a trademark of dimensionally reduced theories). 
Feinstein {\em et al.} provide 
five-dimensional Fonarev spacetimes \cite{5dFonarev}, while higher-dimensional solutions 
are obtained with a solution-generating technique due to Buchdahl \cite{Buchdahl59} and 
Tangen \cite{Tangen07}, see also 
\cite{JanisRobinsonWinicour69}.\footnote{The 
Buchdahl-Tangen generating technique has been 
used also to produce higher-dimensional versions of the BBMB solution of 
the conformally coupled-Einstein-Maxwell field equations 
\cite{WehusRavndal07}.}

The Fonarev solution contains, during different epochs of the ``background'' FLRW universe, 
a naked central singularity or a black hole apparent horizon, and is asymptotically 
FLRW---it can be thought of as a generalization of the FJNWBW solution (to which it is 
conformal) to a FLRW background. The physical nature of the Fonarev class of solutions was 
studied in \cite{Fonarev} and \cite{HidekiFonarev}. At early times, there is no apparent 
horizon and there is a naked central singularity embedded in a FLRW ``background'', but at a 
critical time (dependent on the parameters) a pair of apparent horizons appears, one 
expanding (a cosmological horizon) and one contracting and covering a central singularity 
located at a finite areal radius \cite{Fonarev,HidekiFonarev}. The contracting 
apparent horizon asymptotes to this singularity \cite{Fonarev,HidekiFonarev}. Conformal 
diagrams for the Fonarev spacetime are given in Refs.~\cite{Fonarev,HidekiFonarev}.

A Fonarev solution with a phantom scalar field was discussed  
in \cite{GaoChenFaraoniShen08}.\footnote{A phantom field is unstable and 
should not exist so, at best, it is a signature of a truncated 
effective theory. Its recent popularity 
\cite{Caldwell:1999ew,Caldwell:2003vq,Nojiri:2005sr,Nojiri:2005sx, 
Melchiorri:2002ux} arises from the repeatedly coming and going claims that 
a phantom equation of state of the dark energy is preferred by 
cosmological observations.} It is generated from the Fonarev solution via 
the transformation $ \phi\rightarrow i\phi $,  
$\lambda\rightarrow -i\lambda $ which gives the kinetic 
energy of the scalar the wrong sign in the action and in the field 
equations. These generalized Fonarev metric and phantom scalar are given 
by Eq.~(\ref{Fonarev1}) and 
\begin{eqnarray}
\phi \left( \eta, r \right) &=& \frac{1}{\lambda} \ln \left[ 
\frac{V_0\left(\lambda^2
+2\right)^2}{2\left(\lambda^2+6\right)} \right] -\lambda\ln a -\frac{1}{\sqrt{\lambda^2-2}}\ln \left(1-\frac{2w}{r}\right)  
 \,, \nonumber\\
& &
\end{eqnarray} 
where now 
\begin{eqnarray} 
f(r)&=&\left(1-\frac{2w}{r}\right)^{\alpha/2} \,,\ \ \ \
\alpha=-\frac{\lambda}{\sqrt{\lambda^2-2}} \,, \label{xinoceros}\\ 
S(r) & 
= & r \left(1-\frac{2w}{r}\right)^{\frac{1-\alpha}{2}} \,, \ \ \ a(\eta) 
=\eta^{-\frac{2}{\lambda^2+2}} \,. 
\end{eqnarray} 
Assuming that $ \lambda>\sqrt{2} $, the physical meaning of the constant 
$w$ is recovered as follows.  For $ \lambda \gg \sqrt{2} $ it is $ a 
\approx 1 $ and $\alpha \approx -1 $, while \cite{GaoChenFaraoniShen08}
\begin{eqnarray}
\label{eqq:metric} 
ds^2 \approx -\left(1-\frac{2w}{r}\right)^{-1} 
d\eta^2+\left(1-\frac{2w}{r}\right)dr^2  +r^2 
\left(1-\frac{2w}{r}\right)^{2}d\Omega^2_{(2)} \,.  
\end{eqnarray}
The coordinate change \cite{GaoChenFaraoniShen08} $ y=r\left(1-2w/r 
\right) $ brings the line element (\ref{eqq:metric}) to the form of the 
Schwarzschild solution with mass $ -w$,
\be 
ds^2 = 
-\left( 1+\frac{2w}{y} \right) d\eta^2+ 
\left( 1+\frac{2w}{y} \right)^{-1}dy^2 
+y^{2}d\Omega^2_{(2)} \,, \label{eq:metricsch}  
\ee 
showing that $ w$ corresponds to a negative Misner-Sharp-Hernandez mass $ 
-|M_\text{MSH}|$. The history of the 
apparent horizons of the phantom Fonarev solution is the time reverse of 
that of the apparent horizons of the Fonarev spacetime 
\cite{GaoChenFaraoniShen08}.

\subsubsection{Roberts solution and its generalizations}

Let us  continue our survey within the context of GR with a 
canonical, minimally coupled scalar field without mass or potential. 
The Roberts geometry \cite{Roberts89} is a spherical, continuously 
self-similar spacetime within this context admitting a homothetic 
Killing vector field 
$\xi^c$ that satisfies $\pounds_{\xi} g_{ab}= 2g_{ab}$. The original 
form in \cite{Roberts89} contained an error later corrected in 
\cite{Sussman91,Brady94,Oshiro94,Burko97, 
Hayward00,ClementHayward01}. It  
constitutes a useful example in the study of critical phenomena 
arising in the collapse of scalar fields to black holes  
\cite{Choptuik93, Brady94, Oshiro94, HusainMartinezNunez} and it is of 
interest also for wormhole formation 
\cite{Oliveira96,WangOliveira97,Frolov00,Hideki09,Maeda15,Andre16}. Its 
stability was studied by Frolov \cite{Frolov97,Frolov99}. 

Burko \cite{Burko97} corrected and slightly generalized this solution which, in 
double null coordinates $\left(u,v, \vartheta, \varphi \right)$, reads 
\be
ds^2 =-dudv +r^2 (u,v) d\Omega_{(2)}^2 \,, \label{Robertsmetric}
\ee
\begin{eqnarray}
r^2 \left( u,v \right) &=& 
\frac{1}{4} \left[ \left(1-4\sigma^2 \right) v^2 -2uv +u^2 \right] 
\label{rsquared}\\
&&\nonumber\\
&=&  \frac{1}{4} \left[ (1-2\sigma)v -u \right] \left[ (1+2\sigma) v -u 
\right]  \,,\label{previous}
\end{eqnarray}
\be
\phi =\pm \frac{1}{2} \ln \left[ \frac{(1-2\sigma)v 
-u}{(1+2\sigma)v -u } \right] \,,\label{Robertsscalar}
\ee
where $\sigma $ is a constant scalar charge. The areal radius 
squared~(\ref{rsquared}) and the argument of the logarithm in 
Eq.~(\ref{Robertsscalar}) are positive if $ -1/2< \sigma < 1/2$. The limit 
of vanishing scalar charge $\sigma \rightarrow 0$, in which the matter 
scalar field $\phi$ disappears, produces Minkowski spacetime.  The origin 
$r=0$ corresponds to $u=\left( 1 \pm 2\sigma\right)v$. In the spacetime 
regions where $\nabla^c \phi$ is timelike (in which case the scalar field 
always corresponds to a perfect fluid), the Roberts solution coincides 
\cite{Burko97} with a stiff fluid solution of Gutman and Bespal'ko
\cite{GutmanBespalko67}.

The trace of the Einstein equations $ {\cal R}_{ab}= 8\pi \nabla_a \phi  
\nabla_b \phi  $ yields the Ricci scalar 
\be
{\cal R}=8\pi g^{ab}\nabla_a \phi \nabla_b \phi  
= - 32 \pi \phi_{,u} \phi_{,v} = \frac{ 
8\pi \sigma^2 uv}{r^4} \,,
\ee
which diverges as $r\rightarrow 0$, signaling a spacetime 
singularity.  The latter is naked, as the equation 
\be
g^{ab} \nabla_a r \nabla_b r =-\frac{1}{4 \, r^2} \left[ 
(1-4\sigma^2)v-u \right] (u-v) =0
\ee
admits no positive roots (the only roots are  $v=u$ or 
$v=\frac{u}{1-4\sigma^2} $, which give absurdities: if $v=u$, 
Eq.~(\ref{rsquared}) yields  $r^2 =-\sigma^2 u^2<0$,  
while   $v=u/(1-4\sigma^2)$ gives  
$r^2=-\frac{\sigma^2 u^2}{1-4\sigma^2}<0$).

The Misner-Sharp-Hernandez mass \cite{MSH1,MSH2} is 
\be
M_\text{MSH} = \frac{r}{2} \left( 1- g^{ab}\nabla_a r \nabla_b r 
\right) = -\frac{\sigma^2 \, uv}{2r} 
\ee
and is negative in the light cone of the origin $uv>0$.  Conformal  
diagrams for the Roberts solution were constructed by de Almeida Andr\'e 
\cite{Andre16}.

Roberts proposed also an anti-de Sitter generalization of his original  
solution, which is conformal to the previous one 
\cite{roberts2019hybrid,Maeda15}.  The new field equations
\begin{equation}
    {\cal R}_{ab} = 2\Lambda g_{ab} + 8\pi \nabla_a \phi \nabla_b \phi \, , 
\end{equation} 

\begin{equation}
    \Box \phi=0 \,, 
\end{equation} 
are solved by \cite{roberts2019hybrid,Maeda15} 
\begin{equation}
    ds^2=\left( 1-\frac{\Lambda}{6} \, uv \right)^{-2} \left[
-2dudv+f^2(u,v) d\Omega^2_{(2)} \right] \,, 
\end{equation} 
where 
\begin{equation}
    f^2(u,v) = -kuv+ f_1 v^2 +f_2u^2 \,, 
\end{equation} 
where $f_{1,2} $ are constants. If $k^2-4f_1f_2>0$, the scalar field is 
\cite{Maeda15}
\begin{equation}
    \pm \left( \phi-\phi_0 \right) = \frac{1}{\sqrt{16\pi }} \ln \left[ 
\frac{\sqrt{k^2-4f_1f_2} \, u + \left( ku-2f_1v \right) }{ 
\sqrt{k^2-4f_1f_2} \, u - \left( ku-2f_1v \right)
} \right] 
\end{equation} 
for $f_1 \neq 0$ and   
\begin{equation}
    \pm \left( \phi-\phi_0 \right) = \frac{1}{\sqrt{16\pi }} \, \ln \left( 
f_2-k \, \frac{v}{u} \right) 
\end{equation} 
for $f_1=0$, where $\phi_0$ is a constant. The Ricci scalar is singular at 
$f(u,v)=0$, except when $k^2-4f_1f_2 = 0$.  Also, $v \rightarrow \pm 
\infty$ and $u\rightarrow \pm \infty$ lead to curvature singularities for 
$f_1=0$ and $f_2=0$, respectively, if $k \neq 0$ \cite{Maeda15}.  
Three  cases arise:

\begin{itemize}

\item When $k^2-4f_1f_2 > 0$, the scalar field is real, the equation  
$f(u,v)=0$ has real roots, and the solution can be parametrized as 
\cite{Maeda15} 
\begin{equation}
    f^2 \left( u,v \right) =\frac{1}{4} \left( au-v \right) \left( bu-v 
\right) \,, \label{func} 
\end{equation} 
where the constants $a$ and $b$ satisfy $a+b=4k$.

\item When $k^2-4f_1f_2 < 0$, the scalar field is the phantom given by 
\begin{equation} 
\pm \left( \phi-\phi_0 \right) = i 
\sqrt{\frac{2}{\kappa^2}} \, \arctan \left( 
\frac{ku-2f_1v}{\sqrt{4f_1f_2-k^2u}} \right) \,, 
\end{equation} 
representing a dynamical wormhole \cite{Maeda15}.

\item When $k^2-4f_1f_2 =0$, the scalar field $\phi$ is constant and the 
spacetime is maximally symmetric \cite{Maeda15}. It is 
asymptotically locally AdS with a negative $\Lambda$ and it possesses a 
black hole event horizon depending on the values of the parameters 
\cite{Maeda15}. In this case, for $k=\pm 1$, the constants $f_1$ 
and $f_2$ must be non-negative; for $k=0$, it is either $f_1=0$ and $f_2
>0$, or $f_2=0$ and $f_1>0$ \cite{Maeda15}.

\end{itemize}

Penrose-Carter diagrams for this solution were constructed by Maeda \cite{Maeda15} and de 
Almeida Andr\'e \cite{Andre16}.

\section{Static Kiselev solution}
\label{sec:5}

The static, spherically symmetric Kiselev spacetime \cite{Kiselev03} has 
been the subject of a large literature (over 300 articles at the time of 
writing, which we do not report for lack of space),  ranging from null and 
timelike geodesics to 
horizon thermodynamics and accretion. This interesting metric is popular 
as a toy model for the study of various aspects of gravitation. The line 
element is \cite{Kiselev03}
\begin{eqnarray}
ds^2 &=& - \left[ 1-\frac{2m}{r}-\sum_{n} 
\left( \frac{r_n}{r} \right)^{3w_n+1} \right] dt^2 \nonumber\\
&&\nonumber\\
&\, & +\frac{dr^2}{ 1-\frac{2m}{r}-\sum_{n} \left( \frac{r_n}{r} 
\right)^{3w_n+1} } +r^2 
d\Omega_{(2)}^2 \,, \label{Kiselev} 
\end{eqnarray}
where $r_n>0 $ and $ -1< w_n <-1/3 $  are  constants. 
The Kiselev spacetime describes  a black hole\footnote{Called a ``dirty 
black hole'' in current terminology, because it is surrounded by matter 
and it Hawking-radiates as a graybody.} since it contains 
an event  horizon, a single root of $ \nabla^cr\nabla_cr=g^{rr}=0$, or
\be
r^{3w_N +1} -2m \, r^{3w_N} -\sum_{n=1}^N r_n^{3wn+1} \, r^{3(w_N-w_n)} =0 
\,.
\ee
When the sum $\sum_{n} \left( r_n /r \right)^{3w_n+1} $ vanishes, 
the geometry reduces to  Schwarzschild, while the KSdS metric is formally 
reproduced as $w\rightarrow -1$. The Kiselev metric is not asymptotically 
flat (because the exponents $(3w_n+1)$ are negative) and is not 
asymptotically FLRW, either (if it were, the limit $m\rightarrow 0$ would 
reproduce FLRW space, which is not the case). The Kiselev solution is 
usually reported as describing a ``black hole surrounded by quintessence'' 
but this terminology is extremely misleading. In fact, the matter source 
is not a 
perfect fluid (or a mixture of perfect fluids if the sum extends to 
$n>1$)  \cite{MattKiselev1} because it exhibits anisotropic pressure. 
Moreover, the cosmologist's notion of quintessence,\footnote{Even 
in cosmology, quintessence is usually an effective perfect fluid 
consisting of a scalar field $\phi$ in a suitable potential $V(\phi)$. 
This effective fluid would have a constant equation of state only for very 
special potentials (for example, two tuned exponential terms 
\cite{Bayin:1994nz}), which are not the usual ones encountered in the 
literature on quintessence.  
Nevertheless, most articles on the Kiselev black hole keep referring to it 
as ``a black hole surrounded by quintessence'' and, in this sense, this 
terminology is doubly misleading.} {\em i.e.}, a perfect 
fluid with 
equation of state $P=w\rho$ with $-1 <w<-1/3$, belongs to accelerated 
FLRW universes and the Kiselev geometry has no FLRW ``background''. 
Unfortunately, most of the literature follows the quintessence 
interpretation and only recently (\cite{MattKiselev1,MattKiselev2}, see 
also \cite{CveticGibbonsPope16,Semiz20b}) clarity has been 
made.\footnote{The wrong identification 
of an anisotropic fluid with a perfect one has been all too common in the 
context of exact fluid spheres in GR 
\cite{DelgatyLake98}.}  
Furthermore, the anisotropic energy-momentum tensor sourcing the Kiselev 
geometry can be split into a perfect fluid component plus either an 
electromagnetic or a scalar field \cite{MattKiselev2}. Depending on 
the parameter values, this stress-energy tensor  can violate the null 
energy condition 
\cite{MattKiselev2}, which means that great care is required in the use of 
the 
Kiselev spacetime as a toy model. In the most physically viable of these 
pictures, 
an electrically charged fluid is supported by the pressure gradient and 
by the electric field that it generates \cite{MattKiselev2}.

Contrary to what stated in \cite{Kiselev03}, the line 
element~(\ref{Kiselev}) does not contain the Reissner-Nordstr\"om metric 
when the sum $ \sum_{n} \left( r_n / r \right)^{3w_n+1} $ reduces to 
$ (r_n/r )^2$ for a single $w_q=1/3$. In 
fact the Reissner-Nordstr\"om metric has $ g_{00}=-\left( 1-\frac{2m}{r} 
+\frac{e^2}{r^2} \right)$ and not $ g_{00}=-\left( 1-\frac{2m}{r} 
-\frac{e^2}{r^2} \right)$, which is instead what the Kiselev line 
element~(\ref{Kiselev}) reproduces. Moreover, a radiation fluid is not the 
Kiselev matter source with $w_q=1/3$ (a value which, strictly speaking, is 
not allowed by the Kiselev condition $-1<w<-1/3$ \cite{Kiselev03}) and 
describes an incoherent superposition of electromagnetic waves with random 
phases, directions of propagation, and polarizations and not the  
radial electrostatic field of the Reissner-Nordstr\"om geometry.

Restricting ourselves to a single component for simplicity, the geometry 
\begin{eqnarray} 
ds^2 = -\left(1 -\frac{2m}{r} -\frac{C}{r^{3w+1}}\right) dt^2 + 
\frac{dr^2}{ 
1 -\frac{2m}{r} -\frac{C}{r^{3w+1}}   } +r^2 d\Omega_{(2)}^2 
\end{eqnarray}
has radial and tangential pressures \cite{MattKiselev1}
\begin{eqnarray}
P_r & = & -\rho = -\frac{3wC}{\kappa r^{3(w+1)}} \,,\\
&&\nonumber\\
P_t &=& -\frac{3w(3w+1) }{  2\kappa r^{3(w+1)} } \,;
\end{eqnarray}
the pressure anisotropy is \cite{MattKiselev1}
\be
\frac{\Delta P}{\bar{P} } = \frac{ 3(P_r-P_t) }{ P_r+2P_t} 
=-\frac{3(w+1)}{2w} 
\ee
and the effective average equation of state parameter reads 
\be
w \equiv \frac{ \bar{P}}{\rho} = \frac{ P_r+2P_t}{3\rho} \,.
\ee
This (constant) pressure anisotropy\footnote{The pressure anisotropy  
becomes position-dependent if two or more $w$-components are introduced 
in the matter source \cite{MattKiselev1}.} is always present (except for 
the KSdS limit 
$w\rightarrow -1$) and clearly shows 
that the matter source is not a 
perfect fluid.

The Kiselev metric can be brought to the Kerr-Schild form. Since, for 
Kerr-Schild metrics, the Einstein equations become linear in Kerr-Schild 
coordinates \cite{GursesGursey}, the linearity property expressed by the 
sum $\sum_n \left( r_n/ r \right)^{3w_n+1} $ in the Kiselev line 
element is not too surprising.

The  Misner-Sharp-Hernandez mass  contained  in a sphere of radius $r$ is 
\be
M_\text{MSH}=m + \frac{1}{2} \, \sum_n \frac{ r_n^{3w_n+1} }{r^{3w_n}} 
=m + \sum_n \frac{1}{ |w_n|} \frac{4\pi}{3} \, r^3 \rho_n(r) \,, 
\ee
where
\be
8\pi \rho_n = \frac{6 |w_n| r_n^{3w_n+1} }{ r^{3(w_n+1)} } 
\ee  
(a Puiseux expansion \cite{MattKiselev2}). This decomposition is covariant 
because $M_\text{MSH}$ can be expressed as the sectional curvature of the 
plane tangent to a 2-sphere of symmetry, which is proportional to the 
Riemann tensor \cite{CarreraGiuliniPRD10}.

A pseudo-Newtonian potential for the Kiselev geometry analogous to the 
Paczinski-Wiita potential for the Schwarzschild spacetime was given in 
\cite{Ragtime}.

\section{Fluid spheres}
\label{sec:6}

The main motivation for considering fluid spheres in GR is to model 
(static) relativistic stars. Sometimes it is of interest to consider {\em 
dynamical} asymptotically flat fluid spheres instead of stars. The most 
common situation occurs in the study of spherical collapse to black holes, 
but expanding solutions modelling, {\em e.g.}, exploding wormholes, 
relativistic fireballs, or explosive events have also been explored 
\cite{BarrabesHogan03}, although the literature is naturally much smaller 
than the one on relativistic stellar configurations of equilibrium.

\subsection{Static fluid spheres in GR}

Stars are configurations of equilibrium and the search for static perfect 
fluid spheres amounts to solving the Einstein equations for hydrostatic 
and thermodynamic equilibrium. Aside from rotation, spherical symmetry is 
not just a convenient approximation because static perfect fluid 
configurations on stellar scales are expected to smooth out ``mountains'' 
and other deviations from sphericity (see \cite{Alam07,Pfister11} 
for a partial proof in GR).

Assuming a static and spherically symmetric geometry of the 
form 
\be 
ds^2=- \, \mbox{e}^{-\Phi(r)} dt^2 +\frac{dr^2}{1-2m(r)/r} +r^2 
d\Omega_{(2)}^2 
\ee 
sourced by a perfect fluid with stress-energy tensor 
\be 
T_{ab}=\left(P+\rho\right) u_au_b+Pg_{ab} \,, 
\ee 
the fluid four-velocity has components $u^{\mu}=\, \mbox{e}^{-\Phi(r)} \, 
\delta^{\mu} _{\,\,0} $ and \be m(r)=4\pi \int_0^r dr' r'^2 \rho(r') \ee is the 
Misner-Sharp-Hernandez mass \cite{MSH1,MSH2}. $\Phi(r)$ is determined by 
the Tolman-Oppenheimer-Volkoff equation of hydrostatic equilibrium 
\cite{OppenheimerVolkoff39,Tolman39}
\be 
\frac{d\Phi}{dr} - \frac{m(r) +4\pi r^3 P(r)}{r^2 
\left[ 1-2m(r)/r \right]}=0  
\ee 
(see \cite{ChandraStellar} for a discussion of its Newtonian version).

Exact solutions of the Einstein equations in spherical symmetry with 
perfect fluids as the matter source have a long history, beginning with 
the toy Schwarzschild interior solution with uniform density 
\cite{Schwarzschild16} now reported in all GR textbooks, the Tolman  
solutions \cite{Tolman39}, and other classic ones \cite{Wyman49, 
Buchdahl67, Heintzmann69, FinchSkea89,Durgapal82, 
BergerHojmanSantamarina87, LattimerPrakash01}.  Existence and uniqueness 
of the solutions are discussed in \cite{RendallSchmidt91, Makino98, 
Simon02}.  While it is possible to find analytic solutions, the 
simplifying techniques used \cite{Kuchowicz71, 
StephaniKramerMacCallumHoenselaersHerlt,FinchSkea} very often make the 
solution rather uninteresting from the physical point of view 
\cite{DelgatyLake98}.

Buchdhal \cite{Buchdahl59} proved that all static, spherical perfect fluid 
stars with $ d\rho/dr \leq 0$ satisfy the ``Buchdhal bound'' $R_\text{S} 
/r < 8/9$, where $R_\text{S}=2M$ is the Schwarzschild radius and $M$ is 
the star's mass. Half a century later, Andr\'easson \cite{Andreasson08} 
generalized this bound to the radius-independent upper limit on the 
compactness $2m(r) /r <8/9 $, assuming algebraic inequalities (but 
not $d\rho/dr \leq 0$) on $\rho$ and $P$.

Delgaty and Lake \cite{DelgatyLake98}, Finch and Skea \cite{FinchSkea}, 
and Nambo and Sarbach \cite{NamboSarbach20} review perfect fluid interior 
solutions that are static, spherically symmetric, and asymptotically flat. 
Using computer algebra, Delgaty and Lake \cite{DelgatyLake98} classified 
127 candidate perfect fluid solutions of the Einstein equations that had 
appeared in the literature by 1998. They required:

\begin{enumerate}

\item isotropic pressure $P(r)$;

\item regularity of the curvature scalars at the origin $r=0$;

\item positive-definite energy density $\rho(0)$ and pressure $P(0)$ at 
the origin;

\item vanishing pressure $P(r_0)$ at a finite radius $r_0>0$ (this is a 
necessary and sufficient condition to match the interior metric with a 
Schwarzschild exterior through the usual junction conditions on a 
spacelike surface \cite{LakeMusgrave94});\footnote{Star models with 
density and pressure vanishing asymptotically as $r \rightarrow +\infty$, 
which are excluded by this requirement, are also considered to be physical 
provided that the total mass is finite \cite{LieblingPalenzuela17, 
AnderssonBurtscher19, RammingRein13}, which is the case of common boson star 
models.}

\item the energy density $\rho (r)$ and pressure $P(r)$ are decreasing 
functions of the physical radius;

\item the adiabatic speed of sound $c_s^2=dP/d\rho$ is less than the speed 
of light.

\end{enumerate}

Only 16 candidates satisfied all these physical requirements and, of 
these, only 9 have monotonically decreasing sound speed  
\cite{DelgatyLake98}. A common failure of satisfying the perfect fluid 
criterion is that solutions reported as perfect fluid ones in the 
literature actually have anisotropic pressure instead (in addition, 
several of the putative solutions reported in the literature do not 
actually solve the Einstein equations, or were given with typographical 
errors in the original articles reporting them). As usual with analytic 
solutions, some of them have been discovered more than once, either 
because they were disguised by different coordinates (the most common ones 
are curvature and isotropic coordinates, but Buchdahl, Synge isothermal, 
and Gaussian polar coordinates have also been used, see 
\cite{BoonsermVisser08} for a summary), or because authors were unaware of 
previous discoveries.\footnote{See Appendix~C of \cite{DelgatyLake98} for 
a partial list.}

Lake \cite{Lake03} has provided an algorithm to generate all regular 
static spherically symmetric perfect fluid solutions of the Einstein 
equations. This algorithm is based on the choice of a single monotonic 
generating function which is, however, unknown. In order to extract 
physically meaningful solutions, the generating function must satisfy 
restrictive integro-differential inequalities expressing physical 
requirements external to the algorithm \cite{Lake03}. A similar algorithm 
was proposed in \cite{RahmanVisser02}; Martin and Visser \cite{MartinVisser04}  
proposed a revised version of the Delgaty-Lake algorithm.\footnote{The 
existence of such algorithms was 
pointed out by Wyman in 1949, but they were found only relatively 
recently.} Solution-generating theorems are given in 
\cite{StephaniKramerMacCallumHoenselaersHerlt, 
BoonsermVisserWeinfurtner05, BoonsermVisserWeinfurtner07, 
BoonsermVisserWeinfurtner06, BoonsermVisser08}. The general static and 
spherical perfect fluid solution of the Einstein equations with equation 
of state $P=-\rho/5$ was given by Semiz \cite{Semiz20} using Buchdahl 
coordinates.

Given that exact solution catalogues are extensive and evolving, online 
databases of exact solutions, including perfect fluid spheres, are a 
better heritage repository than journal articles. While the oldest 
repositories are now inaccessible, {\sf grdb.org} stores the metric tensor 
and performs calculations interactively using the {\em GRTensorII} 
software \cite{IshakLake02}. Databases and computer algebra are 
necessarily superseded by new versions.

\subsection{Dynamical fluid spheres in GR}

The search for analytic perfect fluid solutions becomes more 
difficult when the latter are dynamical. The first step consists of 
searching for the analogue of the interior Schwarzschild solution with 
uniform density. Since the interior metric must match the exterior 
Schwarzschild vacuum, in this case the only possibility for the interior 
perfect fluid 
is a dust with vanishing pressure throughout the star 
\cite{OppenheimerSnyder,McVittie,Bondi, Mansouri, 
MashhoonPartovi,Glass,SmollerTemple}. The most well-known solution, which 
describes the gravitational collapse of a FLRW dust is the one by 
Oppenheimer and Snyder \cite{OppenheimerSnyder}, but other contracting and 
expanding solutions with uniform density were studied over the years 
\cite{Vaidyaball,McVittie,Bondi,Mansouri,MashhoonPartovi, ThompsonWhitrow, 
ThompsonWhitrow2, Glass,SmollerTemple}, to which we refer the reader.

Mathematical theorems can be useful not only in the search of exact, but 
also numerical solutions. For example, it was demonstrated in 
\cite{Boonsermetal15} that any arbitrary static anisotropic fluid sphere 
in GR can be mimicked by a perfect fluid, plus an electromagnetic field, 
plus a scalar field.

\section{Inhomogeneities embedded in a FLRW universe with fluid}
\label{sec:7} 

There are a few spherical and time-dependent solutions of the Einstein 
equations that describe central objects (which could possibly be black 
holes, as defined by the notion of apparent horizon instead of event 
horizon \cite{Booth:2005qc,Nielsen:2008cr,Faraoni:2018xwo})   
embedded in FLRW universes. The most well-known is without doubt the 
1933 McVittie solution, which has been the subject of renewed attention 
during 
the past decade and is also a solution of alternative theories of gravity 
\cite{Nolan99CQG, Afshordi07, GibbonsWarnickWerner08,Afshordi09, 
KaloperKlebanMartin10, MimosoLeDelliouMena10, Arakida11, 
LeDelliouMenaMimoso11, LakeAbdelqader11, 
GuarientoFontaninidaSilvaAbdalla12, FaraoniZambranoNandra12, Roshina1, 
Roshina2, FaraoniZambrano13, 
LeDelliouMimosoMenaFontaniniGuarientoAbdalla13, 
daSilvaFontaniniGuariento13, MimosoLeDelliouMena13, 
FaraoniZambranoPrain14, Horndeskisols1, Horndeskisols2, 
MacielLeDelliouMimoso15, MacielGuarientoMolina15,MelloMacielZanchin16, 
Piattella16, AghiliBolenBombelli17, FaraoniLapierreLeonard17}. The charged 
version of the McVittie solution is known 
\cite{ShahVaidya68,MashhoonPartovi79}, as well as generalized McVittie 
solutions \cite{FaraoniJacques07,LiWang07,GaoChenFaraoniShen08, 
CasteloFerreira09, CasteloFerreira10,CasteloFerreira14,CasteloFerreira13}. 
A few other, qualitatively different, solutions are known.

\subsection{McVittie solutions}

In order to study the effect of the cosmic dynamics on local systems, 
McVittie constructed a family of solutions of the Einstein equations 
\cite{McVittie33} describing a central inhomogeneity embedded in a FLRW 
universe.  Although there have been many studies of this family, mostly 
recently \cite{NewmanMcVittie82,Ferrarisetal,Nolan99CQG,NIS,Nolan98PRD,
 CarreraGiuliniPRD10,KaloperKlebanMartin10, 
LakeAbdelqader11,LandryAbdelqaderLake12, 
LeDelliouMimosoMenaFontaniniGuarientoAbdalla13, 
daSilvaFontaniniGuariento13, Anderson11,Roshina1, Roshina2, Arakida11, 
GuarientoFontaninidaSilvaAbdalla12, FaraoniZambranoNandra12}, 
the latter has proved to be rather complex 
and it is 
not yet completely understood in all its aspects. The McVittie spacetime 
is the only perfect ﬂuid solution of the Einstein equations which is 
spherically symmetric, shear free, and asymptotically FLRW with no 
radial energy flow 
\cite{Raychaudhuri79}. This theorem is stated explicitly by 
Raychaudhuri in \cite{Raychaudhuri79} without giving a formal proof. 
However, the proof consists of the step-by-step original 
derivation of this 
solution by McVittie in \cite{McVittie33}. Both electrically neutral and 
charged McVittie 
solutions are special cases of the Kustaanheimo-Qvist family of shear-free 
solutions of the Einstein equations (\cite{KustaanheimoQvist}, see also 
\cite{Krasinski97}). It has been shown that the McVittie metric cannot be 
generated as a scalar field solution of the Einstein equations. However, 
it is a solution of cuscuton theory, a special case of 
Ho\u{r}ava-Lifschitz 
gravity \cite{Horndeskisols2}. The cuscuton theory is the only form of 
$k$-essence that admits McVittie solutions \cite{Horndeskisols2}.  The 
McVittie geometry is also an exact solution of shape dynamics, which 
constitutes a possible approach to quantum gravity \cite{shapedynamics}, 
and is also a non-deformable solution of $f(T)$ gravity, where $T$ is the 
torsion scalar  \cite{BejaranoFerraroGuzman17}.

The McVittie line element in isotropic coordinates is 
\begin{eqnarray}  
ds^2 = -\frac{ \left(1-\frac{m(t)}{2\bar{r}} 
\right)^2}{ \left(1+\frac{m(t)}{2\bar{r}} \right)^2} \, dt^2+ a^2(t) 
\left( 1+\frac{m(t)}{2\bar{r}} \right)^4 \left( d\bar{r}^2 +\bar{r}^2 
d\Omega_{(2)}^2 \right) \,, \label{McVittieisotropic} 
\end{eqnarray} 
where the coefficient $m(t)$ must satisfy the McVittie no-accretion 
condition 
$ G_0^1=8\pi T_0^1=0 $ \cite{McVittie33} forbidding radial flow of cosmic 
fluid and equivalent to
\begin{equation} \label{35} 
\frac{\dot{m}}{m}+\frac{\dot{a}}{a} =0 \,, 
\end{equation} which implies $ m(t)= m_0 / a(t)  $
with $m_0>0 $ a constant.  Then, the McVittie line element becomes
\begin{eqnarray} 
ds^2 = -\frac{ \left[ 
1-\frac{m_0}{2\bar{r}a(t)} \right]^2}{ \left[ 1+\frac{m_0}{2\bar{r}a(t)} 
\right]^2} \, dt^2+ a^2(t) \left[ 1+\frac{m_0}{2\bar{r}a(t)} \right]^4 
\left( d\bar{r}^2 +\bar{r}^2 d\Omega_{(2)}^2 \right) \,. 
\end{eqnarray}

The McVittie geometry reduces to the Schwarzschild one if $ a \equiv 1 $, 
and to the FLRW metric if $m=0 $. Apart from the special case of a de 
Sitter ``background'', there is a spacetime singularity at $ \bar{r}=m/2 $ 
(this 2-sphere reduces to the Schwarzschild horizon if $ a \equiv 1 $)  
\cite{Ferrarisetal, Nolan98PRD, Nolan99CQG, Sussman85}. The singularity is 
spacelike \cite{Nolan98PRD,Nolan99CQG}; another singularity occurs at $ 
\bar{r}=0 $. McVittie's original interpretation of his geometry 
(\ref{McVittieisotropic}) as describing a point mass at $ \bar{r}=0 $ is 
untenable because, in general, the latter would be surrounded by the 
second singularity at $ \bar{r}=m/2 $ \cite{Sussman85, Ferrarisetal, 
Nolan98PRD, Nolan99CQG}, where also the pressure of the cosmic 
fluid\footnote{In the McVittie solution the energy density 
$\rho$ depends only on time, while the isotropic pressure $P$ depends on 
both radius and time 
unless the ``background'' universe is de Sitter with constant Hubble 
function $H$, as can be seen from Eq.~(\ref{pressure}). This radial 
dependence of $P$ disappears asymptotically far from the central object.}
\begin{equation} 
\label{pressure} 
P=-\, \frac{1}{8\pi} \left[ 3H^2+\frac{2\dot{H}\left( 
1+\frac{m}{2\bar{r}} \right) }{1-\frac{m}{2\bar{r}} }\right] 
\end{equation}
and the Ricci scalar $ {\cal R}=8\pi \left( \rho -3P \right) $ diverge 
\cite{Sussman85,Ferrarisetal, Nolan98PRD, Nolan99CQG, McClureDyerCQG, 
McClureDyerGRG}, and the weak and null energy conditions are violated 
(except in a de Sitter ``background'').

In terms of the areal radius 
\be 
R\left(t, \bar{r} \right) \equiv a(t) \, \bar{r} \left( 
1+\frac{m}{2\bar{r}} \right)^2 \,, 
\ee 
the McVittie line element becomes 
\begin{eqnarray}  
& & ds^2 = -\left( 1-\frac{2m_0}{R}-H^2R^2 
\right) F^2 dT^2 +\frac{dR^2 }{1-\frac{2m_0 }{R}-H^2R^2 }  +R^2  d\Omega_{(2)}^2 \,, \nonumber \\
& &  
\label{diagonalMcVittie} 
\end{eqnarray}
where $F(T,R)$ is an integrating factor introduced to guarantee that $dT$ 
is an exact differential \cite{AHbook} (this becomes unity for a de Sitter 
``background'').

The Misner-Sharp-Hernandez mass of a 2-sphere of radius $R$ is 
\be 
M_\text{MSH} =m_0 +\frac{H^2 R^3}{2}= m_0 + \frac{4\pi}{3}\, \rho R^3 \,, 
\ee 
{\em i.e.}, a time-independent contribution $m_0$ from the central object 
plus the mass of the cosmic fluid, with energy density $\rho$, contained 
in this sphere.

The apparent horizons of the McVittie metric are well-known. Restricting 
to the spatially flat case and to dust as an example, the apparent 
horizons are located 
by $ g^{RR}=0 $ or 
\be\label{8} 
H^2(t) \, R^3 -R+2m_0 =0 \,. 
\ee 
This cubic equation is formally the same encountered in the 
Schwarzschild-de Sitter geometry in static coordinates, but now $H$ is not 
constant and the 
apparent horizons are time-dependent. Defining $\psi$ by $ \sin (3\psi) 
\equiv  
3\sqrt{3} \, mH$, two horizons exist if $ 0<\sin ( 3\psi ) <1$, equivalent 
to $ m_0 H(t)<1/(3\sqrt{3}) $, which is only satisfied at certain times. 
In this flat, dust-dominated ``background'', only at $ t_* =2\sqrt{3} \, 
m_0 $ 
it is $ m_0 H(t)=1/(3\sqrt{3}) $.  At early times $ t<t_* $, we have $ 
m_0
>\frac{1}{3\sqrt{3} \,H(t)} $ and no apparent horizons, hence there is a
naked singularity. At $ t_* $, two apparent horizons appear together at $ 
R_1=R_2=\frac{1}{\sqrt{3}\,H(t)} $. Later on, as $ t>t_* $ and $ m_0 < 
\frac{1}{3\sqrt{3}\,H(t)} $, there are two distinct real roots $ 
R_{1,2}(t) $ corresponding to two distinct apparent horizons.

The singularity $ R=2m_0 $ is spacelike and separates the two spacetime 
regions $R<2m_0 $ and $ R>2m_0 $ into two disconnected manifolds 
\cite{Nolan98PRD, Nolan99CQG}.

An interior solution for the McVittie metric was found by Nolan \cite{NIS} 
and describes a relativistic star of uniform density in a FLRW 
``background''. It can be viewed as a generalization of the Schwarzschild 
interior solution to a FLRW ``background''. The Nolan line element in 
isotropic coordinates inside this ``star'' is 
\begin{eqnarray} 
ds^2 &\!\!=\!\!& -\left[ \frac{ 1-\frac{m}{\bar{r}_0 }
+ \frac{m\bar{r}^2}{\bar{r}_0^3} \left(1-\frac{m}{4\bar{r}_0} \right) }{ 
  \left( 1+\frac{m}{2\bar{r}_0} \right) \left( 1+\frac{ m 
  \bar{r}^2}{2\bar{r}_0^3 } \right) } \right]^2 dt^2 \nonumber\\
&&\nonumber\\
&\, & 
+ a^2(t) \, \frac{ \left( 1+ \frac{m}{2\bar{r}_0}\right)^6 }{ \left( 1 
  +\frac{ m\bar{r}^2}{2\bar{r}_0^3} \right)^2 } \left( d\bar{r}^2 
  +\bar{r}^2 d\Omega^2_{(2)} \right)  \label{NIS} 
\end{eqnarray}
where $ \bar{r}_0 $ is the star radius, $ m(t)$ satisfies Eq.~(\ref{35})  
forbidding accretion onto the star surface, and $ 0\leq \bar{r}\leq 
\bar{r}_0 $. The interior metric is regular at the centre and satisfies 
the Darmois-Israel junction conditions at $ \bar{r}=\bar{r}_0 $, matching 
the exterior McVittie geometry. The pressure is continuous there, while 
the otherwise uniform energy density is discontinuous \cite{NIS}:
\begin{eqnarray} 
  \rho(t) &= & \frac{1}{8\pi} \left[3H^2 +\frac{6m}{a^2 \bar{r}_0^3 \left( 
  1+\frac{m}{2\bar{r}_0 }\right)^6 } \right] \,, \label{birba1} \\
&& \nonumber \\
P\left( t, \bar{r} \right)&=& \frac{1}{8\pi} \left[ -3H^2
-2\dot{H}\, \frac{ \left( 1+\frac{m}{2\bar{r}_0 } \right) \left( 1 + 
 \frac{m \bar{r}^2}{ 2\bar{r}^3_0 } \right) }{ 1-\frac{m}{ \bar{r}_0} + 
 \left( 1 - \frac{m}{ 4\bar{r}_0} \right) \frac{m \bar{r}^2}{ \bar{r}_0^3}
 } \right. \nonumber \\ && \nonumber \\ && \left. + \frac{ 
   \frac{3m^2}{\bar{r}_0^4} \left( 1-\frac{\bar{r}^2}{\bar{r}_0^2} 
   \right)}{ a^2 \left( 1+ \frac{m}{2\bar{r}_0} \right)^6 \left[ 
   1-\frac{m}{\bar{r}_0} + \left( 1-\frac{m}{4\bar{r}_0} \right) \frac{ 
   m\bar{r}^2 }{\bar{r}_0^3 } \right]} \right] \,.\nonumber \\
&& 
\end{eqnarray} 
Like the McVittie solution, the Nolan interior solution belongs to 
the shear-free family of  
Kustaanheimo and Qvist \cite{KustaanheimoQvist}. $\rho$ is positive and $P\geq 
0 $ if $\ddot{a}+3\dot{a}^2/2<0 $. The surface of the star is comoving 
with the cosmic substratum. The generalized Tolman-Oppenheimer-Volkoff 
equation inside the star is reported in \cite{FaraoniJacques07}.

Conformal diagrams of the McVittie spacetime were obtained in 
\cite{LakeAbdelqader11,LandryAbdelqaderLake12, 
LeDelliouMimosoMenaFontaniniGuarientoAbdalla13,daSilvaFontaniniGuariento13} 
for various FLRW ``backgrounds''. These diagrams differ 
considerably according to the choice of the scale factor $a(t)$ of the 
FLRW universe in which the central object is embedded. This variety of 
possibilities has caused some debate in the literature, originating from 
different choices of $a(t)$ by different authors and from the fact that 
general statements for  all possible behaviours of the scale factor cannot 
be made.

\subsubsection{Charged and generalized McVittie geometries}

The charged McVittie solution was introduced by Shah and Vaidya
\cite{ShahVaidya68} and generalized by Mashhoon and Partovi
\cite{MashhoonPartovi79}, then rediscovered many years later by 
Gao and Zhang \cite{GaoZhang04}\footnote{Beware of a typographical error in the line 
element of Gao and Zhang \cite{GaoZhang04}.} and generalized to higher dimensions 
\cite{GaoZhangHigherD}. The apparent horizons can be located analytically 
and explicity in the extremal case and numerically otherwise, revealing 
that this geometry contains a black hole or a naked singularity according 
to the values of the mass and charge parameters 
\cite{FaraoniZambranoPrain14}.

The charged McVittie metric was used to disprove the 
universality of certain quantization laws for quantities built from the 
areas of black hole apparent horizons and inspired by string 
theories\footnote{Beware of the same typographical error present in 
\cite{GaoZhang04}. The qualitative results are, however, unchanged.} 
\cite{FaraoniZambrano13}.
 
Generalized McVittie solutions allowing for a radial energy flux onto, or 
from, the central object were introduced in \cite{FaraoniJacques07}. While 
these are rather cumbersome, they have a simpler late-time attractor 
\cite{fatePLB09} that is recognized as the non-rotating Thakurta 
\cite{Thakurta81} solution (see later). A class of solutions of 
Brans-Dicke gravity found in \cite{CMB} has the non-rotating Thakurta 
solution as its limit to general relativity ({\em i.e.}, the limit in 
which the Brans-Dicke parameter $\omega \rightarrow \infty$) 
\cite{AHbook}.

\subsection{Einstein-Straus, Lema\^itre-Tolman-Bondi, and Vaidya 
solutions}

Apparently unaware of the 1933 McVittie solution, Einstein \& Straus 
approached the problem of cosmic expansion versus local systems more than 
a decade later, matching a Schwarzschild vacuum to a FLRW universe 
\cite{EinsteinStraus45, EinsteinStraus46} on a finite radius sphere, while 
the McVittie solution 
interpolates smoothly between a local dynamical Schwarzschild-like object 
and the FLRW universe.  The Einstein-Straus model is strictly dependent on 
the spherical symmetry and does not generalize to axisymmetric spacetimes 
\cite{SenovillaVera97,Mars01,MenaTavakolVera02} (see 
Ref.~\cite{CarreraGiuliniRMP10} for a 
review of the old problem of local dynamics versus cosmic expansion). 
Nevertheless, it is still used occasionally as a toy model to address 
cosmological questions \citep[{\em e.g.},][]{MarraKolbMatarreseRiotto07, 
BiswasNotari08}).

Other well-known spherical and dynamical solutions of the Einstein 
equations are Lema\^itre-Tolman-Bondi geometries 
\cite{Lemaitre33,LTB1,Bondi} that 
can 
describe the collapse of dust and can be used to address the problem of 
local versus cosmological dynamics. The Vaidya solutions, intead, describe 
outgoing or ingoing spherical shells of null dust.

The Einstein-Straus, Lema\^itre-Tolman-Bondi, and Vaidya solutions of the 
Einstein equations have been analyzed in detail for decades and we refer 
the reader to standard books for their discussion 
\cite{StephaniKramerMacCallumHoenselaersHerlt, 
Krasinski97,GriffithsPodolsky09}.

\subsection{Solutions generated by conformal transformations}

Certain spherically symmetric and time-dependent solutions of the Einstein 
equations representing central objects embedded in FLRW universes are 
generated by conformally transforming Schwarzschild, with the scale factor 
of the FLRW ``background'' universe as the conformal factor. These 
solutions include the non-rotating Thakurta \cite{Thakurta81} solution, 
the Sultana-Dyer \cite{SultanaDyer}, and the McClure-Dyer solutions 
\cite{McClureDyerCQG, McClureDyerGRG,McClureAndersonBardhal08}. The 
matter sources are usually complicated and it is common to have spacetime 
regions with negative energy densities, making these regions unphysical. 
However, some of these geometries may turn out to be acceptable in other 
theories of gravity, and they could perhaps be acceptable solutions in GR 
with different matter sources (an issue that has not yet been 
investigated).

\subsubsection{Non-rotating Thakurta spacetime}

The non-rotating Thakurta solution \cite{Thakurta81} is the zero angular 
momentum limit of the GR solution originally introduced to model a Kerr 
black hole embedded in a FLRW universe. The line element is conformal to 
Schwarzschild,
\begin{eqnarray} 
ds^2 &=& a^2(\eta) \left[ -\left( 
1-\frac{2m}{r} \right) d\eta^2
+ \frac{dr^2}{ 1-2m/r } + r^2 d\Omega_{(2)}^2 \right] 
  \label{Thakurta1}\nonumber\\
&&\nonumber\\
&=& -\left( 1-\frac{2m}{r} \right) dt^2
+ \frac{a^2 dr^2}{ 1-2m/r } + a^2r^2 d\Omega_{(2)}^2 
\,,\nonumber\\
&& \label{Thakurta2} 
\end{eqnarray} 
where $a(\eta)$ is the scale factor of the FLRW ``background'', $\eta$ and 
$t$ 
are its conformal and comoving times, respectively, with $dt=ad\eta$, and 
$m$ is a constant mass parameter.  One can switch to the areal radius 
$R(\eta,r)=a(\eta)r$ and the new time coordinate $T$ defined by
\be 
dT= \frac{1}{F} \left( dt +\frac{HR}{ 
  A^2-H^2R^2 }\, dR \right) 
\ee 
where $F(t, R)$ is an integrating factor  satisfying 
\be 
\frac{\partial}{\partial R}\left( \frac{1}{F} \right)= 
  \frac{\partial}{\partial t} \left( \frac{HR}{F\left( A^2-H^2R^2 \right)} 
  \right) \label{Thakurtaintegratingfactor}
\ee 
to guarantee that $dT$ is an exact differential, recasting the 
Thakurta line element in the form 
\cite{MelloMacielZanchin16,FaraoniCardiniChung18} 
\begin{eqnarray} 
& & ds^2 = -\left( 1-\frac{2M}{R}- \frac{H^2R^2}{1-\frac{2M}{R}} \right) F^2
  dT^2 +\frac{dR^2}{1-\frac{2M}{R}
-\frac{H^2R^2}{1- 2M/R } }
  +  R^2 d\Omega_{(2)}^2 \,, \nonumber \\
& &\label{ThakurtaDiag} 
\end{eqnarray} 
where $A(t,R)= 1-2M/R$ and $ M (t) \equiv ma(t) $. Using the 
form~(\ref{Thakurta2}) of the metric, the Einstein equations 
give 
\begin{eqnarray} 
{G_0}^0= 8\pi {T_0}^0 &=& -\frac{3H^2}{ 1-2M/R} 
\,,\\
&&\nonumber\\
{G_1}^0 = 8\pi {T_1}^0 &=& - \frac{2mH}{r^2 \left( 1-2M/R \right)^2 } 
\,,\label{efe10}\\
&& \nonumber\\
{G_1}^1= 8\pi {T_1}^1 &=& 8\pi {T_2}^2 = 8\pi {T_3}^3 \nonumber\\
&&\nonumber\\
&=&  -\frac{1}{ 1-2M/R} 
\left(H^2 + \frac{2\ddot{a}}{a} \right) \,. 
\end{eqnarray} 
Since ${T^0}_r\neq 0$, there is a radial energy 
flow with spacelike current density\footnote{This current density 
is spacelike and non-causal, as in the imperfect fluid stress-energy 
tensor $T_{ab}=\left(P+\rho\right) u_a u_b +P g_{ab} +q_a u_b +q_b u_a$, 
or in Eckart's first order thermodynamics for dissipative fluids, which 
are non-causal.} $q_{\mu}=\left( 0, - 2m\dot{a} a \left( 
1-2M/R\right)^{-3/2}/r^2 , 0, 0 \right)$ in coordinates $\left( t, r, 
\vartheta, \varphi \right)$ \cite{MelloMacielZanchin16}, therefore the 
matter source is an imperfect fluid.

The non-rotating Thakurta solution is the late-time attractor of 
generalized McVittie 
solutions\footnote{The authors of \cite{AHbook,fatePLB09,Culetu13}
did not recognize this geometry as the non-rotating Thakurta solution.} 
\cite{fatePLB09,AHbook}. The non-rotating Thakurta solution is 
also the 
limit to GR of the Clifton-Mota-Barrow family of solutions of Brans-Dicke 
theory \cite{CMB} as the Brans-Dicke parameter $\omega \rightarrow \infty$ 
\cite{fatePLB09, AHbook}. The non-rotating Thakurta solution was studied 
in detail in \cite{MelloMacielZanchin16} (see also 
\cite{Culetu13, McNuttPage17}). Conformal diagrams for the 
non-rotating Thakurta solution can be found in 
Ref.~\cite{MelloMacielZanchin16} for various choices of the scale factor 
$a(t)$.

The non-rotating Thakurta solution has been used \cite{Boehm:2020jwd}, 
together 
with the class of generalized McVittie solutions of which it is an 
attractor \cite{RuizMolinaLima20}, as a toy model for primordial black 
holes for which the size of the black hole horizon is not completely 
negligible in comparison with the Hubble radius (the latter can be very 
small in the early universe). The time-dependent black hole apparent 
horizon is not a null event horizon and the Misner-Sharp-Hernandez mass of 
this black hole is time-dependent, contrary to the Schwarzschild mass. 
These differences with respect to Schwarzschild could be important for the 
evolution of primordial black holes, which can in principle account for a 
substantial fraction (or, hypothetically, even all) of dark matter, 
according to an old hypothesis that is seeing a new lease on life due to 
the unexpectedly large black hole masses measured in black hole mergers by 
the {\em LIGO/VIRGO} detectors \cite{SasakiSuyamaTanaka16, CarrSilk18}.

\subsubsection{Sultana-Dyer spacetime}

The Sultana-Dyer solution of the Einstein equations \cite{SultanaDyer} is 
a Petrov type~D geometry whose physical interpretation, according to their 
proponents, is a black hole embedded in a spatially flat FLRW universe. 
The scale factor of this FLRW universe is $ a(t)\propto t^{2/3}$ in 
comoving time, {\em i.e.}, that associated with a dust. This solution is 
``manufactured'' by conformally rescaling the Schwarzschild metric,  
$g_{ab}^{(S)} \rightarrow g_{ab}= \Omega^2 \, g_{ab}^{(S)}$. The explicit 
intention of Sultana \& Dyer was to map the timelike Killing field 
$\xi^c$ of Schwarzschild into a conformal Killing field (which requires 
$\xi^c \nabla_c \Omega \neq 0$), and the Schwarzschild event horizon into 
a conformal Killing horizon. The line element in its original form is 
obtained by conformally transforming the Schwarzschild metric 
\cite{SultanaDyer} 
\begin{eqnarray} 
ds^2 &\!\!=\!\!& a^2(\eta) \left[ -\left( 
1-\frac{2m}{\tilde{r}} \right) d\eta^2
+ \frac{4m}{\tilde{r}}\, d\eta d\tilde{r} \right. + \left. \left(
1 + \frac{2m}{\tilde{r}} \right) \, d\tilde{r}^2 +\tilde{r}^2d\Omega^2 
\right] \,, \nonumber \\
&& \label{SultanaDyeroriginal} 
\end{eqnarray} 
with $m$ a constant mass parameter\footnote{The metric signature of 
\cite{SultanaDyer} is opposite to ours and their notation for $\eta$ and 
$t$ is switched with respect to ours.} and $a(\eta)=a_0 \eta^2= a_0 
t^{2/3}$.

The Sultana-Dyer metric can be rewritten using the new time defined by 
\be 
dt=d\bar{t}+\frac{2ma\, d\tilde{r}}{\tilde{r}\left( 1-\frac{2m}{\tilde{r}} 
\right)} \,, 
\ee 
which transforms the line element~(\ref{SultanaDyeroriginal}) into 
\begin{eqnarray} 
ds^2& =& -\left( 
1-\frac{2m}{\tilde{r}} \right) d\bar{t}^2 +\frac{a^2 \,d\tilde{r}^2 
}{1-\frac{2m }{\tilde{r}} } +a^2 \tilde{r}^2 d\Omega^2 \label{2} 
\nonumber\\
&&\nonumber\\
&= & a^2\left[
-\left( 1-\frac{2m}{\tilde{r}} \right) d\bar{\eta}^2 
+\frac{d\tilde{r}^2}{1-\frac{2m}{\tilde{r}} } + \tilde{r}^2 d\Omega^2 
\right] \,.\nonumber\\
&& \label{3}
\end{eqnarray} 
This is explicitly conformal to Schwarzschild written in Schwarzschild 
coordinates, with conformal factor $a$ and $ d\bar{t} = a d\bar{\eta}$, 
and it approaches the Einstein-de Sitter geometry as $\tilde{r}\rightarrow 
+\infty$.

The matter source of the Sultana-Dyer metric is a mixture of two 
non-interacting perfect fluids, a null dust and a massive dust 
\cite{SultanaDyer}, with total energy-momentum tensor 
\be 
T_{ab}= T_{ab}^\text{(I)}+T_{ab}^\text{(II)} = \rho \, u_a\, u_b
+ \rho_n \, k_a\, k_b \,, 
\ee 
and $k^c k_c=0$, $k^c \nabla_c k_a=0$ \cite{SultanaDyer}.

The Ricci scalar 
\be
{\cal R}= \frac{12}{ \tilde{\eta}^6} \left( 1- \frac{2m}{\tilde{r}} 
+\frac{2m\tilde{\eta}}{\tilde{r}^2} \right)
\ee
diverges at the central spacetime singularity $\tilde{r}=0$ and at the Big 
Bang $\tilde{\eta}=0$. The equation $\nabla^c R \nabla_cR=0$ 
(where $R$ is the areal radius) locating the 
apparent horizons reads  
\be
2ma+ \frac{ \tilde{r}^2  a^2_{, \tilde{r} }}{a} \left( 
1+\frac{2m}{\tilde{r}} \right) -4m \tilde{r} a_{, \tilde{r}} -a\tilde{r}=0 
\ee
or, since $a(t)=t^2$ \cite{SultanaDyer,Saida:2007ru,AHbook},
\be
4\tilde{r}^3 +8m \tilde{r}^2 -\left( 8m +t\right) t \tilde{r} +2mt^2=0 
\,.
\ee
The positive roots of this cubic, expressed in terms of the areal 
radius $R=a\tilde{r}$, are \cite{SultanaDyer,Saida:2007ru,AHbook}
\begin{eqnarray}
R_1 (t) &=& \frac{t^3}{2} \,,\\
&&\nonumber\\
R_2 (t) &=& 2mt^2 \,,\\
&&\nonumber\\
R_3 (t) &=& -m -\frac{t}{4} +\frac{t^2}{4} \sqrt{ t^2 +24mt +16 m^2} \,.
\end{eqnarray}
$R_1$ is the radius of a cosmological apparent  horizon; $R_2$ is the 
areal radius of the conformally transformed Schwarzschild event horizon 
which, being  a null surface, remains an event horizon for the 
Sultana-Dyer solution; and $R_3$ is the radius of an apparent horizon. The 
Sultana-Dyer construct describes a black hole hiding a central spacetime 
singularity below its horizons \cite{SultanaDyer,Saida:2007ru,AHbook}. 
Conformal diagrams for the Sultana-Dyer geometry are given in 
Ref.~\cite{Saida:2007ru}.

\subsection{Other solutions}

There have been other attempts to construct time-dependent black holes by 
conformally transforming the Schwarzschild geometry. These works have 
generated the McClure-Dyer and similar solutions of the Einstein equations 
\cite{McClureDyerCQG,McClureDyerGRG,McClureAndersonBardhal08, Culetu13}. 
Other inhomogeneous spherical solutions of the Einstein equations include 
the Abbassi-Meissner \cite{Abbassi99, Abbassi02, Meissner09} and the 
Castelo Ferreira \cite{CasteloFerreira10, CasteloFerreira13, 
CasteloFerreira09, CasteloFerreira14} geometries. The former is a fluid 
solution with radial mass flow and the latter is generated by a fluid with 
anisotropic pressure and contains the McVittie metric as a special case. 
They are analyzed in \cite{FaraoniCardiniChung18}. Other spherical 
solutions proposed in the literature include those of 
Carr {\em et al.} \cite{CarrHaradaMaeda10}, Chakrabarti and Banerjee
\cite{ChakrabartiBanerjee17}, Firouzjaee and Mansouri \cite{FirouzjaeeMansouri10},
Kastor and Traschen \cite{KastorTraschen93}, and Sun \cite{Sun09}.

\part{Spherical solutions of scalar-tensor gravities}

Scalar-tensor gravity is the prototypical alternative to GR and contains 
the popular class of $f({\cal R})$ gravities.  The salient feature is that 
the gravitational coupling constant is promoted to the role of a dynamical 
scalar field. Scalar-tensor gravity, in particular $f({\cal R})$ gravity, 
is used as an alternative to dark energy to explain the current 
acceleration of the universe. Theories that achieve this goal contain a 
time-dependent effective cosmological ``constant'' and it is interesting 
to understand how black holes in these theories look like. If  
stationary and asymptotically flat, they are the same as GR 
black holes (apart from mavericks), otherwise they are  time-dependent and 
characterized by 
apparent (instead of 
event) horizons, a subject of considerable interest already in GR. Another 
motivation to study analytic inhomogeneous solutions of scalar-tensor 
gravity is 
exploring the spatial variation of fundamental constants, in this case  
the gravitational coupling.  Furthermore, there is hope that 
deviations from GR may be detectable at astrophysical scales in processes 
involving strong gravity near black holes. Many theoretical efforts are 
devoted to the possibility of detecting black hole hair, which is 
forbidden in vacuum GR \cite{Israel67,Israel68,Chase70, Carter71, 
Wald71, RuffiniWheeler71, Hawkingprevious} but not in the presence of 
matter or in other theories of 
gravity \cite{Charmousis09, Sotiriou15,Sotiriou15b, HerdeiroRadu15}. 
Finally, bosonic string theory reduces to an $\omega=-1$ Brans-Dicke 
theory in its low-energy limit \cite{Callan85,FradkinTseytlin85}, and 
Brans-Dicke spacetimes are often very similar to exact solutions of 
dilaton gravity.

\section{Scalar-tensor gravity}
\label{sec:8} 

The Jordan frame scalar-tensor gravity action is \cite{BD, Bergmann68, 
Wagoner70, Nordtvedt70} 
\begin{eqnarray} 
S_\text{ST} &\!\!=\!\!& \frac{1}{2\kappa} 
\int d^4x \sqrt{-g} \left[ \phi {\cal R} -\frac{\omega(\phi 
)}{\phi} \, 
\nabla^c\phi \nabla_c\phi -V(\phi) \right] +S^\text{(m)} \,, \nonumber \\
&& \label{STaction}
\end{eqnarray} 
where $\phi>0$ is the Brans-Dicke scalar (with the 
effective gravitational coupling approximately equivalent to $\phi^{-1}$), 
the function $\omega(\phi)$ is the ``Brans-Dicke coupling'', $V(\phi)$ is 
a scalar field potential, and $S^\text{(m)}=\int d^4x \sqrt{-g} \, {\cal 
L}^\text{(m)} $ is the matter action.

The variation of the action~(\ref{STaction}) produces the (Jordan frame) 
field equations \cite{BD, Bergmann68, Wagoner70, Nordtvedt70} 
\begin{eqnarray}
 {\cal R}_{ab} - \frac{1}{2}\, g_{ab} {\cal R} &\!\!=\!\!& \frac{8\pi}{\phi} \, 
T_{ab}^\text{(m)} + \frac{\omega}{\phi^2} \left( \nabla_a \phi \nabla_b \phi -\frac{1}{2} 
  \, g_{ab} \nabla_c \phi \nabla^c \phi \right) \nonumber\\ 
&&\nonumber\\
& \, & +\frac{1}{\phi} \left( \nabla_a \nabla_b \phi
- g_{ab} \Box \phi \right) -\frac{V}{2\phi}\, g_{ab} \,, \label{BDfe1} \\
&&\nonumber\\
\Box \phi &\!\!=\!\!& \frac{1}{2\omega+3} \left( \frac{8\pi T^\text{(m)} }{\phi} 
+ \phi \, \frac{d V}{d\phi}
-2V  -\frac{d\omega}{d\phi} \nabla^c \phi \nabla_c \phi 
\right)  \,, \nonumber \\ \label{BDfe2}
\end{eqnarray} 
where $ T^\text{(m)} \equiv g^{ab}T_{ab}^\text{(m)} $ is the trace of the 
matter 
stress-energy tensor $T_{ab}^\text{(m)} $. The original Brans-Dicke theory 
\cite{BD} had $\omega=$~const. and no scalar field potential.

By performing the conformal transformation of the metric and redefining 
non-linearly the Brans-Dicke-like scalar as in 
\begin{eqnarray}
&& g_{ab}\rightarrow \tilde{g}_{ab}=\Omega^2 \, g_{ab},
\,\,\,\,\,\,\, \Omega=\sqrt{\phi} \,, \label{confo1}\\
&&\nonumber \\
&& d\tilde{\phi}=\sqrt{ \frac{\left| 2\omega(\phi)+3
\right|}{2\kappa }} \, \frac{d\phi}{\phi} \label{confo2} 
\end{eqnarray} 
for $\omega\neq -3/2$, the scalar-tensor action assumes the 
Einstein frame form 
\begin{eqnarray}
S_\text{ST} &=& \int d^4x \, \sqrt{-\tilde{g}} \left[
\frac{\tilde{{\cal R}}}{2\kappa } -\frac{1}{2}\, \tilde{g}^{ab} 
\tilde{\nabla}_a 
\tilde{\phi} \tilde{\nabla}_b \tilde{\phi} -U(\tilde{\phi}) 
+ \frac{ {\cal L}^\text{(m)} }{ (\phi)^2} \right]\nonumber\\
&\, &  
\end{eqnarray} 
where a tilde 
denotes quantities associated with the rescaled metric $\tilde{g}_{ab}$, 
${\cal L}^\text{(m)}$ is the matter Lagrangian density, and 
\be \label{questaA} 
U\left( \tilde{\phi} \right)=\frac{V\left[ \phi( \tilde{\phi}) 
\right]}{\left[ \phi ( \tilde{\phi}) \right]^2} \,. 
\ee 
This is formally 
the action of GR with a minimally coupled scalar field $\tilde{\phi}$, but 
with the important difference that this scalar now couples explicitly to 
matter. The Einstein frame field equations are 
\begin{eqnarray}
&&  \tilde{{\cal R}}_{ab}-\frac{1}{2}\,
\tilde{g}_{ab}\tilde{{\cal R}}=\frac{8\pi }{\phi}\, 
T_{ab}^\text{(m)} + 
\, \tilde{T}_{ab}^{(\tilde{\phi})} \,,\\
&&\nonumber\\
&& \tilde{\Box} \tilde{\phi} -\frac{
dU}{ d\tilde{\phi} }=\frac{8\pi T^\text{(m)} }{\phi^2} 
\,,\label{Eframeeqforscalar} 
\end{eqnarray} 
where
\be 
\tilde{ T}_{ab}^{(\tilde{\phi})} =\tilde{\nabla}_a \tilde{\phi} 
\tilde{\nabla}_b 
\tilde{\phi} - \frac{1}{2}\, \tilde{g}_{ab} \, 
\tilde{g}^{cd}\tilde{\nabla}_c \tilde{\phi} \tilde{\nabla}_d \tilde{\phi} 
-\frac{U\left( \tilde{\phi} \right)}{2}\, \, \tilde{g}_{ab} 
\ee 
is the canonical stress-energy tensor for a scalar field minimally coupled 
with the curvature. If the metric $g_{ab}$ is of the spherically symmetric 
form considered below, also the rescaled $\tilde{g}_{ab}$ 
assumes the same form with $\Omega(\phi)=\Omega(t,r)$.

A widely known solution-generating technique in scalar-tensor 
gravity (usually attributed to Bekenstein but anticipated before him 
\cite{Higgs59, Buchdahl72}) consists of 
taking a scalar field solution of GR and regarding it as the Einstein 
frame version of a Jordan frame counterpart, then mapping it back to the 
Jordan frame. In general, a scalar field potential that is physically well 
motivated in the Einstein frame gives rise to a meaningless one in the 
Jordan frame, however this technique works {\em in vacuo} and for certain 
potentials.

A lesser known solution-generating technique restricted to Brans-Dicke 
gravity with electrovacuum or with conformally invariant matter relies on 
a symmetry of the Brans-Dicke field equations. By taking a solution of 
this theory as a seed, one applies the symmetry 
\cite{myBDlimit1,myBDlimit2}
\begin{eqnarray} 
g_{ab} & \rightarrow & \hat{g}_{ab} =  \phi^{2\alpha} g_{ab} \,, 
\label{symmetry1}\\ 
&&\nonumber\\ 
\phi & \rightarrow &\hat{\phi} = \phi^{1-2\alpha} \,, \label{symmetry2}\\ 
&\nonumber\\ 
\omega & \rightarrow & \hat{\omega}( \omega, \alpha) = \frac{ 
\omega +6\alpha(1-\alpha)}{(1-2\alpha)^2} \,, \label{newomega}
\end{eqnarray}
of the corresponding Brans-Dicke field equations for $\alpha \neq 1/2$ to 
generate a new solution.\footnote{The conformal 
transformation~(\ref{symmetry1}) has nothing to do with the conformal 
relation between Jordan and Einstein frames.} Since, as $\alpha$ varies,  
there is a one-parameter Abelian group of transformations 
\cite{myBDlimit1,myBDlimit2}, the result is a family of solutions with 
an extra parameter $\alpha$.

\subsection{Black holes in scalar-tensor gravity: no-hair theorems }

\begin{figure}[h!] 
\begin{center} 
\includegraphics[scale=.4]{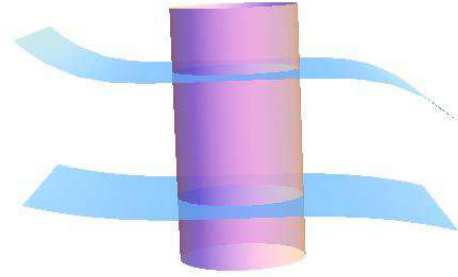} 
\end{center} 
\caption{The 
four-dimensional volume $\mathcal{V}$ bounded by the Cauchy hypersurfaces 
${\cal S}$ and ${\cal S}'$, the black hole event horizon $H$, and spatial 
infinity $i^0$.} \label{figA} 
\end{figure}

All asymptotically flat, stationary, axially symmetric black holes of 
physical relevance in vacuum scalar-tensor gravity reduce to GR black 
holes. A theorem \cite{Hawkingtheorem} states that all 
vacuum, stationary, and asymptotically flat black holes of Brans-Dicke 
gravity ({\em i.e.}, with $\omega=~$const. and $V(\phi) \equiv 0$) must be 
Kerr black holes. The proof, performed in the Einstein frame, does not go 
through the direct integration of the Brans-Dicke field equations, but 
uses an 
integral technique. It can be repeated by allowing the coupling $\omega$ 
to be a function of $\phi$.  A previous result of Hawking \cite{Hawkingprevious} 
states that a stationary, asymptotically flat black 
hole in vacuum GR is necessarily axisymmetric and must have spherical 
topology, if the weak or the null energy conditions hold. The theorem by 
Hawking \cite{Hawkingtheorem} extends this earlier result to Brans-Dicke theory 
proving that the Brans-Dicke scalar field is static. 
The advantage of using the Einstein frame is that the rescaled Brans-Dicke 
scalar
\be 
\tilde{\phi}= \sqrt{ \frac{\left| 2\omega +3\right| }{2\kappa }
}  \, \ln \left( \frac{\phi}{\phi_*} \right) 
\ee 
(where $\phi_*$ is a constant) satisfies the weak and null energy 
conditions \cite{Hawkingtheorem}. Since the spacetime is stationary, it is 
also axisymmetric \cite{Hawkingprevious}. Outside the horizon (with radius 
$r_\text{H}$), there exist a timelike Killing field $t^a$ and a spacelike 
Killing 
field $\psi^a$ and the Einstein frame scalar $\tilde{\phi}$ is constant 
along the orbits of $t^a$ and $\psi^a$. Therefore, 
for $r>r_\text{H}$, 
$\nabla^a \tilde{\phi} $ can only be spacelike or zero.  Consider a 
four-dimensional volume ${\cal V}$ bounded by two Cauchy hypersurfaces 
${\cal 
S}$ and $ {\cal S}'$ representing two consecutive instants of time $t$ 
(where $t^a \equiv \left( \partial/\partial t \right)^a$), a portion of 
the black hole event horizon $H$, and spatial infinity $i^0$ 
\cite{Hawkingtheorem} (fig.~\ref{figA}).
The Einstein frame field 
equation~(\ref{Eframeeqforscalar})  for $\tilde{\phi}$ {\em in vacuo} and 
with $V=0$ becomes $\tilde{\Box} \tilde{\phi}=0$. Multiplying this 
equation by $\tilde{\phi}$, integrating over ${\cal V}$, and using the 
Gauss theorem and the identity $ \tilde{\phi} \tilde{\Box} \tilde{\phi}= 
\tilde{\nabla}^c \left( \tilde{\phi} \tilde{\nabla}_c \tilde{\phi} \right) 
- \tilde{\nabla}^c \tilde{\phi} \tilde{\nabla}_c \tilde{\phi} $, one 
obtains \cite{Hawkingtheorem} 
\be 
\int_{ {\cal V} } d^4x \, \tilde{g}^{ab} 
\tilde{\nabla}_a \tilde{\phi} \tilde{\nabla}_b \tilde{\phi} = \int_{ 
\partial {\cal V} } dS^c \left( \tilde{\phi} \tilde{\nabla}_c \tilde{\phi} 
\right) \, . 
\ee 
The integral over the boundary $\partial {\cal V}$ on the 
right hand side is split into four contributions: 
\be 
\int_{ \partial 
{\cal V}} dS^c \left( \tilde{\phi} \tilde{\nabla}_c \tilde{\phi} \right) = 
\left( \int_{ {\cal S} } + \int_{ {\cal S}' } + \int_{i^0 }
+ \int_{H} \right)  dS^c \left( \tilde{\phi} \tilde{\nabla}_c \tilde{\phi} 
  \right) \,. 
\ee 
The integrals over the Cauchy hypersurfaces ${\cal S}$  and $ {\cal S}' $ 
cancel out because they have the same absolute value  due to time symmetry 
but opposite 
signs because of the opposite  directions that the outgoing unit normal has 
on these hypersurfaces. The integral over $i^0$ vanishes because $  
\tilde{\nabla}_a \tilde{\phi}$ vanishes there.\footnote{It is here that  the 
assumption of asymptotic flatness becomes crucial.} Finally, the  integral over 
the portion of the horizon $H$ vanishes because the  projection of 
$\tilde{\nabla}_a \tilde{\phi} $ along the null vector tangent to the horizon, 
which is a linear combination of $t^a$ and $\psi^a$, vanishes due to the 
symmetries \cite{Hawkingtheorem}. This  argument fails if the Einstein frame 
scalar $\tilde{\phi}$ or its gradient $\tilde{\nabla}_c \tilde{\phi}$ are not 
defined at the horizon.  This is the case of maverick solutions 
violating the 
Hawking theorem, in  which $\tilde{\phi} \propto \ln \phi $ diverges on the 
horizon because $\phi $ vanishes or diverges there; then the conformal 
transformation  becomes ill-defined at the horizon and the Einstein frame 
variables $\left( \tilde{g}_{ab}, \tilde{\phi} \right)$ cannot be used on this  
hypersurface. However, these situations are unphysical.

Hawking's theorem \cite{Hawkingtheorem} has been generalized, with the 
limitation of  spherical symmetry, to various scalar-tensor 
theories with varying $\omega $  
\cite{Bekenstein95,Bekenstein96,MayoBekenstein96,Santos00} and with 
potential $V(\phi)$, 
provided that $V$ has a minimum at the constant value $\phi_0$ at which 
the Brans-Dicke scalar stabilizes outside the horizon, so that the 
dynamics of $\phi$ stops there \cite{SotiriouFaraoniPRL}. A 
generalization to asymptotically de Sitter spaces was given by Bhattacharya {\em et al.}
\cite{BhattaRomanoPRL}. Assuming again spherical symmetry, Hawking's 
theorem can be proved also in the Jordan frame \cite{myJframeproof}. 
Indeed, numerical studies of black hole collapse in Brans-Dicke 
\cite{ScheelShapiroTeukolsky95a,ScheelShapiroTeukolsky95b} and 
scalar-tensor gravity \cite{KerimoKalligas92,Kerimo00} find the 
Schwarzschild black hole as the final state.

No-hair theorems for more general Horndeski and Galileon theories, and 
ways to evade them, are the subject of a large literature 
(\cite{Sotiriou:2013qea,Babichev:2016rlq, Sotiriou15,HerdeiroRadu15} 
and references therein) (see Sec.~\ref{sec:13}).

Weak forms of the Jebsen-Birkhoff theorem, analogous to those of 
\cite{Das60, Isaev76, Bronnikovetal76, BronnikovKovalchuk80} in GR, state 
that the geometry is static if the scalar field $\phi$ (or its effective 
stress-energy tensor) is static \cite{Reddy73, KroriNandy77, 
myJframeproof}. An error in Ref.~\cite{Reddy73} is corrected in 
\cite{KroriNandy77} and the extension of these results to cylindrical and  
planar symmetries is studied by Bronnikov and Kovalchuk \cite{BronnikovKovalchuk80}.

\section{The general non-black hole, spherical, static, asymptotically flat 
solution}
\label{sec:9}

Under the assumptions that: 1)~the vacuum Brans-Dicke field equations 
(with $\omega\neq -3/2$) hold in the Jordan frame; 2)~the geometry is 
spherically symmetric, static, and asymptotically flat; 3)~the 
massless Brans-Dicke 
scalar $\phi$ depends only on the radial coordinate $r$, does not have 
poles or zeros (except possibly for a central singularity), and $\phi(r) $ 
becomes constant as $r\rightarrow +\infty$, the general solution that is 
not a Schwarzschild black hole is known \cite{Bronnikov2,ourALC} and its 
nature depends on a scalar charge parameter. The only possibilities 
for the nature of these spacetimes (apart 
from the Schwarzschild black hole) are wormhole throats and central naked 
singularities. This is the closest known result to the Birkhoff theorem of 
GR in scalar-tensor gravity.

The history of this general solution is a bit confused. A theorem 
of Agnese and La Camera \cite{ALC} stated that the possible solutions 
describe only wormholes or naked singularities, but it is wrong as this 
would exclude the Schwarzschild black hole and contradict Hawking's 
no-hair theorem and its generalizations. The error of Agnese and La Camera \cite{ALC} consists 
of {\em assuming} the  line element and scalar field as
\begin{eqnarray}
ds^2_\text{ALC} &=& -\left( 1-\frac{2\eta}{r}\right)^A dt^2 + \left( 
1-\frac{2\eta}{r}\right)^B dr^2  \nonumber\\
&&\nonumber\\
&\, & + \left( 1-\frac{2\eta}{r}\right)^{1+B} r^2 d\Omega_{(2)}^2 \,, 
\label{ALC1}\\
&&\nonumber\\
\phi_\text{ALC}(r) &=& \phi_0 \left( 
1-\frac{2\eta}{r}\right)^{\frac{-(A+B)}{2}} \,,\label{ALC2}
\end{eqnarray}
with  
\be
1-\frac{ \omega+1}{\omega+2} =\frac{ (A+B)^2}{2(1+AB)} \,,\label{stocz}
\ee
where $A,B$, and $\eta$ are real constants. This is assumed 
to be  a permissible gauge choice in \cite{ALC} but it is instead a 
specific solution, which nowadays is called the Campanelli-Lousto 
solution. The general form of the Campanelli-Lousto solution is 
\cite{CampanelliLousto1, CampanelliLousto2} 
\begin{eqnarray}
ds^2_\text{CL} &=& -\left( 1-\frac{2\eta}{r}\right)^{b_0+1}dt^2 + 
\left( 1-\frac{2\eta}{r}\right)^{-a_0-1} dr^2  \nonumber\\
&&\nonumber\\
&\, & + \left( 1-\frac{2\eta}{r}\right)^{-a_0}r^2 d\Omega_{(2)}^2 \,, 
\label{CL1}\\
&&\nonumber\\
\phi_\text{CL}(r) &=& \phi_0 \left( 
1-\frac{2\eta}{r}\right)^{\frac{a_0-b_0}{2}} \,,\label{CL2}
\end{eqnarray}
where the parameters $a_0$ and $b_0$ satisfy 
\be
\omega =\frac{-2\left(a_0^2 +b_0^2 -a_0b_0 
+a_0+b_0\right)}{\left(a_0-b_0 \right)^2} \,. \label{omegaab}
\ee
The values $ a_0 = -B-1 $ and $ b_0 = A-1$ reproduce the Agnese-La Camera 
choice. Therefore, their results can only be true for this particular 
solution, which resolves the conflict with the no-hair theorems.  The 
Campanelli-Lousto solution was originally advertised as a family of black 
holes \cite{CampanelliLousto1,CampanelliLousto2}, but it describes only 
wormhole throats or naked singularities instead \cite{Vanzo}.

The general solution was instead obtained by Bronnikov \cite{Bronnikov2} 
(see also\footnote{A casual remark to this regard is found in 
\cite{BhadraSarkar}.} \cite{ourALC}). The main idea consists of going 
to the Einstein frame, in which the Brans-Dicke scalar field becomes a 
standard, minimally coupled, massless, matter scalar field and the general 
solution is well-known to be the FJNWBW metric~(\ref{FJNWBW}), (\ref{F1}) 
of GR. The general Jordan frame solution is then obtained by inverting the 
conformal transformation from the Jordan to the Einstein frame.

When the scalar charge $\sigma \neq 0$, the conformal image of the FJNWBW 
geometry of GR is the Jordan frame solution
\begin{eqnarray}
ds^2 &=&  -\, \mbox{e}^{ (\alpha+\beta)/r } dt^2
+ \, \mbox{e}^{ ( \beta-\alpha)/r  }
\left( \frac{ \gamma/r }{ \sinh ( \gamma/r ) } \right)^4 dr^2  \nonumber\\
&&\nonumber\\
&\, & + \, \mbox{e}^{ (\beta-\alpha)/r }  
 \left(
\frac{ \gamma/r }{ \sinh( \gamma/r) } \right)^2 r^2 d\Omega_{(2)}^2 
\,, \label{new1}
\end{eqnarray}
\be
\phi (r) = \phi_0 \, \mbox{e}^{-\beta/r}  \,,
\;\;\;\;\;\;\; \beta= \frac{\sigma}{\sqrt{|2\omega+3|} }  \,,
\label{new2}
\ee
when $\gamma\neq 0$ (the special case $\gamma=0$ is discussed later).

There is another form of the solution for imaginary $\gamma$, for which 
the hyperbolic sine in Eq.~(\ref{new1}) is replaced by a sine 
\cite{Bronnikov2}. 
This possibility (called a ``cold black hole'' 
in \cite{Bronnikov3,ColdBH2}) 
occurs for $2\omega+3<0$ and makes the Einstein frame $\tilde{\phi}$ 
purely imaginary, giving the wrong sign to its kinetic energy density in 
what can be regarded as the ghost counterpart of the FJNBW 
solution~(\ref{FJNWBW}), (\ref{F1}) discussed by Bergmann and Leipnik \cite{BergmanLeipnik57}. 
In this case, $\sigma$ becomes imaginary and the 
Wyman relation~(\ref{WymanRelation}) between parameters becomes  $ 
-4\gamma^2 = \alpha^2 
-2\sigma^2$ \cite{Bronnikov2}, where the negative signs can be absorbed  
making both $\sigma$ and $\gamma$ imaginary. This anomalous general 
solution is
\begin{eqnarray}
&&ds^2 = -\, \mbox{e}^{\frac{\alpha+\beta}{r}} dt^2 \nonumber\\
&&\nonumber\\
& & + \, \mbox{e}^{\frac{\beta-\alpha}{r}}
\!\!\left(\frac{\gamma/r}{\sin(\gamma/r)} 
\right)^2\left[\left(\frac{\gamma/r}{\sin(\gamma/r)}\right)^2dr^2   + r^2 
d\Omega_{(2)}^2\right] \,.\nonumber\\
&&
\end{eqnarray}
Its special cases $\alpha=\beta$, $\alpha=(2\omega+3)\beta$, and 
$\alpha=-(\omega+1)\beta$ were reported by Van den Bergh \cite{Bergh}.  A 
later and 
more complete analysis \cite{Bronnikov3} in the wider Bergmann-Wagoner 
class of scalar-tensor gravities with $\omega=\omega(\phi)$ found that 
black hole geometries (``cold black holes'') can occur in these anomalous 
solutions. We will not discuss this ghost solution further, focusing 
on~(\ref{new1}), (\ref{new2}) instead.

Next, one wonders about the relation of~(\ref{new1}), (\ref{new2}) with 
the previous Campanelli-Lousto (and other) solutions. To elucidate this 
question when 
$\gamma\neq 0$, one performs the two consecutive coordinate changes
\begin{equation}  \label{CoordTransformations}
\mbox{e}^{\gamma/r} = \frac{1+B/\rho}{1-B/\rho} \,,\qquad
\bar{r} = \rho \left( 1+\frac{B}{\rho} \right)^2 \,,
\end{equation}
introduces  $\eta=2B=\sqrt{m^2+\sigma^2} \,,  m/\eta 
=-\alpha/(2\gamma)$, 
$ \sigma/\eta=\beta \sqrt{|2\omega+3|}/(2\gamma)$,  and rescales the 
time coordinate by $|\gamma/(2B)|$, which recasts the pair~(\ref{new1}), 
(\ref{new2}) as
\begin{eqnarray}
ds^2 &=& -\left( 1-\frac{2\eta}{\bar{r}} \right)^{ \frac{1}{\eta} \left( 
m-\frac{\sigma}{ \sqrt{|2\omega+3|}}\right)} dt^2  + \left( 1-\frac{2\eta}{\bar{r}} \right)^{ \frac{-1}{\eta} \left(
m+\frac{ \sigma}{ \sqrt{|2\omega+3|} } \right)} d\bar{r}^2 \nonumber\\
&&\nonumber\\
&\, & + \left(
1-\frac{2\eta}{\bar{r}} \right)^{1 - \, \frac{1}{\eta} \left(
m+\frac{\sigma}{ \sqrt{|2\omega+3|}}\right)} \bar{r}^2 d\Omega_{(2)}^2 
\,,\label{CL2_1}\\ 
&&\nonumber\\
\phi & = & \phi_0 \left(
1-\frac{2\eta}{\bar{r}} \right)^{ \frac{\sigma}{\eta \sqrt{ |2\omega+3|} } 
} \,.\label{CL2_2}
\end{eqnarray}
This is exactly the form of the Campanelli-Lousto solution with 
parameters\footnote{The anomalous case $2\omega+3<0 $ cannot be obtained  
here because $\gamma$ is imaginary and the new coordinates 
(\ref{CoordTransformations}) 
are not real.}
\begin{eqnarray}
a_0 &=& -1+\frac{1}{\eta} \left( m+\frac{\sigma}{ 
\sqrt{|2\omega+3|}}\right) \,,\\
&&\nonumber\\ 
b_0 &=&  -1+\frac{1}{\eta} \left( 
m-\frac{\sigma}{\sqrt{|2\omega+3|}}\right) \,.
\end{eqnarray}
The Campanelli-Lousto metric (\ref{CL1}) \cite{CampanelliLousto1, 
CampanelliLousto2} is just another form of the general solution, but it is 
restricted by the validity of the new coordinates (and to $2\omega+3>0$). 
Other solutions appearing in the literature, for example the Brans 
Class~I-IV solutions\footnote{Originally, Brans \cite{Brans} reported four 
classes of static spherical solutions, but later 
on it was found that pairs of these classes are related by dualities 
\cite{BhadraSarkar,BhadraNandi15,VFS}.} \cite{Brans} fall in the general 
form~(\ref{new1}, 
(\ref{new2}). All this is consistent with independent realizations 
\cite{HeKim02,BhadraSarkar,Bhadraetal05,Vanzo,VFS} that 
Campanelli-Lousto and Brans solutions describe wormhole throats or naked 
singularities, but not black holes.

In order to assess the physical nature of the solutions, one classifies 
the possible roots of $ \nabla^c R \nabla_c R=0$  
\cite{MSH1,NielsenVisser, AHbook} where
\be
R(r)= \gamma\, \frac{ \mbox{e}^{\frac{\beta-\alpha}{2r}} 
}{\sinh(\gamma/r)}  \label{arealradius}
\ee
is the areal radius.  We have 
\begin{eqnarray} 
&& \nabla^c R \nabla_c R = g^{rr} \left( \frac{dR}{dr} \right)^2 
=  \sinh^2(\gamma/r)  \left[  \frac{\alpha -\beta}{2\gamma} + \coth 
\left( \gamma/ r \right)   \right]^2  =0 \,.\nonumber\\
&&  
\end{eqnarray}
Real roots 
\be
r_\text{H}= \frac{2\gamma}{\ln \left( \frac{\beta-\alpha 
+2\gamma}{\beta-\alpha-2\gamma}\right)} = \frac{\gamma}{ \tanh^{-1} \left( 
\frac{2\gamma}{\beta-\alpha}\right)} \,, 
\ee
exist if $(\beta-\alpha)/\gamma >0$, and  are always double roots  
corresponding to wormhole throats. 
If $(\beta -\alpha)/\gamma<0$, instead, there is a naked singularity at 
$R=0$, highlighted by the divergence of the Ricci scalar 
\be
{\cal R}= \frac{\omega}{\phi^2} \nabla^c \phi \nabla_c\phi 
=\left\{ 
\begin{array}{cc}
 \frac{ \omega \beta^2}{\gamma^4}  \, \mbox{e}^{\left( 
\alpha-\beta\right)/ r }  
\sinh^4(\gamma/r) & \;\; \mbox{if} \; \gamma\neq 0  \,,\\
&\\
\frac{ \omega \beta^2}{r^4}  \, \mbox{e}^{\left( 
\alpha-\beta\right)/ r }  & \;\; \mbox{if} \; \gamma= 0  \,.
\end{array} \right.
\ee
In fact, if $\gamma\neq 0$ then when $r\rightarrow 0$ we have
\be
{\cal R}= \frac{\omega \beta^2}{16\gamma^4} \, \mbox{e}^{\left( 
\alpha-\beta \pm 4\gamma\right)/r} \,,
\ee
where the upper [lower] sign applies if $\gamma>0  $ [$\gamma<0$]. 
${\cal R}$  diverges as $r \rightarrow 0$ 
only for $\beta -\alpha<4\gamma$ or for $\alpha-\beta > 4\gamma$, 
respectively.

Let us discuss now the case $\gamma=0$:  the Einstein frame FJNWBW metric 
degenerates into the Yilmaz 
geometry \cite{Yilmaz58}, while  its Jordan  frame  image~(\ref{new1}), 
(\ref{new2}) becomes the Brans Class~IV solution  
\begin{eqnarray}
ds^2 &=& -\, \mbox{e}^{ -2B/r }dt^2 + \, \mbox{e}^{ 2B(C+1)/r }
\left( dr^2+r^2 d\Omega_{(2)}^2 \right) \,,\\
&&\nonumber\\
\phi&=& \phi_0 \, \mbox{e}^{-BC/r} \,,
\end{eqnarray}
found by Brans \cite{Brans} immediately after the introduction of Brans-Dicke 
theory, and where $B=-(\alpha+\beta)/2, C=-2\beta/(\alpha+\beta)$. 
The apparent horizons are now located by the roots of $ \left( 1- 
\frac{\beta-\alpha}{2r} \right)^2=0 $. There is a double root 
$r_\text{H}=(\beta-\alpha)/2$ (and, therefore, a wormhole throat) if 
$\beta>\alpha$ and a naked central singularity otherwise, which matches 
the results of studies of the specific Brans~IV 
class \cite{Wormholes,VFS}.

The most general electrovacuum solution of the generalized scalar-tensor 
theory which is static, spherical, and asymptotically flat was found by 
Bronnikov {\em et al.} \cite{Bronnikov3}.

\section{A selection of dynamical solutions of Brans-Dicke theory}
\label{sec:10} 

In the following we review analytic and spherically symmetric solutions 
of scalar-tensor 
theories. These solutions are not general solutions in any sense. Usually, 
their stability with respect to perturbations has not been studied and 
they are probably fine-tuned. Their physical significance is often 
questionable, especially considering that many of them describe naked 
singularities or wormhole throats, nevertheless they provide some insight 
on the physical properties of these theories and are useful as examples, 
the catalogue of which is rather meagre.

\subsection{Clifton-Mota-Barrow spacetimes}

A family of solutions of the Brans-Dicke field equations with a perfect 
fluid as the matter source was found by Clifton {\em et al.} \cite{CMB}. 
This fluid has energy density $\rho^\text{(m)}$, pressure 
$P^\text{(m)}$, and constant barotropic equation of 
state $ P^\text{(m)}=\left( \gamma -1 \right) 
\rho^\text{(m)} $. The line element and Brans-Dicke scalar are spherically 
symmetric and dynamical \cite{CMB}
\be\label{CMBmetric}
ds^2=-e^{\nu (\bar{r} )}dt^2+a^2(t) e^{\mu 
(\bar{r} )}(d\bar{r} ^2+\bar{r}^2d\Omega^2_{(2)}) \,,
\ee
where 
\begin{eqnarray} 
e^{\nu (\bar{r} )} & = &  
\left(\frac{1-\frac{m}{2\alpha \bar{r} }}{1+\frac{m}{2 \alpha \bar{r} 
}}
\right)^{2\alpha 
}\equiv A^{2\alpha} \,,\\
&&\nonumber\\
e^{\mu (\bar{r} )} & = & \left(1+\frac{m}{2\alpha \bar{r} }\right)^{4} 
A^{\frac{2}{\alpha}( \alpha-1)(\alpha +2)} \,, \\
&&\nonumber\\
\label{abeta}
a(t) & = & a_0\left(\frac{t}{t_0}\right)^{\frac{ 
2\omega (2-\gamma)+2}{3\omega \gamma(2-\gamma)+4}}\equiv 
a_{\ast}t^{\beta} \,,\\
&&\nonumber\\
\label{scalart}
\phi(t, \bar{r} ) &= & 
\phi_0\left(\frac{t}{t_0}\right)^{\frac{2(4-3\gamma)}{ 
3\omega \gamma(2-\gamma)+4}}A^{-\frac{2}{\alpha }(\alpha^2-1)} \,,\\
&&\nonumber\\
\alpha & = & \sqrt{ \frac{ 2( \omega +2 )}{2\omega +3} } 
\,,\label{7}\\
&&\nonumber\\
\rho^\text{(m)}(t, \bar{r} ) & = & \rho_0^\text{(m)} \left( \frac{ 
a_0}{a(t)} 
\right)^{3\gamma} A^{-2\alpha} \,. \label{density}
\end{eqnarray}
 $m$ is a mass parameter, while $\alpha, 
\phi_0, a_0$, $\rho^\text{(m)}_0$ and 
$t_0$ are positive constants ($\alpha>0$ requires $\omega <-2$ or  $ 
\omega > -3/2$), and  $\bar{r} $ is the  isotropic radius 
(the  Schwarzschild or curvature radial coordinate $r$ is given by 
$r \equiv \bar{r} \left(1+\frac{m}{2\alpha \bar{r} }\right)^{2}$). We 
assume that $ \omega>-3/2 $ and $ \beta \geq 0 $.  The 
line element~(\ref{CMBmetric}) reduces to spatially flat FLRW if $m=0$. If 
$ \gamma \neq 2$, then  
$ \omega =\left( \gamma -2 \right)^{-1} $ gives  
$ \beta =0$ and the spacetime becomes static with a 
time-dependent scalar field. If instead $ \gamma=2 , 4/3 $, then $ 
\beta =1/2 $ and  $ a(t) \sim \sqrt{t} $ regardless of 
the value of  $ \omega$.

The solution can be rewritten in terms of the areal radius  
$R  =  A^{\frac{1}{\alpha}( \alpha -1)( \alpha +2)} $, obtaining 
\cite{FVSL2012cmbblackholes, AHbook}
\begin{eqnarray}
&& ds^2= -A^{2\alpha}DF^2d\bar{t}^2+\left(\frac{H^2}{B^4D}R^2 
A^{2(2-\alpha )}+\frac{A^2}{B^2}\right)dR^2 + R^2d\Omega^2_{(2)}  \,, \nonumber \\
& &
\end{eqnarray}
where $ H \equiv \dot{a}(t)/a(t) $, $F$ is an integrating 
factor, 
\be \label{B(R)}
B(\bar{r} ) \equiv A^2(\bar{r} )+ \frac{( \alpha -1)( \alpha + 
2)}{\alpha^2} \, 
\frac{m}{r}  >0 \,,
\ee
\be
\psi=\frac{\dot{a}(t) r}{B^2} \, 
\frac{A^{\frac{-\alpha^2+3\alpha -2}{\alpha}}}{D(t,\bar{r} )} \,,
\ee
and
\be
D(t, \bar{r} )\equiv 1-\frac{\dot{a}^2(t) 
r^2}{B^2}A^{\frac{4}{\alpha}( 
\alpha -1)}\,,
\ee
The apparent horizons which (when they exist) are the positive
roots of   ${\displaystyle g^{RR}=0 }$ have been identified in
\cite{FVSL2012cmbblackholes} for various values of the parameters. A rich 
variety of phenomenologies is obtained, including appearing and 
disappearing pairs of apparent horizons, no horizons, single horizons, and 
collapsing horizons. When $m=0$ there is, of course, a single FLRW 
apparent horizon of areal radius  $R =H^{-1}$, and this horizon is found 
also as $\bar{r} \rightarrow +\infty$.

The static case  $ \beta=0$ is obtained for $ \omega=\left( \gamma 
-2\right)^{-1} $ with $ \gamma\neq 2$ (with $ 
 \omega<-2$ or $ \omega >-3/2 $ implying $ \gamma >3/2 $ or $ \gamma<4/3 $ 
 when $ \beta=0$). This static spacetime always contains a naked central 
singularity \cite{FVSL2012cmbblackholes}.

The limit to GR $\omega \rightarrow \infty $ when $ 
\gamma \neq 0 , 2 $ produces  $ \alpha \rightarrow 1$, 
$ \phi \rightarrow \phi_0 $, and the spacetime 
\begin{eqnarray} 
ds^2 & = & - \left( \frac{ 1-\frac{m}{2\bar{r} } }{ 
1+\frac{m}{2\bar{r} } 
}\right)^2 dt^2 +  
a^2(t) \left( 1+\frac{m}{2\bar{r} } \right)^4  \left( 
d\bar{r}^2+\bar{r}^2 d\Omega^2_{(2)} \right) \,, \nonumber\\ 
&& \label{grlm}\\
&&\nonumber\\
a(t) & = & a_0 \left( \frac{t}{t_0} \right)^{ \frac{2}{3\gamma}} 
\,,\\
&& \nonumber\\
\rho^\text{(m)}(t) & = & \rho_0^\text{(m)} \left( \frac{t_0}{t} 
\right)^2 
A^{-2}\,.
\end{eqnarray}
which is the generalized McVittie metric in isotropic coordinates. The 
mass 
function $ M(t)\equiv ma(t) \geq 0 $ is an arbitrary function of time and $ 
G^1_0 \neq 0 $, corresponding to a radial energy flow. The non-rotating 
Thakurta solution of the Einstein equations with $ M(t)=M_0 \, a(t) $ and 
$M_0=$~const., which is a late-time attractor in the class of generalized 
McVittie solutions, is the GR limit of the Clifton-Mota-Barrow family of 
Brans-Dicke spacetimes.

For $ \gamma=0 $ the cosmic fluid reduces to a cosmological 
constant, $\beta \rightarrow \infty$ as 
$ \omega \to \infty $ and $a(t)$  
becomes a power law. If $ \gamma=2 $, then the  $ 
\omega\to \infty $ limit 
yields $ \alpha\to 
1$, $ \phi\propto t^{-1} $,  
$ a(t)\propto \sqrt{t} $, and  
$ \rho^\text{(m)}\propto t^{-3} A^{-2} $.

\subsection{Conformally transformed Husain-Martinez-Nu\~nez 
spacetime}

Clifton {\em et al.} \cite{CMB} generated another two-parameter 
family of dynamical and spherically symmetric solutions of the Brans-Dicke 
field equations by mapping the Husain-Martinez-Nu\~nez geometry of GR to 
the Jordan frame. The result is \cite{CMB}
\begin{eqnarray} 
ds^2 &=& -A^{\alpha \left( 
1-\frac{1}{\sqrt{3}\, \beta}\right)} (r) \, dt^2  +A^{-\alpha \left( 
1+\frac{1}{\sqrt{3}\, \beta}\right)} (r) \, t^{\frac{2 \left( 
\beta-\sqrt{3}\right)}{3\beta - \sqrt{3}} }  \nonumber\\ 
&&\nonumber\\
&\,&  \times \left[ dr^2 + r^2 A(r) 
d\Omega_{(2)}^2 \right] \,,\\ 
&&\nonumber\\
\phi( t,r) &=& A^{\frac{\pm 1}{2\beta}} (r)\, 
t^{\frac{2}{\sqrt{3}\, \beta -1}} \,, \label{cousinmetric}
\end{eqnarray} 
where  $\omega>-3/2$ and 
\be 
 A(r) = 1-\frac{2C}{r} 
\,,\;\;\; \beta =\sqrt{2\omega+3} \,, \; \;\;
\alpha=\pm \sqrt{3}/2 \,.
\ee 
This is an inhomogeneous spacetime with a spatially flat FLRW 
``background'' with scale factor 
\be
a(t)= t^{\frac{\beta-\sqrt{3}}{3\beta-\sqrt{3}} } \,. 
\ee 
The apparent horizons and singularities of this rather involved solution 
are reported in \cite{FaraoniZambranocousin,AHbook}.

\subsection{Geometry conformal to Fonarev}

The Fonarev solution of GR, used as a seed and mapped to the Jordan frame, 
generates a four-parameter family of solutions of Brans-Dicke theory with 
a power-law potential, which are dynamical, spherically symmetric, and 
asymptotically FLRW. It contains as special cases two previously known 
classes of solutions and solves also the field equations of $f({\cal 
R})={\cal R}^n$ gravity. The Jordan frame scalar field is 
\cite{confonarev}
\be
\phi (t,r)= \phi_0 \, \mbox{e}^{ \frac{4\alpha at}{\sqrt{ 
|2\omega+3|}} }\left(1-\frac{2m}{r} \right)^{ \frac{1}{ 
\sqrt{|2\omega+3|(1+4\alpha^2)}} }  \label{21}
\ee
and the corresponding  scalar field potential is $ V(\phi)=V_0 \, 
\phi^{2\beta} $ with
\be
\beta=1-\alpha \sqrt{|2\omega+3|} \,, \;\;\;\;\;\;
V_0=\tilde{V}_0 \, \phi_0^{2\alpha \sqrt{|2\omega+3|}} 
\label{23} \,.
\ee
In this case, the potential is physically motivated and well studied in 
cosmology and particle physics, including the pure mass, quartic, and 
many quintessence potentials \cite{Linde90,LiddleLyth, 
PeeblesRatra88, PeeblesRatra03, WandsCopelandLiddle93,   
AmendolaTsujikawabook}. The usual relation $g_{ab}=\phi^{-1} 
\tilde{g}_{ab} $ gives the Jordan frame line element
\begin{eqnarray}
&&ds^2 = - {A(r)}^{\frac{1}{\sqrt{1+4\alpha^2}} \left( 
2\alpha  -\frac{1}{\sqrt{ |2\omega+3|}} \right) } 
\mbox{e}^{ 4\alpha 
at \left( 2\alpha -\frac{1}{\sqrt{|2\omega+3|}} \right)} 
dt^2 \nonumber\\
&&\nonumber\\
& & + \, \mbox{e}^{ 
2at \left(1-\frac{2\alpha}{\sqrt{|2\omega+3|} } \right)}
\left[ {A(r)}^{ -\frac{1}{\sqrt{1+4\alpha^2} } 
\left( 2\alpha +\frac{1}{\sqrt{|2\omega+3|} } \right)}  
dr^2 \right. \nonumber\\
&&\nonumber\\
& & \left. + {A(r)}^{1-\frac{1}{ \sqrt{1+4\alpha^2}} 
\left( 
2\alpha + \frac{1}{\sqrt{|2\omega+3|} } \right) } r^2 
d\Omega_{(2)}^2 \right] \,. \label{24}
\end{eqnarray}
This is a family of solutions of the vacuum 
Brans-Dicke field equations parametrized by the four 
parameters $\left( \omega, m, a, \alpha \right)$ (the last three being 
parameters of this specific family).  Special cases include the $V=0$  
conformal counterpart of the Husain-Martinez-Nu\~nez solution and the 
general static solution in the Campanelli-Lousto form. This solution is, 
again, complicated and contains naked singularities or wormhole throats 
for most values of the parameters \cite{confonarev}. 

A special subcase of the Brans-Dicke solution conformal to Fonarev, which 
is static but has time-dependent scalar field, was used to generate new 
dynamical solutions using the symmetry (\ref{symmetry1})-(\ref{newomega}) 
of the vacuum field equations \cite{DilekShawn}.

\section{Nonminimally coupled scalar fields}
\label{sec:11}

The action of gravity with a nonminimally coupled scalar field is
\begin{eqnarray}  
S_\text{(NMC)} &=& \int d^4x \sqrt{-g} \Bigg[ \left( \frac{1}{2\kappa} 
-\frac{\xi}{2} \, \phi^2 \right) {\cal R}  -\frac{1}{2}
\nabla^c \phi \nabla_c \phi -V( \phi ) \Bigg] + 
S^\text{(m)} \, , \nonumber \\
& &
\label{nmaction} 
\end{eqnarray}
where $\xi$ is a dimensionless coupling constant.  The value $\xi=1/6$ 
(conformal coupling) makes the physics of $\phi$
conformally invariant if $ V=0 $ or $ V=\lambda \phi^4 $, in which case 
$V$ is scale-free   
\cite{Waldbook}. Varying  the action~(\ref{nmaction}) produces the field 
equations 
\begin{eqnarray}
&& \left( 1-\kappa \xi  \phi^2 \right) G_{ab} = \kappa \Big\{
\nabla_a\phi
\nabla_b \phi -\frac{1}{2} \, g_{ab} \, \nabla^c\phi \, \nabla_c \phi  \nonumber\\
&& -V\,g_{ab}  +  \xi \, \left[ g_{ab} \Box \left( \phi^2 \right)
-\nabla_a \nabla_b
\left( \phi^2 \right) \right] \Big\} \,,  \label{pippa2} \\
&& \nonumber\\
&& \Box \phi  -\frac{dV}{d\phi} -\xi {\cal R} \phi =0 \,.   \label{nmKG}
\end{eqnarray}
They describe a scalar-tensor gravity; in 
fact, by redefining the scalar field and its potential as 
\be   \label{abba76}
\varphi = 1-\kappa \xi \phi^2  \,,
\ee
\be     \label{abba79}
U \left( \varphi \right) =16\pi V \left[ \phi \left( \varphi \right)
\right] = 16 \pi V \left( \pm \, \sqrt{ \frac{ 1-\varphi}{\kappa \xi } }
\,\, \right) \,,
\ee
the nonminimally coupled action~(\ref{nmaction}) becomes
\begin{eqnarray}
S_\text{(NMC)} &=& \int d^4x \, \frac{ \sqrt{-g}}{16\pi} \left[ \varphi 
{\cal R}
-\frac{ \omega \left( \varphi \right)}{\varphi} \, \nabla^c \varphi \,
\nabla_c \varphi -U \left( \varphi \right) \right] \nonumber\\
&&\nonumber\\
&\, & +S^\text{(m)}  \,,
\end{eqnarray}
where 
\be 
\omega \left( \varphi \right) = \frac{ \varphi}{4\xi \left( 1- \varphi
\right)} \,.
\ee

The nonminimal coupling of the scalar field to the Ricci scalar was 
originally introduced by Chernikov and Tagirov \cite{ChernikovTagirov68} 
in the context of classical 
radiation problems. It became widely known when Callan {\em et al.}
\cite{CCJ70} used it to 
renormalize a scalar  with quartic self-interaction in curved space. A 
nonzero $\xi$ is generated by first loop corrections 
even if absent at the classical level 
\cite{BirrellDaviesbook,BirrellDavies80,NelsonPanangaden82,FordToms82, 
ParkerToms85,Ford87}. At the classical level, conformal coupling is 
necessary to avoid causal pathologies in the propagation of $\phi$, {\em 
i.e.}, to avoid any possibility that a massive scalar field propagates 
strictly along the light cone \cite{SonegoFaraoni93}. More 
precisely, waves propagating in a curved spacetime suffer 
backscattering from the background curvature and sharp propagation is the 
exception rather than the norm \cite{EllisSciama}. For scalar waves 
$\phi$, assuming the 
potential to consist of a mass term $V(\phi)=m^2\phi^2/2$,  
the wave equation~(\ref{nmKG}) admits a retarded Green function 
$G_R(x',x)$ as the solution of
\be     \label{3}
\left[ g^{a'b'}(x') \nabla_{a'}\nabla_{b'}-m^2-\xi R(x') \right]
G_R(x',x)=-\, \delta(x',x)    
\ee
with an impulsive source. Here $\delta(x',x)$ is the delta function 
on spacetime \cite{DeWitt:1960fc} with the boundary condition  
$G_R(x',x)=0$ if $x$ is in the future of $x'$. Let $x$ and $x'$ be 
points in a normal domain and let  $\Gamma(x',x)$ be the 
square of the proper distance  along the unique geodesic 
connecting these points. Then, it is well known that 
$G_R(x',x)$ decomposes as \cite{Friedlander}
\be         \label{xi4}
G_R(x',x)=\Sigma(x',x) \,\delta_R(\Gamma(x',x))+V(x',x)
\,\Theta_R(-\Gamma(x',x)) \; ,    
\ee
where 
$\delta_R$ and $\Theta_R$ are the Dirac delta distribution and the 
Heaviside step function with support in the past of $x'$. The coefficients  
 $\Sigma$ and $V$ are functions of the spacetime position    
specified  uniquely in a given geometry \cite{DeWitt:1960fc,Friedlander}.  
Taylor-expanding  around $x$ gives \cite{SonegoFaraoni93} 
\begin{eqnarray}
\Sigma \left(x',x \right) &=& \frac{1}{4\pi}+ \ldots \,, \label{xi7}\\
&&\nonumber\\                             
V \left( x',x \right) &=& -\,\frac{1}{8\pi}\left[ 
m^2+\left(\xi-\frac{1}{6}\right)R(x)\right]  + \ldots \,.\label{xi9} 
\end{eqnarray}
Locally, the propagation of $\phi$-waves  must be as in Minkowski 
space: if $m=0$, there should be no tails and, since in general $R(x)\neq 
0$,  this property is guaranteed only if $\xi=1/6$.
If $\xi\neq 1/6$ there is the very disturbing possibility that a massive 
field propagate {\em exactly} on the light cone where  $ 
m^2+\left( \xi-\frac{1}{6}\right) R(x)=0 $ ({\em i.e.}, where the tail due 
to the field mass cancels exactly the tail due to backscattering off the 
background spacetime curvature). Indeed, one 
can even tailor constant curvature spaces where this happens at every 
spacetime point \cite{Faraoni:1999us}. The only way to prevent this 
unphysical feature once and for all is by requiring that $\xi=1/6$.

 It has been argued 
that nonminimal coupling is necessarily present at high curvatures 
\cite{FordToms82,Ford87}, that the approach to a classical universe in 
quantum cosmology requires $\xi \neq 0$ \cite{Okamura98}, and that 
nonminimal coupling could solve potential problems of primordial 
nucleosynthesis \cite{Chenetal01}. Nonminimal coupling has been used 
extensively in cosmology 
\citep[{\em e.g.},][]{Abbott81,LucchinMatarresePollock86, 
FutamaseMaeda89,Futamaseetal89,FaraoniPRD96}. 
Various arguments are used to advocate for different values of the 
coupling constant $\xi$ \cite{VoloshinDolgov82, HillSalopek92, Reuter94, 
Hosotani85, BuchbinderOdintsov83,BuchbinderOdintsov85, 
ElizaldeOdintsov94,book, Buchbinderetal89, Bonanno95, FaraoniPRD96, 
FutamaseTanaka99}.

The map to the Einstein conformal frame\footnote{Often referred 
to as Bekenstein transformation.} assumes the form  
\begin{eqnarray}  
g_{ab}  & \rightarrow & \tilde{g}_{ab} = \Omega^2 \, g_{ab} \,, 
\quad \quad \Omega=\sqrt{1-\kappa \xi \phi^2} \,,\label{nmc500}\\
&&\nonumber\\
d \tilde{\phi}& =& \frac{\sqrt{ 1-\kappa \xi ( 1-6\xi ) \phi^2 } 
}{1-\kappa
\xi\phi^2} \,\, d\phi  \, .   \label{nmc501}
\end{eqnarray}
The integration of Eq.~(\ref{nmc501}) gives
\be  \label{nmcphif}
\tilde{\phi}= \sqrt{ \frac{3}{2\kappa}} \ln \left[ 
\frac{ \xi \sqrt{6\kappa\phi^2} + \sqrt{1-\xi \left( 1-6\xi \right) \kappa
\phi^2}}
{\xi \sqrt{6\kappa\phi^2} -\sqrt{1-\xi \left( 1-6\xi \right) \kappa
\phi^2}}  \, \right] + f \left( \phi \right) \, ,
\ee
where 
\be 
f \left( \phi \right) = 
\left( \frac{1-6\xi}{\kappa \xi} \right)^{1/2} \arcsin
\left( \sqrt{ \xi \left( 1-6\xi \right)  \kappa\phi^2} \right)
\ee
for $0 < \xi < 1/6 $ and
\be  \label{nmcphi2}
f \left( \phi \right) = 
\left( \frac{6\xi -1 }{\kappa \xi}  \right)^{1/2} 
\mbox{ arcsinh}
\left( \sqrt{ \xi \left( 6\xi -1 \right)  \kappa \phi^2}  \, \right) 
\ee
for $ \xi > 1/6 $.  When $\xi=1/6$, we have
\be  
\tilde{\phi}= \sqrt{ \frac{3}{2\kappa}} \ln  \left(  \frac{ 
 \sqrt{ 6/ \kappa} + \phi }{
 \sqrt{ 6/ \kappa} - \phi }  \right) \;\;\;\;\;\;\;\;\;\;
\mbox{if} \;\; \left| \phi \right| < \sqrt{ \frac{6}{\kappa} } \,,
\ee
or
\be  
\tilde{\phi}= \sqrt{ \frac{3}{2\kappa}} \ln  \left(  \frac{ 
 \phi - \sqrt{ 6/ \kappa} }{
 \phi+  \sqrt{ 6/ \kappa}  }  \right) \;\;\;\;\;\;\;\;\;\;
\mbox{if} \;\; \left| \phi \right| > \sqrt{ \frac{6}{\kappa} } \, .
\ee
The Einstein frame scalar $\tilde{\phi} $ is minimally coupled and
satisfies 
\be
\widetilde{\Box} \tilde{\phi} - \frac{d \tilde{V}}{d\tilde{\phi}}=0 \, ,
\ee
where 
\be   \label{nmcVtilde}
\tilde{V} \left( \tilde{\phi} \right) = \frac{ V \left[  \phi \left(
\tilde{\phi} \right) \right]}{\left( 1-\kappa\xi\phi^2 \right)^2} \,,
\ee
with $\phi=\phi \left( \tilde{\phi} \right) $ as in 
Eq.~(\ref{nmcphif}).

\subsection{Conformal coupling}

The value $1/6$ of the nonminimal coupling constant $\xi$ is probably the 
most well studied, also in terms of exact solutions of the field 
equations.

\subsubsection{BBMB solution}  

The Bocharova-Bronnikov-Melnikov-Bekenstein (BBMB) solution for gravity, 
the conformally coupled scalar, and the Maxwell field was found early on 
by Bocharova {\em et al.} \cite{BBM70} and rediscovered by Bekenstein
\cite{Bekenstein74}.\footnote{It was rediscovered again by 
Fr{\o}yland \cite{Froyland82} by integrating directly the Jordan frame field 
equations instead of using the conformal mapping trick  (a similar, but 
rather confused, attempt was made in Ref.~\cite{AgneseCamera85}).} The 
BBMB 
solution written in terms of the areal radius is \cite{Bekenstein74}
\begin{eqnarray} 
ds^2 &=& - \left(1-\frac{m}{r} \right)^2 dt^2
+ \frac{dr^2}{ \left(1-m/r \right)^2} +r^2 d\Omega_{(2)}^2 \,, \nonumber\\
&&\\
\phi(r) &=& \sqrt{ \frac{6}{\kappa}} \, \frac{m}{r-m} \,, \\
&&\nonumber\\
F_{ab} &=& \frac{Q}{r^2} \left( \delta^1_{a} \, \delta^0_{b} - 
\delta^0_{a} \, \delta^1_{b} \right)\,,   \label{BBMBmetric}
\end{eqnarray}
where $m$ and $Q$ are the mass and electric charge and $F_{ab}$ is the 
Maxwell tensor. The geometry is that of an extremal Reissner-Nordstr\"om 
black hole, but the scalar field $\phi$ is singular on the horizon $r=m$. 
This property is unphysical, as originally remarked by Bekenstein himself 
\cite{Bekenstein74} and contrary to what later argued in 
\cite{Bekenstein75}, but reconsidered in 
\cite{XanthopoulosZannias91, XanthopoulosDialynas92, 
Zannias95}.\footnote{Curiously, 
in higher dimension the (unique) 
analogue of the BBMB solution describes a naked singularity 
\cite{XanthopoulosZannias89, 
XanthopoulosDialynas92,Klimcik93,WehusRavndal07}.} It has also been 
argued that, 
in advanced Eddington-Finkelstein coordinates extending beyond the 
horizon, the BBMB geometry fails to satisfy the field equations at the 
horizon \cite{Zannias95,SudarskyZannias98}. The usual black 
hole thermodynamics becomes impossible \cite{Zavlaski}.  Moreover, the BBMB 
solution is unstable to linear perturbations 
\cite{BronnikovKireyev78,ZouMyung20}.

Following the derivations of Bocharova {\em et al.}
\cite{BBM70} and Bekenstein \cite{Bekenstein74},  
Xanthopoulos and Zannias \cite{XanthopoulosZannias91}, and later Klimc{\'i}k \cite{Klimcik93}, 
proved explicitly that the BBMB construct is the unique solution of the 
Einstein-conformal scalar field equations which is static, spherical, 
asymptotically flat, and does not have constant $\phi$. A new proof of the 
uniqueness of the BBMB solution outside the photon surface (the surface 
composed of the unstable circular photon orbits) was given in Tomikawa {\em et al.} 
\cite{TomikawaShiromizuIzumi17a,TomikawaShiromizuIzumi17}. This proof 
shows also that no static multiple disconnected photon surfaces exist, and 
it does not assume the existence and connectedness of the photon sphere.

Rotating extensions of the BBMB black hole are derived by Astorino 
\cite{Astorino15} (a previous attempt to build a slowly rotating BBMB 
solution in \cite{BatthacharyaMaeda14} failed near the horizon because it 
was subject to the excessively restrictive assumption of separability for 
the $ g_{t\varphi}$ metric coefficient).  The BBMB solution has also been 
generalized by including a cosmological constant plus a quartic potential 
$V(\phi)=\lambda \phi^4$, a Maxwell field, and different horizon 
topologies \cite{MartinezTroncosoZanelli03, VirbhadraParikh94a, 
MartinezStaforelliTroncoso06}. Adding a cosmological constant moves the 
horizon outward and the scalar field singularity remains hidden within it 
\cite{MartinezTroncosoZanelli03}, making thermodynamics meaningful again 
\cite{BarlowDohertyWinstanley05}. An accelerating BBMB black hole has also 
been found by Charmousis {\em et al.} 
\cite{CharmousisKolyvarisPapantonopoulos09}. The Pleba\'nski-Demia\'nski 
family of solutions of the Einstein equations (the most general Petrov 
type D solution of the Einstein-Maxwell equations 
\cite{PlebanskiDemianski76,GriffithsPodolsky06}, which describes an 
accelerating and rotating black hole), generalized to include a scalar 
field \cite{AnabalonMaeda10}, contains the BBMB geometry and its 
accelerated version of Charmousis {\em et al.} 
\cite{CharmousisKolyvarisPapantonopoulos09} as special cases. Since 
the geometry is that of an extremal Reissner-Nordstr\"om black hole, the 
Penrose-Carter diagram is well known (see, {\em e.g.}, 
Ref.~\cite{Bengtsson:2014fha}.

The field equations for trace-free matter have been studied recently by  
Carranza {\em et al.} \cite{CarranzaHursitValienteKroon19}. The solution of the coupled 
equations for gravity, a conformally coupled scalar field, and the 
source-free Maxwell field or a radiation fluid is again the BBMB 
geometry~(\ref{BBMBmetric}) of the extreme Reissner-Nordstr\"om black hole 
already discussed.

The BBMB geometry is also a solution of a conformally invariant gravity 
theory, in which different conformal frames are pure gauge choices and the 
singularity of the conformal scalar field at the horizon is automatically 
resolved by changing gauge \cite{Prester14}.

\subsubsection{Abreu {\em et al.} generalization to scalar field 
potential}

Bekenstein's solution-generating technique for a free conformally coupled 
scalar field \cite{Bekenstein74} was used by Abreu {\em et al.} 
\cite{Abreuetal94} for a scalar subject to a non-zero potential 
$V(\phi)$. 
Although their interest was in FLRW cosmology, the technique can be 
used for inhomogenous spherical spacetimes. Starting from a solution of 
the minimally coupled ($\xi=0$)  Einstein-Klein-Gordon equations
 $\left( \tilde{g}_{ab}, \tilde{\phi}, U(\tilde{\phi}) 
\right)$ as a seed, one generates a solution of the conformally coupled 
($\xi=1/6$) equations $\Big( g_{ab}, \phi, V(\phi ) \Big)$, where
\begin{eqnarray} 
g_{ab} &=& \Omega^{-1} \tilde{g}_{ab} \,, 
\;\;\;\;\;\;\; \Omega^{-1}=\cosh (\zeta \tilde{\phi} ) \,,\\
&&\nonumber\\
\phi &=& \zeta^{-1} \tanh ( \zeta \tilde{\phi} ) \,,\\
&&\nonumber\\
V(\phi) &=& \left(1-\zeta^2\phi^2 \right) U \left[ \frac{1}{2\zeta} \ln 
\left( \frac{ 1+\zeta \phi}{1-\zeta\phi} \right) \right] \,, 
\end{eqnarray} 
and $\zeta = \sqrt{\kappa/6}$ \cite{Abreuetal94}.

\subsection{Other time-dependent solutions}

Other time-dependent solutions of scalar-tensor gravity that we cannot 
analyze here for lack of space include those of Shaw and Barrow 
\cite{SakaiBarrow01,ShawBarrow06}, and Banijamali {\em et al.} \cite{AliBehnaz}. 
A common generating technique 
for time dependent Einstein-scalar solutions takes known seed solutions 
with minimally coupled massless scalar fields and, using the conformal map 
to the Jordan frame, produces solutions with a nonminimally coupled field 
and with a perfect fluid interpretation ({\em e.g.}, \cite{RobertsNMC, 
Sultana15, AliBehnaz, DilekShawn,Fahim:2020uua}).

\section{$\bm{f ({\cal R})}$ gravity}
\label{sec:12} 

Quantum corrections to the Lagrangian density of GR introduce  
quadratic terms in the curvature. The simplest such scenario is 
Starobinsky's model 
of inflation in $f({\cal R})={\cal R}+\alpha {\cal R}^2$ gravity 
\cite{Starobinsky80}. This was 
the first inflationary scenario proposed and is currently the one favoured 
by cosmological observations. Since 2003, there has been a resurgence of 
interest in $f({\cal R})$ gravity as a way to explain the present 
acceleration of the universe without invoking an {\em ad hoc} dark energy 
\cite{CCT,CDTT} (see \cite{review1,review2,review3} for reviews). The 
idea is to modify gravity at large scales and to model the universe in the 
context of $f({\cal R})$ theories which incorporate an effective 
time-dependent cosmological ``constant''. This approach has the advantage 
of bypassing the extreme fine-tuning associated with the standard 
cosmological constant $\Lambda$ of the $\Lambda$CDM model (but does not 
solve the cosmological constant problem, of course).

Metric $f({\cal R})$ gravity is nothing but an $\omega = 0$ Brans-Dicke 
theory with a potential in disguise. The action is \cite{review1, 
review2, review3} 
\be
S = \int d^4 x \, \frac{\sqrt{-g}}{16\pi}  \, f({\cal R}) +S^\text{(m)}  
\,,
\label{f(R)action}
\ee
where $f({\cal R})$ is a non-linear function of the Ricci scalar ${\cal 
R}$ and $S^\text{(m)}$ is the matter action. The field equations are the  
fourth order  
ones
\be
f'({\cal R}) {\cal R}_{ab}-\frac{f({\cal R})}{2}\, g_{ab}=\nabla_a\nabla_b 
f'({\cal R})-g_{ab} \Box f'({\cal R}) +8\pi T_{ab}^\text{(m)} \,.
\ee
If one uses the new scalar field  $\phi \equiv  f'({\cal  R})$ with  potential 
\be \label{f(R)potential}
V(\phi)= \phi {\cal R} (\phi) -f\left( {\cal R}(\phi)  
\right) \Big|_{\phi=f'({\cal R})} \,,
\ee
one can show \cite{review1, review2, review3} that the 
action~(\ref{f(R)action}) is equivalent 
to 
\be
S = \int d^4 x \, \frac{ \sqrt{-g}}{16\pi}  \left[ \phi 
{\cal R}-V(\phi) \right]  +S^\text{(m)} \,,
\ee
which is a Brans-Dicke action with Brans-Dicke parameter 
$\omega=0$ and the  potential~(\ref{f(R)potential}).

Since the main motivation for $f({\cal R}) $ gravity comes from cosmology, 
the most well-studied analytic solutions of these theories should describe 
central objects embedded in FLRW universes. Very often the locally static 
KSdS geometry is a solution, but it is not the only one.

\subsection{Black holes and no-hair theorems in $\bm{f(\mathcal{R})}$ gravity}

If a metric $f({\cal R})$ theory of gravity admits an electrovacuum 
solution with ${\cal R}=0$ ({\em i.e.}, if $f(0)=0$), the no-hair theorems of 
scalar-tensor gravity apply {\em in vacuo}. Strictly speaking, 
such a theory misses the  motivation of explaining 
the present acceleration of the universe without dark energy, which motivated  
the resurgence of $f({\cal R})$ theories. In these theories and also in 
theories 
with $f(0)\neq 0$, to the extent that the 
cosmological asymptotics can be neglected, the usual scalar-tensor no-hair 
theorems apply and then the only physical black holes which are stationary 
and stable are the Kerr ones of GR. The argument  
relies on the equivalence between metric $f({\cal R})$ gravity and an 
$\omega=0$ Brans-Dicke theory with a complicated potential.  
Furthermore,\footnote{With the exception of 
Ref.~\cite{myJordanBirkhoffproof}.} it is performed in the Einstein 
conformal frame.  An alternative proof was obtained by Ca{\~n}ate
\cite{Canate18} by analyzing directly the fourth order electrovacuum 
field 
equations of $f({\cal R})$ gravity (with $f$ a function continuous with 
its first and second derivatives and satisfying $f(0)=0$) without 
resorting 
to the equivalence with Brans-Dicke theory. The proof mirrors the standard 
one in scalar-tensor gravity using integrals, and does not apply to the 
special theory $f({\cal R})={\cal R}^2$.\footnote{Care must be taken, 
where the map between conformal frames breaks down (in ${\cal R}^2$ 
gravity, where ${\cal R}=0$ \cite{Rinaldi18}).} This purely 
quadratic theory is pathological since 
it exhibits a restricted conformal invariance and does not admit a Newtonian 
limit \cite{PechlanerSexl66}, however it is a good approximation to the 
Starobinsky \cite{Starobinsky80} inflationary scenario based on $f({\cal 
R})={\cal R}+\alpha {\cal R}^2$  
in the early universe. The special case $f({\cal 
R})={\cal R}^2$ 
was covered separately by Sultana and Kazanas \cite{SultanaKazanas18} 
(their proof uses the equivalence with $\omega=0$ Brans-Dicke 
theory and generalizes the Jordan  frame proof\footnote{This proof is 
repeated, but specifically for $f({\cal R})$ gravity instead of general 
scalar-tensor theory, in \cite{Reddyetal18}.} of 
Faraoni \cite{myJordanBirkhoffproof} for vanishing potential to the case of a 
quadratic potential, which is well-known to disappear from the field 
equation for the Brans-Dicke scalar).  More important, the proof of 
Ca{\~n}ate \cite{Canate18} is extended to asymptotically de Sitter electrovacuum 
stationary black holes in theories which admit constant curvature 
solutions ${\cal R}=\mbox{const.}>0$.

Separate no-hair theorems restricted to specific $f({\cal R})$ theories 
were given earlier, including the Starobinsky model $f({\cal R})={\cal 
R}+\alpha {\cal R}^2$ \cite{Bhattacharya16} and $f({\cal R})={\cal R}^n$ 
\cite{Schmidt90, CliftonBarrowCQG06b, CliftonCQG06, 
CarloniDunsbyCapozzielloTroisi05, LeachCarloniDunsby06}.

\subsection{Clifton-Barrow static solution of $\bm{f({\cal R}) = {\cal 
R}^{1+\delta}}$ gravity}

This solution of $f({\cal R}) = {\cal R}^{1+\delta}$ gravity reads 
\cite{CliftonBarrowPRD05,CliftonCQG06}
\be
ds^2 = - A_1(r) dt^2 +\frac{dr^2}{B_1(r)} +r^2 d\Omega_{(2)}^2 \,,
\ee
where
\begin{eqnarray}
A_1(r) &=& r^{ \frac{2\delta (1+2\delta)}{1-\delta} } +\frac{C_1}{r^{ 
\frac{1-4\delta}{1-\delta} } } \,,\\
&&\nonumber\\
B_1(r) &=& \frac{ (1-\delta)^2}{ (1-2\delta +4\delta^2)\left[ 
1-2\delta(1+\delta)\right]} \left( 1+\frac{C_1}{ 
r^{\frac{1-2\delta+4\delta^2}{1-\delta} } } \right) \,.
\end{eqnarray}
This static geometry is conformal to the special case of a solution of 
Einstein-Maxwell-dilaton gravity found by Chan {\em et al.}
\cite{ChanHorneMann} obtained when the electric charge vanishes 
\cite{CliftonCQG06}. It reduces to Schwarzschild when $\delta \rightarrow 
0$. It has been shown to describe  a black hole by Cognola {\em et al.}
\cite{Cognolaetal11}.

\subsection{Clifton’s inhomogeneous cosmology in $\bm{{\cal 
R}^{1+\delta}}$ gravity}

A spherically symmetric and dynamical solution of vacuum $ f( {\cal 
R})={\cal R}^{1+\delta} $ gravity is due to Clifton \cite{CliftonCQG06}. 
Solar System tests of gravity constrain the parameter $\delta $ to the 
range $ \delta= \left( -1.1 \pm 1.2 \right) \cdot 10^{-5} $ 
\cite{CliftonBarrowPRD05, CliftonBarrowCQG06b, 
Zakharovetal06}. Furthermore, it must be $f''({\cal R}) \geq 0$ for 
stability 
\cite{mattmodgrav}, which implies $\delta \geq 0$ and we take the mass 
parameter $C_2>0$.

The line element of the Clifton solution in isotropic 
coordinates reads \cite{CliftonCQG06}
\be\label{Clifton1}
ds^2=-A_2(\bar{r})dt^2+a^2(t)B_2(\bar{r})\left( d\bar{r}^2 
+\bar{r}^2 d\Omega^2_{(2)} \right) 
\,,
\ee
where 
\begin{eqnarray}
A_2(\bar{r}) &=& \left( 
\frac{1-C_2/\bar{r}}{1+C_2/\bar{r}}\right)^{2/q} \,, 
\label{Clifton2} \\
&&\nonumber \\
B_2(\bar{r}) &=& \left( 1+\frac{C_2}{\bar{r}} 
\right)^{4}A_2(\bar{r})^{\, q+2\delta 
-1} \,,\label{Clifton4}\\
&&\nonumber \\
a(t) &= & t^{\frac{ \delta ( 1+2\delta)}{1-\delta}} 
\,,\label{Clifton3}\\
&&\nonumber \\
q^2 &= & 1-2\delta+4\delta^2 \,. \label{Clifton5}
\end{eqnarray}
It reduces to the FLRW line element when $C_2 =0$. In the GR limit $ 
\delta \rightarrow 0 $, the geometry~(\ref{Clifton1}) reduces to the 
Schwarzschild one. 
The Clifton spacetime~(\ref{Clifton1})-(\ref{Clifton5}) is conformal 
to the Fonarev solution of GR which is conformally static 
\cite{HidekiFonarev}, therefore it is also conformally static (they are 
both conformal to the FJNWBW geometry).  The 
presence of a central spacetime singularity and of dynamical apparent 
horizons have been investigated in \cite{myClifton}. At least for 
some 
values of the parameters $C_2$ and $\delta$, the phenomenology of apparent 
horizons involves a cosmological horizon accompanied by the creation  
of a pair of apparent horizons, which then merge and disappear 
\cite{myClifton}. This apparent horizon phenomenology is qualitatively the 
same as that of the Husain-Martine-Nu\~nez solution of the Einstein 
equations \cite{HusainMartinezNunez}.

\subsection{Conformal image of the Fonarev solution}

By setting $\omega=0$ in the Brans-Dicke analogue of the Fonarev 
solution, one obtains a solution of $f({\cal R})$ gravity 
since the power-law potential $V(\phi)$ corresponds to the effective 
potential
\be
V_0 \left[ f'({\cal R})\right]^{2\beta}= {\cal R}f'({\cal 
R}) -f({\cal R}) \,, \;\;\;\;\;\;\;
\beta= 1-\alpha \sqrt{3} \,. \label{form}
\ee 
for $f({\cal R})= \mu {\cal R}^n$ and 
\begin{eqnarray}
\beta &=&\frac{n}{2(n-1)} \,, \label{beta}\\
&&\nonumber\\
V_0 &=& \frac{n-1}{n^{2\beta}} \, \mu^{1-2\beta} 
\,,\label{Vquesta}
\end{eqnarray}
with $n\neq 1$. Then, the parameter $\alpha$ of 
the family of solutions~(\ref{21}), (\ref{24}) is 
\be\label{alpha}
\alpha =  \frac{n-2}{2\sqrt{3} \left(n-1\right)} \,.
\ee
The conformal image of the 
Husain-Martinez-Nu\~nez solution is obtained for $\alpha=\pm 
\sqrt{3}/2$ and is, therefore, a solution of $f( {\cal R})=\mu {\cal R}^n$ 
gravity for $n=1/2, 5/4$ (however,  these values of  $n$ are  ruled out by  
Solar System  experiments \cite{CliftonCQG06,  
CliftonBarrowCQG06b, Zakharovetal06}). 
Any $f({\cal R})$ 
theory must satisfy $f'>0$ (in order for the graviton to 
carry positive energy) and $f''\geq 0$ (to ensure local stability)
 \cite{review1, review2, review3}. Then, $\delta \geq 0$,  
\be
\alpha=-\frac{ (1-\delta)}{2\sqrt{3} \, \delta} \,, 
\;\;\;\;\;\;\;\;\;
\beta=\frac{1+\delta}{2\delta}  \label{quella}
\ee
(with $\alpha<0$), and 
\begin{eqnarray}
ds^2&=&-{A(r)}^{ -\, \frac{1}{\sqrt{ 1-2\delta+4\delta^2}}} 
\, \mbox{e}^{ \frac{2\left(1-\delta\right) at}{\sqrt{3} \, 
\delta} } dt^2 \nonumber\\
&&\nonumber\\
&\,&  + \, \mbox{e}^{\frac{2(1+2\delta)at}{3\delta}} \left[
{A(r)}^{\frac{1-2\delta}{ \sqrt{1-2\delta+4\delta^2}}} 
dr^2  
+ {A(r)}^{\frac{1-2\delta}{ \sqrt{1-2\delta+4\delta^2}}-1 }
r^2 d\Omega_{(2)}^2 \right] \,. \nonumber \\
&&
\end{eqnarray}

\subsection{Other solutions}

In addition to the previous spacetimes, various other analytic and 
spherically symmetric solutions of various $f({\cal R})$ theories of 
gravity have been proposed over the years. Sometimes the form of the 
function $f({\cal R})$ is not fixed and is found together with the 
specific spherical solution, which means that physical motivation is 
lacking---indeed many forms of $f({\cal R})$ are already ruled out by the 
existing experimental constraints and other forms are overly complicated. 
As a general rule, it is easier to find static solutions with constant 
Ricci curvature, an avenue which has been explored (see 
\cite{booksalv,CapozzielloLaurentis11} 
for a summary of approximate solutions with constant ${\cal R}$) but the 
resulting catalogue of physically motivated solutions is still 
surprisingly scarce. Due to space limits, we can only list references 
without analyzing these solutions in detail, including \cite{Whitt85, 
MignemiWiltshire92,BronnikovChernakova05a, BronnikovChernakova05b, 
BronnikovChernakova05c, MultamakiVilja06, MultamakiVilja07, 
MultamakiVilja08, BusteloBarraco07, Capozzielloetal08, Nziokietal10, 
SebastianiZerbini11, MyrzakulovSebastianiZerbini13,GaoShen16, HoldomRen17, 
CalzaRinaldiSebastiani18, Elizalde:2020icc}.

Sometimes exact solutions can be obtained by matching interior and 
exterior spacetime regions through appropriate matching conditions: these 
matching conditions have been studied specifically for $f({\cal R})$ 
gravity by Deruelle {\em et al.} \cite{DeruelleSasakiSendouda08}, 
Senovilla \cite{Senovilla13}, Clifton \cite{Clifton:2015sra}, 
Reina {\em et al.} \cite{ReinaSenovillaVera16}, and Chakrabarti {\em et al.} \cite{Chakrabarti:2018qsh}.

\section{Horndeski gravity}
\label{sec:13}

The problematic ultraviolet limit of GR and the need to include {\em ad 
hoc} exotic forms of matter to account for the late-time 
acceleration of the universe suggest another generalization of 
Einstein gravity. A minimal modification of GR is obtained by relaxing 
some of the assumptions of Lovelock's theorem 
\cite{Lovelock:1971yv, Lovelock:1972vz}. This can be achieved by adding 
new degrees of freedom to the theory, altering the number of spacetime 
dimensions, including higher order derivatives of the metric in 
the action, or adding non-local terms to the theory. Nonetheless, most of 
these modifications of GR lead to effective low-energy models that 
merely involve a scalar degree of freedom added to the metric. This 
general result embodies the power of scalar-tensor theories, 
combining their (apparent) simplicity and minimality with the ability to 
capture the main features of more complex scenarios as effective models.

	When formulating a classical field theory, it is necessary to 
avoid phenomena that would make it ill-posed from the start. An example of 
such issues is ghost-like instability. To this regard, a general theorem 
by Ostrogradsky \cite{Ostrogradsky:1850fid} (see also \cite{Woodard:2015zca}) states 
that, in classical mechanics, a non-degenerate Lagrangian containing time 
derivatives higher than first order leads to a linearly unstable 
Hamiltonian. Likewise, higher order non-degenerate classical field 
theories are affected by ghost-like (or Ostrogradsky) instabilities. Thus, 
limiting ourselves to local theories with equations of motion containing 
field derivatives not higher than second order constitutes a sufficient 
condition to avoid Ostrogradsky instabilities.

Horndeski gravity \cite{Horndeski:1974wa} is the most general scalar-tensor 
theory with second order equations of motion. In its modern formulation, 
this  theory is expressed in terms of the generalized Galileon Lagrangian 
that reads
\begin{eqnarray}
\label{eq:HorndeskiLagrangian}
\mathcal{L} &=& \mathcal{L}_2 + \mathcal{L}_3 + \mathcal{L}_4 + 
\mathcal{L}_5 \, , \\
&&\nonumber\\
\nonumber \mathcal{L}_2 &=& G_2 \, , \\
&&\nonumber\\
\nonumber \mathcal{L}_3 &=& - G_3 \, \Box \phi \, ,\\
&&\nonumber\\
\nonumber \mathcal{L}_4 &=& G_4 \, R + G_{4 \, X} \left[ (\Box \phi)^2 - 
(\nabla_a \nabla_b \phi)^2 \right] \,,\\
&&\nonumber\\
\nonumber \mathcal{L}_5 &=& G_5 \, G_{ab} \, \nabla^a \nabla^b 
\phi -  \frac{G_{5 \, X}}{6} \Big[  (\Box \phi)^3 \\
&&\nonumber\\
\nonumber
& \, &  - 3 \, \Box \phi \, (\nabla_a \nabla_b \phi)^2 
+ 2 \, (\nabla_a \nabla_b \phi)^3 \Big] \,,
\end{eqnarray}
where the $G_i (\phi, X)$ ($i=2,3,4,5$) are arbitrary functions of 
the scalar field $\phi$  and of the canonical kinetic term $X \equiv - 
(1/2)  \partial _\mu \phi \partial ^\mu \phi$. According to standard 
notation, if $f$ is a function of $X$ then $f_X \equiv \partial f / 
\partial  X$, 
\be 
(\nabla_a  \nabla_b \phi)^2 \equiv \nabla_a 
\nabla_b \phi \nabla^a \nabla^b \phi \,,
\ee
and 
\be
(\nabla_a \nabla_b \phi)^3 \equiv \nabla_a \nabla_c \phi 
\nabla^c \nabla^d \phi \nabla_d \nabla^a \phi \,.
\ee
Then, given the action
\be
\label{eq:HorndeskiAction}
S[g_{ab} , \phi] = \int d^4 x \, \sqrt{-g} \, \Big( \mathcal{L}_2 + 
\mathcal{L}_3 + \mathcal{L}_4 + \mathcal{L}_5 \Big) \, ,
\ee
we define
\be
\mathcal{E}_\phi \equiv \frac{1}{\sqrt{-g}} \frac{\delta S}{\delta \phi} 
\, ,\quad
\mathcal{E}_{ab} \equiv \frac{2}{\sqrt{-g}} \frac{\delta S}{\delta 
g^{ab}} \, .
\ee
The field equations of the theory defined by the action  
\eqref{eq:HorndeskiAction} then read  $\mathcal{E}_\phi=0$ and 
$\mathcal{E}_{ab}=0$.

The connection with the Galileon model is particularly instructive and 
offers a rather pedagogical path toward the derivation of the Horndeski 
Lagrangian. The Galileon theory was originally proposed by Nicolis {\em et al.}
\cite{Nicolis:2008in} as a generalization of the Dvali-Gabadadze-Porrati 
(DGP)  \cite{Dvali:2000hr} four-dimensional effective theory. Galileon 
theory 
represents the most general ghost-free scalar field theory on a 
four-dimensional Minkowski spacetime, which is symmetric under the field 
transformation $\phi (x)  \to \phi (x) + b_\mu x^\mu + c$ (where $b_\mu$ 
and $c$ are constants) and leads to a second order field equation. 
Deffayet {\em et al.} \cite{Deffayet:2009wt} extended the Galileon model to curved space 
(``covariant Galileon''). For the covariant Galileon, the symmetry 
$\phi (x)  \to \phi (x) + b_\mu x^\mu + c$ must be dropped in favor of 
general covariance. Kobayashi {\em et al.} \cite{Kobayashi:2011nu} showed that 
the generalized Galileon Lagrangian \eqref{eq:HorndeskiLagrangian} of 
Deffayet {\em et al.} \cite{Deffayet:2011gz}, representing a natural extension 
of the covariant Galileon and preserving the second order nature of the field 
equations, is equivalent to Horndeski's old and forgotten theory.

	One of the main motivations for Horndeski's theory is 
providing an alternative to dark energy for modelling the cosmic 
evolution.  To this end, one needs to make 
sure that such modifications of GR, while properly 
accounting for the cosmic expansion, do not affect the predictions for 
Solar System experiments. In other words, Solar System physics should be 
screened from the effects of the new scalar degree of freedom $\phi$. 
In scalar-tensor theories, this goal can be achieved by either making 
$\phi$  effectively massive at short scales through the  
chameleon 
mechanism \cite{Khoury:2003aq, Khoury:2003rn}, or else the theory itself 
exploits its nonlinearities to screen physics below a certain scale from 
the effects of $\phi$ through the Vainshtein mechanism  
\cite{Vainshtein:1972sx} (see also the 
review \cite{Babichev:2013usa}  and the references 
therein). The latter, in particular, plays a fundamental role for 
cosmological models in Horndeski gravity.

	Another important theoretical reason for studying more general 
scalar-tensor theories relies on potential violations of the no-hair 
theorems of GR. Indeed, while these theorems can be easily extended to 
simpler scalar-tensor theories as discussed earlier, 
since most of these 
models turn out to be singular at the black hole horizon unless the scalar 
field is in a trivial configuration, the situation is significantly more 
complicated in Horndeski gravity and its generalizations. We refer the 
interested reader to the excellent review by Kobayashi 
\cite{Kobayashi:2019hrl} for a discussion of the literature on this 
subject.

	It is also important to mention that the combined detection of 
gravitational waves (GW170817) and of a $\gamma$-ray burst (GRB170817A) 
from a binary neutron star merger \cite{TheLIGOScientific:2017qsa, 
Monitor:2017mdv} has put severe constraints on viable scalar-tensor 
models ({\em e.g.}, \cite{Bettoni:2016mij}) by setting the bounds
\be
-3 
\times 10^{-15} <    c_{\rm GW} - 1 < 7 \times 10^{-16}
\ee
on the speed of 
gravitational waves $c_{\rm GW}$ (\cite{Kobayashi:2019hrl} 
and references therein).

From a field-theoretic perspective, Horndeski gravity is an 
effective field theory (EFT) defined in such a way that certain higher 
derivative operators become as relevant as lower order ones within the 
regime of validity of the low-energy model. For such an EFT to be 
physically consistent one has to require, within the regime of 
applicability of the model, that: a) the leading higher derivative terms 
do not introduce ghost instabilities; b) quantum corrections must be 
suppressed in order not to spoil the computation of physical observables. 
While in the case of Horndeski gravity a) is fulfilled by construction, 
the point b) requires the theory to exhibit some ``non-renormalization'' 
properties. For a detailed discussion of the EFT approach to 
higher-derivative scalar-tensor theories we refer the interested reader to 
\cite{Luty:2003vm, deRham:2012ew, Brouzakis:2013lla, Pirtskhalava:2015nla, 
Santoni:2018rrx, Heisenberg:2020cyi} and the references therein.

We begin by reviewing some generalities of the no-hair theorems in Horndeski gravity, 
focusing on the case of shift-symmetric theories and on possible ways to evade them. We then 
discuss two important classes of non-trivial hairy solutions exploiting the loopholes in the 
no-hair argument of Hui and Nicolis \cite{Hui:2012qt}. This discussion is not meant to be an 
in-depth analysis of the literature on exact solutions of Horndeski gravity (a topic that 
would deserve to be surveyed on its own), but it is rather a summary of the results 
capturing the main features of hairy black hole solutions within this theory. We refer the 
interested reader to \cite{HerdeiroRadu15,Babichev:2016rlq,Kobayashi:2019hrl,Silva:2016smx} 
for a discussion of the relevant literature.

\subsection{Hair or no-hair?}

Shift-symmetric Horndeski theories are a subclass of Horndeski 
gravity equipped with a residual Galileon symmetry from the flat space 
case. Specifically, they preserve the invariance under the transformation 
$\phi \to \phi + c$ (with $c$ a constant) even for curved manifolds. This 
class is defined by the Lagrangian \eqref{eq:HorndeskiLagrangian} with the 
condition that $G_2, \ldots, G_5$ be arbitrary functions of $X$ and 
do not depend explicitly on $\phi$. Clearly, this shift symmetry is  
associated with a covariantly conserved current $J^a$ (see, {\em 
e.g.}, \cite{Babichev:2016rlq} for its explicit form).

The most stringent theorem for shift-symmetric Horndeski theories 
\cite{Babichev:2016rlq} states that assuming: (i)~spherically symmetric 
spacetime with a static scalar field; (ii)~asymptotic flatness, $\phi' 
\equiv  \partial _r \phi \to 0$ as $r\to \infty$, and $J^2 \equiv J^a 
J_a$ finite 
at the horizon; (iii)~there is a canonical kinetic term $X$ in the action 
and the $G_i$ are such that their $X$-derivatives contain only positive or 
zero powers of $X$, then the scalar field is constant and the spacetime 
is isometric to the Schwarzschild spacetime.

Taking a deeper look at (i--iii) one has that:

\begin{itemize}

\item (i)~implies that the spacetime metric and the scalar field take the 
form
\begin{eqnarray}
\label{eq:metricspherical}
ds^2 &=& - {\cal A}(r) \, dt^2 + {\cal B}(r)^{-1} \, dr^2 + R^2(r) \, d \Omega_{(2)} 
^2 \, , \nonumber\\ 
&&\\
\phi &=& \phi (r) \, ,
\end{eqnarray}
with $R(r)$ denoting the areal radius. Hence, the norm of the current 
reduces to $J^2 = (J^r)^2 / {\cal B}(r)$. 

\item Since the location $r=r_{\rm H}$  of the horizon is determined by 
the largest root of ${\cal B}(r) = 0$, the 
requirement (ii) concerning the regularity of $J^2$ at the horizon  
suggests that $J^r$ should vanish at $r_{\rm H}$. Furthermore, the 
conservation of $J^a$ reduces to $ \left( R^2 \, J^r \right)'=0$ (with a  
prime denoting differentiation with respect to $r$), leading to $R^2 \, 
J^r 
= \mbox{constant}$. However, since $R^2 \, J^r|_{r=r_{\rm H}} = 0$, then 
$J^r (r) = 0$, $\forall r$. 

\item Specializing the general result for the conserved current $J^a$ in 
\cite{Babichev:2016rlq} to the symmetries of the problem, one has that 
$J^r$ takes the general form
\be
\label{eq:currentHN}
J^r = {\cal B} \, \phi' \, F(\phi', g, g', g'')
\ee
with $F$ a function of the first radial derivative of the field and of 
the metric and its first and second radial derivatives.

\item (iii) and the requirement of asymptotic flatness then imply that 
$\phi'(r) = 0$, $\forall r$, leading to a solution isometric to the 
Schwarzschild spacetime. 

\end{itemize}

The first no-hair argument, by Hui and Nicolis \cite{Hui:2012qt}, 
for black holes in Horndeski gravity relied heavily on the idea that the 
function $F$ in \eqref{eq:currentHN} would approach a non-zero constant at 
infinity, provided that one requires the theory to have a standard 
 kinetic term in the weak field regime. Thus, according to 
Hui and Nicolis \cite{Hui:2012qt}, asymptotic flatness implies that 
${\cal A}, {\cal B} \to 1$ and 
$\phi'  \to 0$ at infinity. If one then moves continuously towards the 
horizon, $\phi'$ can become non-zero while ${\cal B}$ and $F$ remain 
non-zero, contradicting the condition $J^r = 0$ everywhere. This 
should then imply 
that $\phi' (r) = 0$ everywhere, exploiting a much weaker condition than 
(iii). However, this argument has two major loopholes allowing for hairy 
black hole solutions \cite{Sotiriou:2013qea, Babichev:2013cya} as we  
discuss in detail in the following sections. Furthermore, one can also 
find hairy solutions by relaxing some of the conditions in (i--iii).

To conclude this brief section on no-go results for black holes in 
shift-symmetric Horndeski theories, we mention that black hole solutions 
of non-shift-symmetric models have been obtained analytically as effective 
models emerging from the Kaluza-Klein reductions of specific truncations 
of Lovelock gravity \cite{Charmousis:2012dw}. Hence, one can arguably 
consider these solutions as peculiar to Horndeski gravity and, for this 
reason, they will not be discussed here.

\subsection{Perturbative and numerical solutions}

We begin by recalling the action of Einstein-dilaton-Gauss-Bonnet (EdGB) 
gravity with linear coupling \cite{Sotiriou:2014pfa}
\be
\label{eq:EdGBAction}
S = \frac{1}{8 \pi} \int d^4 x \, \sqrt{-g} \, 
\left( \frac{\mathcal{R}}{2} - \frac{1}{2} \, \nabla_a \phi 
\nabla^a \phi + \alpha \, \phi \, \mathcal{G} \right) \, ,
\ee
where $\alpha$ has the dimensions of a squared length and 
\be
\mathcal{G} \equiv \mathcal{R}^2 - 4 \mathcal{R}_{a b}\mathcal{R}^{a 
b} 
+ \mathcal{R}_{a b c d}\mathcal{R}^{a b c d} 
\ee
is the Gauss-Bonnet integrand. This Lagrangian is a 
shift-symmetric Horndeski Lagrangian with
\be
\label{eq:Horndeski-EdGB}
G_2 = X \, , \,\, G_3 = 0 \, , \,\, G_4 = \frac{1}{2} \, , \,\, G_5 = - 4 
\alpha \ln |X| \, . 
\ee
The variation of the action \eqref{eq:EdGBAction} yields the field 
equations \cite{Sotiriou:2014pfa}
\be
\label{eq:EdGBEq-1}
\mathcal{E}_{a b} = G_{a b} - \mathcal{T}_{a b} = 0 \, ,
\ee
\be
\label{eq:EdGBEq-2}
\mathcal{E}_{\phi} = \Box \phi + \alpha \, \mathcal{G} = 0 \, , 
\ee
with 
\begin{eqnarray}
\label{eq:EdGBEffTensor}
\nonumber
\mathcal{T}_{a b} &=& \nabla _a \phi \nabla _b \phi - 
\frac{1}{2}\, g_{ab} \,  \nabla_c \phi \nabla^c \phi\\
&&\nonumber\\
&\,  &  - \alpha (g_{d a} g_{e b} + g_{d b} 
g_{e a}) \nabla_f (\nabla _m \phi \, \epsilon^{m 
e p q}  \epsilon^{d f r s} \mathcal{R}_{rs p q})   
\end{eqnarray}
denoting the effective stress-energy tensor, and where $\epsilon^{abcd}$ 
is the Levi-Civita pseudotensor.  
The covariantly conserved current $J^a$ associated with the shift 
symmetry is \cite{Sotiriou:2014pfa}
\be
\label{eq:EdGBCurrent}
J^a = \sqrt{-g} \,  (\nabla ^a \phi + \alpha \, 
\tilde{\mathcal{G}}^a) \, , 
\ee
where the explicit form of $\tilde{\mathcal{G}}^a$ is found 
by taking the general form of $J^a$ for a shift-symmetric Horndeski 
Lagrangian (reported in \cite{Babichev:2016rlq}) and setting 
the functions $G_2, \, \ldots \, , G_5$ as in \eqref{eq:Horndeski-EdGB}.

With the ansatz \eqref{eq:metricspherical} parametrized in terms 
of the areal radius $R$, one finds that \eqref{eq:EdGBCurrent} reduces to 
\cite{Silva:2016smx}
\be
J^R = - {\cal B} \, \phi' - 4 \alpha \, \frac{{\cal A}'}{{\cal A}} \,  
\frac{{\cal B} ({\cal B}-1)}{R^2} \, ,
\ee
where now the prime denotes differentiation with respect to $R$. Assuming 
asymptotic flatness, $J^R$ vanishes at infinity, though now 
the second term allows for a non-trivial $\phi'$ when $J^R = 0$.

Because of the complexity of the field equations \eqref{eq:EdGBEq-1} and \eqref{eq:EdGBEq-2} 
even assuming \eqref{eq:metricspherical} and asymptotic flatness, no closed form expression 
is known for the solution. The problem can be treated numerically and perturbatively.  A 
perturbative solution of stringy gravity found by Campbell, Kaloper, and Olive 
\cite{Campbell:1991kz} contains, as a special case, the more recent proposal of Sotiriou and 
Zhou \cite{Sotiriou:2014pfa}. A close dilatonic gravity theory with exponential potential 
has been solved numerically by Kanti, Mavromatos, Rizos, Tamvakis, and Winstanley 
\cite{Kanti:1995vq}. The Campbell-Kaloper-Olive-Sotiriou-Zhou solution harbours a curvature 
singularity at a finite areal radius and can be cloaked by a horizon if the black hole is 
sufficiently massive \cite{Sotiriou:2014pfa}.

\subsection{Babichev-Charmousis class of solutions}

Consider the subset of shift-symmetric Horndeski theories characterized 
by the Lagrangian \cite{Babichev:2013cya}
\be
\label{eq:LBC}
\mathcal{L} = \mathcal{R} - \eta \, \nabla^c \phi \nabla_c \phi 
+ \beta \, G^{ab} \nabla_a \phi \nabla_b \phi - 2 \,\Lambda \, ,
\ee
where $\eta, \beta, \Lambda$ are constants, which is obtained from 
\eqref{eq:HorndeskiLagrangian} by setting $G_3=G_5=0$, $G_2 = -2 \Lambda 
+ 2 \eta X$, and $G_4 = 1+\beta X$. This Lagrangian contains an explicit  
non-minimal coupling between the Einstein tensor and the scalar field in  
 $G^{ab} \nabla_a \phi \nabla_b \phi$, which  corresponds to a specific 
realization of the ``John Lagrangian'' 
of the ``Fab Four'' subclass of Horndeski gravity 
\cite{Charmousis:2011bf}. The corresponding field equations are 
\cite{Babichev:2016rlq, Babichev:2013cya}
\begin{eqnarray}
\label{eq:BCEq-1}
\nonumber
\mathcal{E}_{ab} &=& G_{ab} - \eta \left[ \nabla_a \phi 
\nabla_b \phi - \frac{g_{ab}}{2} \, \nabla^c \phi \nabla_c \phi \right] 
+ \Lambda \, g_{ab} \\
&&\nonumber\\
& & \nonumber + \frac{\beta}{2} \Big[ (\nabla^c \phi \nabla_c  
\phi) \, G_{ab} + 2 P_{a c b d} \nabla^c \phi 
\nabla^d \phi\\  
&&\nonumber\\
& \, &  + g_{a c} \delta^{cef} _{b gh } \nabla^g 
\nabla_e \phi \nabla^h \nabla_f 
\phi \Big] = 0 \,, \nonumber\\
&&
\end{eqnarray}
while the equation for the scalar field can be recast in terms of the 
current conservation associated with the shift symmetry of the theory
\be
\label{eq:BCEq-2}
\nabla_a J^a = 0 \, , \quad J^a = (\eta g^{ab} - \beta G^{ab}) \nabla_b 
\phi \,.
\ee
Here 
\be
\delta ^{abe} _{cdf} \equiv 3! \, 
\delta^a _{[c} \delta^b_d  \delta^e_{f]}
\ee
and 
\be
P_{abcd} \equiv (-1/4) \,  \epsilon _{abef} 
\mathcal{R}^{efgh}  \epsilon_{cdgh}
\ee 
denotes the double dual of the Riemann tensor.  The form of the current 
$J^a$ offers a natural way to evade the Hui-Nicolis no-hair argument. In 
fact, in spherical symmetry and assuming the ansatz 
\eqref{eq:metricspherical} parametrized in terms of the areal radius $R$, 
the condition $J^R = 0$ is satisfied by either a constant scalar field or, 
more interestingly, by configurations such that $\eta g^{RR} - \beta 
G^{RR}=0$ allowing for a non-vanishing gradient $\nabla_c \phi$.

Considering the ansatz \eqref{eq:metricspherical}, again parametrized in 
terms of the areal radius $R$ and $\phi = \phi (t, R)$, then the only 
ansatz for $\phi$ compatible with Eqs.~(\ref{eq:BCEq-1}) and 
(\ref{eq:BCEq-2}) is \cite{Babichev:2016rlq, Babichev:2013cya}
\be
\label{eq:BC-Phi}
\phi = q \, t + \psi (R) \, , 
\ee
with $q=$~const. This condition is evaded by self-tuning flat space 
solutions \cite{Babichev:2016rlq} that are not of interest here.

	Setting $\eta=1/2$, according to the canonical normalization of 
the scalar field, Babichev and Charmousis \cite{Babichev:2016rlq} proved that, assuming 
\eqref{eq:metricspherical} and \eqref{eq:BC-Phi}, the general solution of  
the theory \eqref{eq:LBC} can be expressed in terms of the solution of the 
algebraic equation for $k(R)$
\begin{eqnarray}
\label{eq:orcamadonna}
\nonumber 
& & (q \beta)^2 \left(1 + \frac{R^2}{2 \beta} \right)^2 - \left( 2 + 
\frac{1-2 \beta \Lambda}{2 \beta} \, R^2 \right) k(R) 
+ C_0 \, k^{3/2} (R) =0 \, , \\
&& 
\end{eqnarray}
with $C_0$ is an integration constant.\footnote{The solution of  
Babichev and Charmousis \cite{Babichev:2016rlq} is a bit more general 
than the one discussed here since it also includes the cases of planar and 
hyperbolic symmetries, however here we are only interested in spherical 
solutions.} Then, the metric components are
\begin{eqnarray}
{\cal A}(R) &=& -\frac{\mu}{R} + \frac{1}{\beta R} \int^R \frac{k(\bar{r})}{1 + 
(\bar{r}^2/2\beta) } \, d\bar{r} \, ,\\
&&\nonumber\\
{\cal B}(R) &=& \left(1+\frac{R^2}{2 \beta}\right)^2 \frac{\beta {\cal A}(R)}{k(R)} \, ,
\end{eqnarray}
while the radial dependence of the scalar field is given by
\begin{eqnarray}
\nonumber
\psi'(R) &=& \pm \frac{\sqrt{R}}{\left( 1 + \frac{R^2}{2 \beta} \right) 
{\cal A}} \left[q^2 \left( 1 + \frac{R^2}{2 \beta} \right) {\cal A}' 
- \frac{1+2\beta \Lambda}{4 \beta^2} (R^2 {\cal A}^2)' \right]^{1/2} \, , \\
&&
\end{eqnarray}
where $\mu$ is an integration constant.

We highlight three relevant examples in this general class of solutions:

\begin{itemize}

\item \textbf{Self-tuned Schwarzschild-de-Sitter solution}  
\cite{Babichev:2013cya}: If one looks for solutions with de Sitter or 
anti-de-Sitter asymptotics, assuming ${\cal A}(R) = {\cal B}(R)$ and taking $\beta q^2 
 = 1 + 2 \beta \Lambda$ and $C_0 = (1-2\beta\Lambda)/\sqrt{\beta}$ 
yields
\begin{eqnarray}
{\cal A}(R) &=& {\cal B}(R) = 1 - \frac{\mu}{R} -\frac{\Lambda_{\rm eff}}{3} \, R^2 \, 
,\\
&&\nonumber\\
\psi'(R) &=& \pm \frac{q}{{\cal A}(R)}\sqrt{1-{\cal A}(R)} \, ,
\end{eqnarray}
where the effective cosmological constant is $\Lambda_{\rm eff} = 
-1/2\beta$, which is positive if $\beta<0$. The self-tuning nature of this 
solution is due to the fact that the vacuum value of $\Lambda$ does not 
affect the geometry of the solution.

\item \textbf{Rinaldi and generalized Rinaldi solutions} 
\cite{Rinaldi:2012vy, Anabalon:2013oea, Minamitsuji:2013ura}: Without 
entering into the details of the derivation, these solutions correspond to 
$q=0$ and $\beta>0$ and are given by
\begin{eqnarray}
{\cal A}(R) &=&  1 - \frac{\mu}{R} -\frac{\Lambda_{\rm eff}}{3} \, R^2 \nonumber 
\\
&&\nonumber\\
& \, &  - \frac{(1+2\beta\Lambda_{\rm eff})^2}{8\beta\Lambda_{\rm 
eff}} \, \frac{\arctan \left( R/\sqrt{2 \beta}\right)}{R/\sqrt{2 \beta}} 
\,, \nonumber\\
& &\\
{\cal B}(R) &=&  \frac{2 \beta + R^2}{2 \beta} \frac{{\cal A}}{(R{\cal A})'}\, ,\\
&&\nonumber\\
\phi'^2  &=&  \frac{R^2 (1 +2 \beta \Lambda_{\rm 
eff})(1-2\beta \Lambda_{\rm eff}-2\Lambda_{\rm eff}R^2)^2}{\Lambda_{\rm 
eff} (1-2\beta \Lambda_{\rm eff})(2\beta + R^2)^3 {\cal A}(R)} \,, \nonumber\\
&&
\end{eqnarray}
with 
\be
\Lambda_{\rm eff} = \frac{2 \beta \Lambda - 1}{2 \beta (2 \beta 
\Lambda +3)} \, . 
\ee
The original Rinaldi solution \cite{Rinaldi:2012vy} was obtained for 
$\Lambda = 0$ and was affected by an unphysical behavior of the scalar 
field beyond the horizon ({\em i.e.}, $\phi$ becomes imaginary inside the 
horizon). It was then pointed out by Anabal{\'o}n {\em et al.} 
\cite{Anabalon:2013oea} and Minamitsuji \cite{Minamitsuji:2013ura} that 
this behavior of $\phi$ can be healed with the addition of a bare 
cosmological constant $\Lambda \neq 0$.
 
\item \textbf{Stealth Schwarzschild black hole} \cite{Babichev:2013cya}: 
Let $\Lambda = \eta = 0$, then \eqref{eq:orcamadonna} is solved by $k(R) 
= \mbox{constant}$. If one sets $k(R) =\beta$, then 
\be
{\cal A}(R) = {\cal B}(R) = 1 - \frac{\mu}{R} \, ,
\ee
and
\be
\psi'(R) = \pm q \, \frac{\sqrt{\mu R}}{R - \mu} \, .
\ee
Integrating the latter and substituting the result in 
\eqref{eq:BC-Phi}, one finds
\begin{eqnarray}
\phi_{\pm} (t,R) &=& q t \pm q \mu \left[ 2 \sqrt{\frac{R}{\mu}} + \ln 
\left( 
\frac{\sqrt{R}-\sqrt{\mu}}{\sqrt{R}+\sqrt{\mu}} \right) \right] + \phi_0 
\,,
\end{eqnarray}
with $\phi_0$ an integration constant.

The defining feature of this solution is that, while the geometry 
coincides with that of the Schwarzschild solution of GR, it is equipped 
with a non-trivial, time-dependent, scalar field.

\end{itemize}

\section{Beyond Horndeski: DHOST gravity}

In the spirit of keeping only one scalar field $\phi$ in the theory in 
addition to the GR graviton ({\em i.e.}, three degrees of freedom), 
attempts to generalize scalar-tensor gravity beyond Horndeski theory were 
made by allowing for derivatives of higher order than second. As expected, 
these attempts usually run into the Ostrogradsky instability. However, it 
was discovered that certain higher derivative theories are, rather 
surprisingly, healthy in this respect 
\cite{Gleyzes:2014dya,Gleyzes:2014qga}. A wider class of theories with 
this feature was then discovered, in which the Ostrogradsky ghost is 
exorcised by imposing a phase space constraint which makes the Lagrangian 
density degenerate \cite{Langlois:2015cwa, Langlois:2015skt, 
Achour:2016rkg, Crisostomi:2016czh, Motohashi:2016ftl, 
BenAchour:2016fzp,Crisostomi:2017aim}.  These Degenerate Higher Order 
Scalar-Tensor (DHOST) theories have been the subject of much attention, 
especially in view of their confrontation with observations. The 
multi-messenger event GW170817--GRB170817 \cite{Langlois:2017dyl} and, in 
addition, the study of stability against 
perturbations\footnote{Metric perturbations around a 
spherically symmetric background are decomposed into polar ({\em 
i.e.}, even parity) and axial ({\em i.e.}, odd parity) modes, as in GR. 
The scalar field $\phi$ appears only in the even parity perturbations.} 
\cite{Creminelli:2018xsv, Babichev:2017lmw, 
Babichev:2018uiw,Takahashi:2019oxz, deRham:2019gha} impose severe 
constraints on DHOST theories (see 
Refs.~\cite{Langlois:2018dxi,Kobayashi:2019hrl} for reviews).

Denoting with $\bar{X}\equiv \nabla^c \phi \nabla_c \phi$ the kinetic term, the 
action of general DHOST gravity is
\begin{eqnarray}
S_{(\text{DHOST})} \left[ g_{ab}, \phi \right] &=& \int d^4 x \, \sqrt{-g} 
\left\{ f_0\left( \phi, \bar{X}\right) +f_1\left( \phi, \bar{X} \right) \Box\phi 
+ f_2\left( \phi, \bar{X} \right) {\cal R} \right.\nonumber\\
&&\nonumber\\
&\, & \left. + {\cal A}_{(2)}^{abcd}\nabla_a\nabla_b\phi  
\nabla_c\nabla_d\phi
+f_3\left( \phi, \bar{X} \right) G_{ab} \nabla^a\nabla^b\phi \right.\nonumber\\
&&\nonumber\\
&\, & \left. + {\cal A}_{(3)}^{abcdef}\nabla_a\nabla_b\phi  
\nabla_c\nabla_d\phi 
\nabla_e\nabla_f \phi \right\} \,,
\end{eqnarray}
where the quadratic terms are written as 
\be
{\cal A}_{(2)}^{abcd} \nabla_a\nabla_b \phi \nabla_c\nabla_d \phi  
=\sum_{i=1}^5 \, \alpha_i \left( \phi, \bar{X} \right) {\cal 
L}_{(2)}^i \,,
\ee
and 
\begin{eqnarray}
{\cal L}_{(2)}^1 &=& \nabla_a\nabla_b \phi \nabla^a\nabla^b \phi \,,\\
&&\nonumber\\
{\cal L}_{(2)}^2 &=& \left( \Box\phi\right)^2 \,,\\
&&\nonumber\\
{\cal L}_{(2)}^3 &=&  \left( \Box\phi\right)  
\nabla^a \phi \nabla^b \phi \nabla_a\nabla_b \phi \,,\\
&&\nonumber\\
{\cal L}_{(2)}^4 &=& \nabla^a \phi \nabla_b \phi \nabla_a\nabla_c \phi
\nabla^c\nabla^b \phi \,,\\
&&\nonumber\\
{\cal L}_{(2)}^5 &=& \left( \nabla^a \phi \nabla^b \phi \nabla_a\nabla_b 
\phi \right)^2 \,.
\end{eqnarray}
The  cubic terms are written as 
\be
{\cal A}_{(3)}^{abcdef} \nabla_a\nabla_b\phi \nabla_c\nabla_d\phi 
\nabla_e\nabla_f\phi =
\sum_{i=1}^{10} \, \beta_i \left( \phi, \bar{X} \right) {\cal L}_{(3)}^i \,,
\ee
where 
\begin{eqnarray}
{\cal L}_{(3)}^1 &=& \left( \Box\phi \right)^3 \,,\\
&&\nonumber\\
{\cal L}_{(3)}^2 &=&\left( \Box \phi \right) \nabla_a\nabla_b \phi  
\nabla^a\nabla^b \phi  \,,\\
&&\nonumber\\ 
{\cal L}_{(3)}^3 &=& \nabla_a\nabla_b \phi \nabla^b \nabla^c \phi \nabla^a 
\nabla_c\phi \,,\\
&&\nonumber\\
{\cal L}_{(3)}^4 &=& \left( \Box\phi \right)^2 \nabla_a\phi \nabla_b\phi 
\nabla^a\nabla^b \phi  \,,\\
&&\nonumber\\
{\cal L}_{(3)}^5 &=& \left( \Box\phi \right) \nabla_a\phi \nabla^c\phi 
\nabla^a\nabla^b \phi \nabla_b\nabla_c\phi \,,\\
&&\nonumber\\
{\cal L}_{(3)}^6 &=& \left( \nabla_a\nabla_b\phi \nabla^a\nabla^b\phi 
\right) \nabla_c\phi \nabla_d\phi \nabla^c\nabla^d \phi \,,\\
&&\nonumber\\
{\cal L}_{(3)}^7 &=& \nabla_a\phi \nabla_d\phi \nabla^a\nabla^b\phi 
\nabla_b\nabla_c\phi \nabla^c\nabla^d\phi \,,\\
&&\nonumber\\
{\cal L}_{(3)}^8 &=& \nabla_a\phi \nabla_c \phi \nabla_d\phi \nabla_e\phi 
\nabla^a\nabla^b\phi \nabla_b\nabla^c\phi 
\nabla^d\nabla^e\phi \,,\\
&&\nonumber\\
{\cal L}_{(3)}^9 &=& \left(\Box\phi \right) \left( \nabla_a\phi 
\nabla_b\phi 
\nabla^a\nabla^b \phi \right)^2 \,,\\
&&\nonumber\\
{\cal L}_{(3)}^{10} &=& \left( \nabla_a\phi \nabla_b \phi \nabla^a\nabla^b 
\phi \right)^3 \,.
\end{eqnarray}
The $\alpha_i\left( \phi, \bar{X} \right)$ and $\beta_i\left( \phi, \bar{X} \right)$, 
and therefore the $ {\cal A}_{(2)}^{abcd} $ and $ {\cal A}_{(3)}^{abcdef} 
$  are arbitrary (regular) functions of their arguments. The requirement 
that the Ostrogradsky ghost is avoided imposes relations between these 
functions \cite{Langlois:2018dxi}.  

The EFT approach to scalar-tensor theories briefly mentioned in 
Sec.~\ref{sec:13} can also be applied to the more general scenario 
discussed here in order to investigate the quantum stability of the 
proposed class of theories. Again, we refer the reader interested in the 
details on this topic mostly to \cite{Santoni:2018rrx} and references 
therein.

Naturally, spherically symmetric solutions of DHOST theories describing 
compact objects (and, more recently, also cylindrical solutions describing 
rotating objects \cite{Charmousis:2019vnf, Minamitsuji:2020jvf, 
Long:2020wqj, BenAchour:2020fgy, Anson:2020trg}) have been studied. 
Compact star configurations were discussed in Refs. 
\cite{Babichev:2016jom, Sakstein:2016oel, Kobayashi:2018xvr, 
Babichev:2018rfj, Cisterna:2015yla,Cisterna:2016vdx, Lehebel:2017fag}. 
Subject to certain hypotheses, no-hair theorems for 
black holes were stated in \cite{Babichev:2017guv}, but soon conditions 
were found that allow one to circumvent them and to construct hairy 
black holes. The most notable way to do this consists of a scalar (hair) 
field that violates the staticity of 
the geometry through a linear dependence on time, $\phi(t,r)=qt +\psi(r)$, 
with $q$ a constant \cite{Babichev:2016rlq, Babichev:2016kdt, 
Motohashi:2018wdq, Minamitsuji:2018vuw} (a trick known from Horndeski 
\cite{Babichev:2012re, Anabalon:2013oea, Babichev:2013cya} and beyond 
Horndeski \cite{Charmousis:2014zaa, Kobayashi:2014eva} gravity). Most of 
these DHOST geometries are solutions of the Einstein equations with 
vanishing stress-energy tensor of the scalar field, {\em i.e.}, stealth 
solutions. The kinetic term $\bar{X}$ is usually 
constant \cite{BenAchour:2018dap, Motohashi:2019sen,Minamitsuji:2019shy, 
Motohashi:2019ymr}. Some of these analytic solutions arise in the earlier 
context of ghost condensation and are known to be immune from the strong 
coupling problem \cite{Mukohyama:2005rw, Mukohyama:2009rk, 
ArkaniHamed:2003uy, ArkaniHamed:2003uz} that usually affects constant $\bar{X}$ 
solutions. Stealth solutions include the 
Schwarzschild and the Schwarzschild-(Anti-)de Sitter spaces 
\cite{Motohashi:2019sen, Babichev:2013cya,Kobayashi:2014eva, 
BenAchour:2018dap, Takahashi:2019oxz, Minamitsuji:2019shy, 
Babichev:2017lmw} (neutron star models were also developed 
\cite{Kobayashi:2018xvr}). These stealth black hole solutions with 
constant kinetic term $\bar{X}$ are plagued by the problem of the strong 
coupling of scalar perturbations \cite{deRham:2019gha}.

A step beyond stealth solutions was taken in Ref.~\cite{BenAchour:2019fdf} 
by using disformal transformations as a solution-generating technique to 
map seed solutions into new DHOST solutions. The new solutions thus 
obtained have non-constant kinetic term $\bar{X}$. The metric is transformed as 
\be
g_{ab} \rightarrow \tilde{g}_{ab} = A\left( \phi, \bar{X} \right) g_{ab} 
+B\left( \phi, \bar{X}\right) \nabla_a\phi \nabla_b \phi \,.
\ee
For example, taking the FJNWBW geometry (\ref{becomes1}) and 
(\ref{becomes2}) as a seed, one obtains the static 
and spherically symmetric DHOST solution \cite{BenAchour:2019fdf} 
\begin{eqnarray}
d\tilde{s}^2 &=& \frac{1}{A(r)} \Big\{ -\left(1-\frac{r_S}{r} 
\right)^{\gamma} dt^2 + 
\left(1-\frac{r_S}{r} \right)^{-\gamma} \nonumber\\
&&\nonumber\\
&\, & \times \left[ 
1-\frac{q^2}{4\pi r^4} 
\left(1-\frac{r_S}{r} \right)^{\gamma-2} B\left(\phi(r), \bar{X}(r) \right) 
\right] dr^2 \nonumber\\
&&\nonumber\\
&\, & +r^2 \left(1-\frac{r_S}{r} \right)^{1-\gamma} d\Omega_{(2)}^2 
\Big\} \,,\\
&&\nonumber\\
\bar{X}(r) &=& \frac{q^2}{4\pi r^4} \frac{A(r) \, \left(1-\frac{r_S}{r} 
\right)^{\gamma-2}}{ 
1-\frac{q^2}{4\pi r^4} 
\left(1-\frac{r_S}{r} \right)^{\gamma-2} B\left(\phi(r), \bar{X}(r) \right)} \,,
\end{eqnarray} 
where $q$ and $r_S>0$ are constants. This solution does not admit 
horizons and, like FJNWBW, it exhibits a naked singularity. In general, 
by disformally transforming a naked singularity one cannot obtain a black 
hole, but produces another naked singularity \cite{BenAchour:2019fdf}.

Other analytic black hole solutions with nonconstant $\bar{X}$ in quadratic 
DHOST theories, which are static and spherically symmetric but not  
Schwarzschild-de Sitter, were found in Ref.~\cite{Minamitsuji:2019tet} by 
direct integration of the equations of motion, without using the 
disformal transformation generating technique (see also 
\cite{Takahashi:2019oxz, Minamitsuji:2019shy}). For 
example, in the quadratic DHOST gravity with
\begin{eqnarray}
f_0(\bar{X})&=&\eta_1 \bar{X}^n \,, \quad f_2(\bar{X})=\zeta \,,\\
&&\nonumber\\
\alpha_{1} (\phi, \bar{X})&=& \frac{\beta_1 \zeta}{\bar{X}} \,, \quad
\alpha_{2} (\phi, \bar{X}) = -\alpha_{1} (\phi, \bar{X}) \, ,
\quad \,
\alpha_{3} (\phi, \bar{X})= \frac{\gamma_1 \zeta}{\bar{X}^2} \,, \qquad\\
\nonumber\\
\alpha_{4} (\phi, \bar{X}) &=& -\frac{
\left[ -12\beta_1^2 +16\beta_1^3 -12\beta_1\gamma_1 +\gamma_1(8+\gamma_1) 
\right]\zeta}{
8(1-\beta_1)^2\bar{X}^2}       \,,\\
\nonumber\\
\alpha_{5} (\phi, \bar{X}) &= & \frac{ \left(2\beta_1-\gamma_1\right)\left( 
2\beta_1^2 
-4\gamma_1 +3\beta_1 \gamma_1\right) \zeta}{
8\left(1-\beta_1\right)^2 \bar{X}^3} \,,
\end{eqnarray}
with $\eta_1, \beta_1, \gamma_1$ constants, $n>0$, $\eta_1>0$, 
$\beta_1<0$, $4n(1-\beta_1)>3(\gamma_1 -2\beta_1)>0$, and all the other 
coefficients vanishing, the black hole is \cite{Minamitsuji:2019tet} 
\begin{eqnarray}
ds^2 &=& - r^{ 2\sqrt{ \frac{ \beta_1}{\beta_1 -1}}} \left( 
1-\frac{2m}{r^{1+\sqrt{\frac{\beta_1}{\beta_1-1}} } }\right) dt^2 
+
\left( 1-\frac{2m}{r^{1+\sqrt{\frac{\beta_1}{\beta_1-1}} } }\right)^{-1} 
dr^2 \nonumber\\
&&\nonumber\\
&\, & +r^2 d\Omega_{(2)}^2   \,,\label{BH1-1}\\
&&\nonumber\\
\bar{X} &=& \left[ 4\zeta \frac{-\beta_1+ \sqrt{\beta_1(\beta_1 -1)}}{3\eta_1 \, 
r^2} 
\right]^{1/n} \,.
 \label{BH1-2}\end{eqnarray}

In the other DHOST theory with
\begin{eqnarray}
f_0(\bar{X}) &=& - \Lambda \,, \quad\quad f_2(\bar{X})= \zeta -\beta_2 \bar{X}^n \,, \\
&&\nonumber\\
\alpha_{1} (\phi, \bar{X})&=&-\alpha_{2} (\phi, \bar{X})=0 \, , \qquad \alpha_{3} (\phi, \bar{X})= \gamma_2 \bar{X}^{n-2} \,, \\
&&\nonumber\\
\alpha_{4} (\phi, \bar{X}) &=& \frac{ -48n^2\beta_2^2 \bar{X}^{2n} +8n\beta_2\gamma_2 
\bar{X}^{2n} 
+ \gamma_2\bar{X}^n \left[ \bar{X}^n\left( -8\beta_2+\gamma_2\right) +8\zeta\right]}{
8\bar{X}^2 \left( \beta_2 \bar{X}^n-\zeta\right)} \,,\nonumber\\
&&\\
\alpha_{5} (\phi, \bar{X}) &=& \frac{ \gamma_2 \bar{X}^{2n-3} \left( 4n\beta_2 
-\gamma_2\right)}{2\left( \beta_2 \bar{X}^n -\zeta\right)}  \,,
\end{eqnarray}
where $\Lambda$, $\beta_2>0$, $\gamma_2 =4n\beta_2/\sqrt{3} >0$, $\zeta$, 
and $n $ are constants satisfying $-\left(3\gamma_2 -8n\beta_2 
\right)\Lambda>0$, the DHOST black hole is \cite{Minamitsuji:2019tet}
\begin{eqnarray}
ds^2 &=& - r^{ 2(\sqrt{3}-1)} \left( 
1-\frac{2m}{r^{2+\sqrt{3} } } \right) dt^2 
+
\left( 1-\frac{2m}{r^{2+\sqrt{3} } } \right)^{-1} dr^2 
+r^2 d\Omega_{(2)}^2   \,,\nonumber\\
&& \label{BH2-1}\\
\bar{X} &=&\left( \frac{  4\zeta +\left( 2\sqrt{3}-3\right) \Lambda r^2}{ 
4\beta_2 } \right)^{1/n} \,. \label{BH2-2}
\end{eqnarray}
A third solution for a different quadratic DHOST model is provided in 
\cite{Minamitsuji:2019tet}. 

In all these models, the equation locating the event horizons $\nabla^c r 
\nabla_c r=g^{rr}=0$ (where $r$ is the areal radius) has a single root, 
identifying clearly a black hole horizon (as is clear from the dimensions, 
the constant $m$ is not the black hole mass, but a power of it). However, 
these black holes are not asymptotically flat \cite{Minamitsuji:2019tet}, 
which can be seen as a consequence of the fact that the black hole 
exterior is not vacuum but hosts the scalar hair. Stability 
against odd 
parity perturbations is studied in \cite{Minamitsuji:2019tet}, with the 
result that the black hole (\ref{BH1-1}) is stable, while the 
geometry~(\ref{BH2-1}) is not.

The search for static and spherical DHOST solutions continues ({\em e.g.}, 
\cite{Alinea:2020sei, Gorji:2020bfl, Khoury:2020aya, Khodadi:2020jij, 
Deffayet:2020ypa, Takahashi:2020hso}). Non-constant $\bar{X}$ solutions are 
useful when attempting to confront DHOST theories with observations, which 
has already led to significant restrictions on these theories.  The reader 
should be warned that this area of research is very recent and  
developing fast.

%

\section{Summary and conclusions} 
\label{sec:14}

It is apparent from our analysis that, even in the simplest theory of 
relativistic gravity, GR, and under an assumption as simplifying as 
spherical symmetry (which is indeed, an oversimplification in most 
situations of astrophysical interest), only a restricted number of  
analytic solutions have 
been discovered, which leads to the usefulness of compiling a small 
catalogue. This richness is due to the variety of matter sources and of 
possible boundary conditions (asymptotically flat, FLRW, de Sitter, {\em 
etc.}). What is more, our current knowledge of exact solutions is 
incomplete and there are very few cases in which a geometry can be called 
``generic'' in some sense, like the Schwarzschild solution in vacuum and 
asymptotically flat GR. Here we have refrained from entering into great 
detail about the concept of ``generic'', which will leave the more 
mathematically oriented reader dissatisfied, and we have focussed on the 
physical aspects of these solutions. Likewise, we have privileged the 
description of the geometries examined in the most well-known and used 
coordinate systems and we have not entered the discussion of special 
algebraic properties and classifications of the solutions examined, as 
done instead in the well known 
book by Stephani {\em et al.} \cite{StephaniKramerMacCallumHoenselaersHerlt} 
discussing solutions of the Einstein equations of many different 
natures. This kind of book-length analysis is beyond the scopes of the 
present work and, although necessary for an invariant characterization 
of the geometries, ultimately does not serve well the purposes of the 
physicist 
or the cosmologist. Some of the spacetimes described here have some kind 
of ``generic'' character as described in the text, whereas those that do 
not have been reported as interesting examples that still show 
certain properties of the non-linear theories that they solve. A 
curious feature of certain geometries is that they solve the field 
equations of different theories, or of the same theory with different 
matter sources. This largely unexplored (beyond the 
KSdS and the McVittie spaces) aspect deserves more 
attention in future analyses.

Most likely, some of the analytic solutions presented here do not describe 
objects found in nature. Nevertheless, they provide useful insight into 
the nature of the field equations that they satisfy, for which it is 
difficult to build intuition due to their non-linearity.  Other times, 
these geometries are useful as toy models for theoreticians (for example, 
the McVittie and non-rotating Thakurta solutions as toy models for 
primordial black holes \cite{Boehm:2020jwd, RuizMolinaLima20}, the subject 
of 
renewed attention in relation with the dark matter problem and black hole 
mergers detected by {\em LIGO}).

Given the enormous motivation for the study of alternative theories of 
gravity coming from cosmology \cite{AmendolaTsujikawabook}, quantum 
corrections to gravity \cite{UtyamaDeWitt62, Stelle77, Stelle78}, the 
low-energy limit of string theories \cite{Callan85,FradkinTseytlin85}, and 
the experimental efforts to test gravity at various astrophysical scales, 
we have included scalar-tensor \cite{BD,Bergmann68, Wagoner70, 
Nordtvedt70}, as well as Horndeski gravity and DHOST 
theories beyond Horndeski in our discussion. The reasons for 
choosing those theories include the fact that scalar-tensor gravity is the 
prototypical and longest known alternative to GR. Furthermore, it 
includes $f({\cal R})$ gravity as a subclass which is extremely popular 
to explain the current acceleration of the universe without invoking dark 
energy \cite{CCT, CDTT, review1, review2, review3} and is also intimately 
connected with early universe inflation \cite{Starobinsky80}.  Whether 
$f({\cal R})$ gravity is a toy model or, as many authors believe, a 
realistic model, or an effective field theory \cite{BelSirousseZia85, 
Simon90, DeDeoPsaltis08}, is an open question.
 
Horndeski theory is arguably the most well studied recent extension of 
``simple'' scalar-tensor gravity which is still well motivated from the 
physical point of view and useful for cosmology, while much work is 
presently being done on the more recent DHOST theories, for which it is 
fair to say that we do not yet have a complete picture. The recent 
opening of gravitational wave astronomy \cite{LIGO1, LIGO2, LIGO3} and the 
imaging of the near-horizon region of black holes \cite{EHT1, EHT2, EHT3, 
EHT4, EHT5, EHT6}, together with the birth of multi-messenger astrophysics 
\cite{MMA} offer new tools to explore the nature of gravity, and a clear 
picture of what one should expect beyond Einstein theory could help or 
even guide these observational efforts.

The presence of a scalar degree of freedom, either as a matter field or as 
part of the gravitational field, has been a recurring element in this 
work. Unfortunately, even for this simplest degree of freedom of field 
theory, our knowledge is still limited.  Matter scalar fields in GR can 
collapse to black holes, which are then indistinguishable from any other 
black hole with the same mass and angular momentum, according to the 
no-hair theorems. The exact spherical, static, and asymptotically flat 
solution of the Einstein equations which does not contain a black hole, 
the FJNWBW solution, exhibits a naked singularity. Generally speaking, 
asymptotically flat analytic solutions of GR with non-trivial scalar field 
configurations tend to generate geometries that contain only naked 
singularities or wormhole (horizon) throats, but not black hole horizons. 
This conclusion seems to apply to both static and time-dependent 
geometries, when gravitational collapse does not occur. Non-asymptotically 
flat solutions can have a rich structure of apparent horizons which can 
change in number and usually appear/disappear in pairs. The same 
conclusions apply, in general, to scalar-tensor gravity, although there is 
no specific theorem to this regard. A sore point is the lack of general 
theorems for GR or scalar-tensor (including Horndeski and DHOST)  
solutions which are spherical and asymptotically FLRW.

Changing source type, many exact solutions sourced by perfect or imperfect 
fluids are known, especially in the form of perfect fluid stellar 
configurations, but only relatively few are physical. Then, there are 
families of solutions of GR and scalar-tensor gravity interpreted as 
central objects embedded in FLRW universes, some of which are black 
holes (as defined by their apparent horizons) at least part of the time. 
Other solutions contain naked singularities for at least part of the 
history of the universe. Apparent horizons that are also the throats of 
(non-traversable) wormholes are ubiquitous. 

We have pointed out existing relations between many pairs of geometries 
listed here. Usually, these relations are conformal dualities, due to the 
frequent use of conformal transformations as solution-generating 
techniques. Other times, researchers have sought to generalize previously 
known solutions, which results in some of the spacetimes examined being  
special cases of previous ones.

In spite of the fact that, in the space available, it is impossible to 
make justice to all the physical and mathematical aspects involved, or  
even to compile a complete list of all related references, this review is 
meant to provide a reference for the researcher and an introduction to the 
subject for the less experienced reader. A review can never replace a 
thorough study of the references relevant to a particular aspect or a 
specific geometry. The usefulness of simpler situations is that they can 
be kept in mind and used as paradigms, or for contrast, when studying more 
complicated ones. They are particularly useful in the attempt to discover 
what a new theory of gravity adds to the existing ones, and as 
examples or counterexamples. We have learned 
much physics in the past decades of studying analytic solutions of GR and 
scalar-tensor gravity but, remarkably, there are many areas in which we 
still know very little. We hope that highlighting them will renew research 
efforts that will bring a better understanding of important aspects of 
gravity.

\section*{Acknowledgments}

We are grateful to two referees for numerous helpful suggestions and comments. We are 
indebted with many colleagues---too many to list---for discussions, comments, and references 
and to Vincenzo Vitagliano for drawing Fig.~\ref{figA}. This work is supported by the 
Natural Sciences \& Engineering Research Council of Canada (Grant No. 2016-03803 to V.F.) 
and by Bishop's University. This research was supported, in part, by Perimeter Institute for 
Theoretical Physics. Research at Perimeter Institute is supported by the Government of 
Canada through the Department of Innovation, Science and Economic Development Canada and by 
the Province of Ontario through the Ministry of Research, Innovation and Science.

A.G. is supported by the European Union's Horizon 2020 research and 
innovation programme under the Marie Sk\l{}odowska-Curie Actions (grant 
agreement No. 895648 -- CosmoDEC). His work has been carried out in the 
framework of the activities of the Italian National Group for 
Mathematical Physics [Gruppo Nazionale per la Fisica Matematica (GNFM), 
Istituto Nazionale di Alta Matematica (INdAM)].

\bibliography{ssreview_rev2}{}
\bibliographystyle{elsarticle-num}

\end{document}